\documentclass[twocolumn,nofootinbib]{aastex631}
\usepackage{graphicx}	
\usepackage{amsmath}
\usepackage{amssymb}
\usepackage{subfigure}
\usepackage{color}
\usepackage{txfonts}
\usepackage{tabto}
\usepackage{braket}
\usepackage{xcolor}
\usepackage{physics}

\newcommand\simlt{\lower.5ex\hbox{$\; \buildrel < \over \sim \;$}}
\newcommand\simgt{\lower.5ex\hbox{$\; \buildrel > \over \sim \;$}}

\usepackage{slashed}
\usepackage{lipsum}
\usepackage{subfigure}
\usepackage{multirow}
\usepackage{amsmath}
\usepackage{array} 
\usepackage{varwidth} 
\usepackage{romannum}
\usepackage{graphicx}
\usepackage{bm}
\usepackage{times}

\usepackage{soul}

\usepackage{aas_macros}
\usepackage{hyperref}
\usepackage{breqn}
\hypersetup{
     colorlinks   = true,
     citecolor    = blue,
     urlcolor     = blue,
     linkcolor    = blue
}

\newcommand{\be}{\begin{equation}}
\newcommand{\ee}{\end{equation}}
\newcommand{\bea}{\begin{eqnarray}}
\newcommand{\eea}{\end{eqnarray}}

\begin{document}

\title{Reacceleration of Galactic Cosmic Rays Beyond the Knee at the Termination Shock of a Cosmic-Ray-Driven Galactic Wind }

\author[0000-0002-3954-2005]{Payel Mukhopadhyay}
\affiliation{SLAC National Accelerator Laboratory, Stanford University, Menlo Park, CA 94025, USA; \href{mailto:pmukho@berkeley.edu}{\rm{pmukho@berkeley.edu}}} 
\affiliation{Physics Department, Stanford University, Stanford, CA 94305, USA}
\affiliation{Department of Physics, University of California, Berkeley, CA, 94720, USA}

\author[0000-0003-0543-0467]{Enrico Peretti}
\affiliation{Niels Bohr International Academy, Niels Bohr Institute, University of Copenhagen, Blegdamsvej 17, DK-2100 Copenhagen, Denmark}

\author[0000-0001-9011-0737]{No\'emie Globus}
\affiliation{Department of Astronomy and Astrophysics, University of California, Santa Cruz, CA 95064, USA}
\affiliation{Astrophysical Big Bang Laboratory, RIKEN, Wako, Saitama, Japan}

\author[0000-0001-7763-4405]{Paul Simeon}
\affiliation{Independent Researcher}

\author[0000-0002-1854-5506]{Roger Blandford}
\affiliation{Physics Department, Stanford University, Stanford, CA 94305, USA}
\affiliation{Kavli Institute for Particle Astrophysics and Cosmology, 452 Lomita Mall, Stanford, CA 94305, USA}

\begin{abstract}

The origin of cosmic rays above the \textit{knee} in the spectrum is an unsolved problem. We present a wind model in which interstellar gas flows along a non-rotating, expanding flux tube with a changing speed and cross-sectional area. Cosmic rays from Galactic sources, such as supernova remnants, which are coupled to the plasma via Alfv\'{e}n waves, provide the main pressure source for driving this outflow. These cosmic rays are then subject to diffusive shock reacceleration at the Galactic wind termination shock, which is located at a distance $\sim200\,{\rm kpc}$. Some of the highest-energy reaccelerated particles propagate upstream against the wind and can contribute to the PeV--EeV range of the spectrum. We analyze the conditions under which efficient reacceleration can occur and find that rigidities $\sim$ 10--40~PV can be obtained and that the termination shock may account for half of the proton spectrum measured in IceCube/IceTop experiment. The highest-energy particles that escape downstream from our termination shock, and similar shocks surrounding most galaxies, can be further accelerated by intergalactic shock fronts.

\end{abstract}

\section{Introduction}
\label{sec:intro}

The origin of cosmic rays (CRs) is a long-standing open question since their discovery. 
The most established interpretation ascribes a Galactic origin to CRs populating the spectrum up to the ``knee'', with rigidity (momentum per unit charge) $R=R_{15}\,{\rm PV}$ satisfying $R_{15}\sim 3$, where the spectrum steepens \citep{2021:TIBET,2021:lhasso,2020:hawcsnr,2007:HESS_SNR}. By contrast, cosmic rays with energy above the ``ankle'' in the spectrum at $R_{15}\gtrsim3000$, where the spectrum flattens, are widely believed to have an extragalactic origin because the average Galactic magnetic field is not strong enough to confine such energetic particles. Furthermore, the arrival flux does not feature any relevant anisotropy correlating with the Galactic disk \citep[see e.g.,][]{PAO_2018}.

Similar to extragalactic cosmic rays, the origin of cosmic rays between the knee and the ankle, also referred to as the ``shin'' of the cosmic-ray spectrum, is highly uncertain. We call these ``intermediate-energy'' cosmic rays. The most widely discussed sources are pulsars, magnetars, circumgalactic and extragalactic shock fronts, and unidentified ``pevatrons'' \citep{Abeysekara:2021yum,2021:lhasso,2021:Amenomori,2016:HESS,2021:Ore}.

In this paper, we focus on one particular type of source of high energy CRs that could possibly populate the shin region of the spectrum: the Galactic wind termination shock (GWTS). We consider Galactic cosmic rays as being injected by sources located in the Galactic disk from the lowest energies up to the knee. These cosmic rays are assumed to escape the disk and be advected in the Galactic wind on timescales shorter than the Galactic rotation period. The spectral index of the escaping cosmic rays is essentially that produced by the Galactic accelerators like supernova remnants (SNRs), not what is observed at Earth, assuming the distribution of Galactic cosmic rays is approximately stationary, with the time-averaged source flux equal to the escaping flux. The Galactic wind advects cosmic rays in an accelerating flow that passes through a critical point and becomes supersonic. Eventually, this flow will pass through a strong shock front, behind which it will decelerate and eventually become incorporated into the intergalactic medium. The shock reaccelerates Galactic cosmic rays by diffusive shock (re)acceleration (DSA), contributing to the shin region of the spectrum. Some of these higher-energy particles will propagate upstream back to the Galactic disk, affecting the spectrum, composition, and anisotropy observed at Earth.

Of course, all other galaxies, especially disk galaxies, where there is active star formation and a relatively high supernova rate, should behave similarly to our Galaxy. The high-energy cosmic rays that are advected downstream from the termination shock escape their original galaxy and provide an input for larger shock fronts. Galaxies with active nuclei, notably starburst galaxies, likely emit an outsized flux of these intergalactic cosmic rays \citep[see e.g.,][]{Starburst_Zhang,Starburst_wind}. The largest intergalactic shock fronts, especially those associated with clusters of galaxies and the filaments that connect them in the ``cosmic web,'' may be the source of the extragalactic cosmic rays. This possibility will be discussed in a future publication (Simeon et al. in preparation). 

Galactic wind termination shocks were first proposed as a cosmic ray acceleration site in the papers by \cite{1985:Jokipii} and \cite{198:Jokipii}. Since then, the idea has been explored by different authors \citep{2004:Volk,2006:Zirakashvilli,2016:Thoudam,2017:Bustard,Merten:2018qoa}. The injected cosmic rays that are accelerated at the GWTS can either come from particles injected at the shock front itself \citep{2017:Bustard,Merten:2018qoa} or from Galactic cosmic rays that are transported along the Galactic wind \citep{2006:Zirakashvilli,2016:Thoudam}. \citet{2006:Zirakashvilli} and \citet{2016:Thoudam} argued that cosmic rays reaccelerated at the GWTS that propagate back to the disk can fully explain the observed all-particle spectrum at Earth between the knee and the ankle region. These studies were limited in their treatment of the cosmic-ray transport equation. Specifically, the maximum energies achieved in the reacceleration process were not self-consistently computed from the transport equation but were prescribed in the solution. \citet{2016:Thoudam} treated the maximum energies as free parameters chosen such that the spectrum of reaccelerated particles propagating back to the disk reasonably agrees with the measured all-particle spectrum. \citet{2006:Zirakashvilli} treated the GWTS as a reflecting boundary and defined the maximum energy as the highest energy beyond which the particles can cross the termination shock diffusively. Additionally, these studies neglected the transport equation downstream of the shock and did not explore the possibility of a fraction of the reaccelerated cosmic rays escaping into the intergalactic medium, seeding it with intermediate-energy cosmic rays.

In this paper, we consistently solve the CR transport equation both upstream and downstream of the GWTS. We find the maximum energies of the reaccelerated cosmic rays self-consistently as a solution to the transport equation, instead of prescribing it externally. We also calculate the flux of reaccelerated cosmic rays that propagate back to the Galaxy, as well as the flux of particles escaping into the intergalactic medium. Additionally, we provide an analysis of the conditions under which efficient CR reacceleration can happen at the GWTS. To this end, we explore a physically motivated model of the Galactic wind where the wind is driven by cosmic rays and the pressure of self-excited Alfv\'{e}n waves streaming away from the disk. This model follows the treatment presented by \citet{1991:Breitschwerdt}, albeit with an updated treatment of the dark-matter potential of the Milky Way. We will call this model, the ``Cosmic Ray--Alfv\'{e}n wave driven'' wind model. We then solve the transport equation by prescribing the velocity profiles obtained from the wind model. The transport equation is computed for two different prescriptions of the upstream diffusion coefficient to test the conditions under which efficient reacceleration can occur and to understand what conditions are suitable for efficient reacceleration of cosmic rays at the GWTS. In this model, we neglect the rotation of the disk for computing the wind. In a follow-up paper (Blandford et al. 2023 in preparation), we will develop our model, by including rotation through a magnetocentrifugally-driven wind that becomes a significant part of the mass, energy, and angular momentum budget of the interstellar medium and which may accelerate cosmic rays more efficiently to higher energy than in the present model while being subject to significant observational constraints.

We review the current state of measurements in Sec.~\ref{sec:observations}. We describe a relatively simple, though quite general, formalism for describing the outflow in Sec.~\ref{sec:model1CR}, emphasizing the choice that we argue is most appropriate for the solar neighborhood. We also describe the transport of Galactic CRs in the wind and the reacceleration of these CRs at the GWTS. In Sec.~\ref{sec:results_CR}, we present our results. Finally, in Sec.~\ref{conclusions_CRs}, we summarize and present our conclusions

\section{Cosmic-Ray Observations}
\label{sec:observations}

It is widely believed that cosmic rays below $\sim10^{15}$~eV are accelerated by Galactic accelerators, possibly SNRs, and that the ankle of the cosmic-ray spectrum marks the end of the transition from Galactic to  extragalactic cosmic rays \citep[e.g.,][]{globus2015}. The explanation of the intermediate-energy cosmic rays is still unsolved but increasingly constrained by the many detailed observations of the past decade. We summarize some of these observations below. 

\paragraph{Spectrum} The AMS-02, CALET, DAMPE, HAWC, and ISS-CREAM experiments have measured the individual spectra of various intermediate energy species with unprecedented accuracy \citep{2015:AMS,2017:AMS,2020:AMS,2019:CALET,2021:CALET,Alemanno:2021gpb,HAWC:2021zmq}. These spectra show  ankles and breaks. 
A spectral softening, or a ``knee," can be caused by the maximum energy of the acceleration process, a leakage from the Galaxy, or a combination of both. An ``ankle" is a hardening of the cosmic-ray spectrum, a natural feature marking the transition between two components of the spectrum.

Above a few GeV/nucleon, the energy spectrum follows a power law proportional to $E^{-2.7}$. A first softening in the spectrum occurs at around $1.2 \times 10^{13}$~eV \citep{Alemanno:2021gpb}. It is then followed by a hardening at $\sim 10^{14}$~eV \citep{HAWC:2021zmq}. The cosmic-ray spectrum then has another stronger knee at around $3 \times 10^{15}$~eV --- a well-known feature of the cosmic-ray spectrum known since the 1950s. Beyond that energy, the cosmic-ray spectrum then follows a power law proportional to $E^{-3.0}$. At $\sim10^{17}$~eV, the spectrum shows another softening only in the heavy component, the so-called ``heavy knee" \citep{apel2011}. At the same energy of about $\sim10^{17}$~eV, the light component (proton and helium) shows a hardening, a ``light ankle" \citep{apel2013}, that has been interpreted as the beginning of the light extragalactic component of cosmic rays \citep{globus2015}. At $4 \times 10^{18}$~eV, the total cosmic-ray spectrum gets harder again. This final ankle marks the end of the transition between Galactic and extragalactic cosmic rays.

\paragraph{Composition}
Several experiments, e.g., ARGO-YBJ, KASCADE-Grande, Tunka, IceCube, and IceTop \citep[e.g.,][]{andeen2019,kang2019}, have measured the composition of cosmic rays between the knee and the ankle of the spectrum. The average composition is light (predominantly protons and helium nuclei) at the knee, i.e., at energy $\sim10^{15.5}$~eV. Above the knee, the average composition becomes heavier as a function of energy up to about $10^{17}$~eV, and then, although the statistical errors become significant at these energies, the average mass seems to become lighter up to the ankle at $\sim4$~EeV. The different interaction models predict a different average mass, and the differences increase with primary energy. This trend is naively expected when the maximum energy attained during the acceleration process is proportional to the nuclear charge; if the proton knee occurs at few PeV, then the iron knee would occur at around $10^{17}$~eV, which is consistent with the evolution of the composition. The heavy knee reported by KASCADE-Grande at $10^{17}$~eV could mark the end of the heavy component from the SNR contribution. 

 \paragraph{Anisotropy} \citet{Ahlers:2019} reports some hints of intermediate-scale anisotropy at 33~PeV, although the level of anisotropy in the PeV--EeV energy range  is low. The upper limit to the amplitude of the first harmonic (dipole) is $10^{-3}$ at $2\times 10^{14}$~eV and increases to $\sim10^{-2}$ at $10^{17}$~eV. KASCADE-Grande,  ESA-TOP, IceCube, and IceTop observed a change of the phase of the first harmonic in the direction of the Galactic center at energies of $2\times 10^{14}$~eV. After that, the phase remains roughly flat until $\sim10^{17}$~eV and then changes again (to $\sim100$ degrees). The first phase change could indicate a transition between local sources located in the neighboring arms to a more global contribution of the whole Galaxy, while the second phase change could be the sign of an extragalactic origin of cosmic rays. 

\paragraph{Impact of reaccelerated CRs on the observables} 

In this paper, we will compute the flux of Galactic cosmic rays that are reaccelerated at the GWTS and propagate back to the Galactic disk. We will compare how the spectrum of the reaccelerated CRs compare with the measured spectrum at the disk beyond the knee. We will see that these backstreaming particles can form spectral bump features as a mark of transition between the Galactic and the GWTS components. We will find that these bump features are consistent with existing measurements of different elements and that they can be further constrained in future by more precise experiments. 

\section{Cosmic ray--Alfv\'{e}n wave driven wind model}
 \label{sec:model1CR}
 
\subsection{General Considerations}
\label{subsec:generalmodel1}

Galactic winds driven by thermal pressure, cosmic rays, or magnetic fields have been studied for more than 50 years \citep{1968:Burke,1971:Johnson,1975:Ipavich,Chevalier:1985,1991:Breitschwerdt,2008:Everett,2016:Recchia,2018:Mao}. If such a wind for the Galaxy were to be thermally driven, the sound speed at the solar circle would be $v_{\rm s} \sim \sqrt{\frac{2}{3}} v_{\rm esc} \sim 470~{\rm km\,s}^{-1}$, where $v_{\rm esc}$ is the local escape speed, which is $\sim$580~km~s$^{-1}$ at the solar circle \citep{2018:Monari,Necib:2021vxr}. This sound speed corresponds to a temperature higher than $10^7$~K, which is very high considering the warm interstellar medium (ISM) temperature of the Galaxy is between $10^5$~K and $10^6$~K. Therefore, the scenario of a purely thermally driven wind in the Galaxy is in tension with the observations of the ISM.

One alternative is that cosmic rays assist in driving the wind, a possibility that has been discussed by a number of authors \citep{1975:Ipavich,1991:Breitschwerdt,2008:Everett,2012:Dorfi,2016:Recchia,2019:dorfi,2018:Mao}. One plausible scenario, first described by \citet{1991:Breitschwerdt}, is that the cosmic-ray sources in the Galactic disk produce energetic particles that cannot freely escape from the Galaxy but rather amplify Alfv\'{e}n waves \citep{1974:Wentzel}. Such waves lead to an efficient coupling of the thermal gas to energetic particles through the cosmic-ray resonant streaming instability \citep{1967ApJ...147..689L,1969ApJ...156..445K}, and the pressure gradient of cosmic rays drives a wind outflow. These winds start at subsonic speeds of a few km~s$^{-1}$ at the base of the wind and become supersonic at $\mathcal{O}(10 ~\mathrm{kpc})$ distances. The flow forms a wind termination shock at $\mathcal{O}(10^2 ~\mathrm{kpc})$ distances \citep{1991:Breitschwerdt}. We adopt the simple model of the wind used by \citet{1991:Breitschwerdt} and modify it by including an updated dark-matter potential of the Galaxy, consistent with GAIA observations.

\subsection{Wind geometry and dynamics}
\label{B1991:geom}

Our reference wind model treats cosmic rays as a rarefied plasma in dynamic interaction with the magnetic field in the interstellar medium. This approximation holds because rapid pitch-angle scattering of cosmic rays on the fluctuations (waves) of the field makes the cosmic-ray momentum distribution isotopic to the lowest order in the wave frame. The residual cosmic-ray diffusion thorough the gas, along the mean magnetic field, adds itself to any convective motions of the thermal plasma. Propagation of waves along the mean magnetic field will, in addition, give rise to a drift speed of cosmic rays equal to the wave phase velocity. The cosmic ray motion is therefore composed of convection, wave drift, and diffusion.

\begin{figure*}
    \centering
    \includegraphics[width=0.8\textwidth]{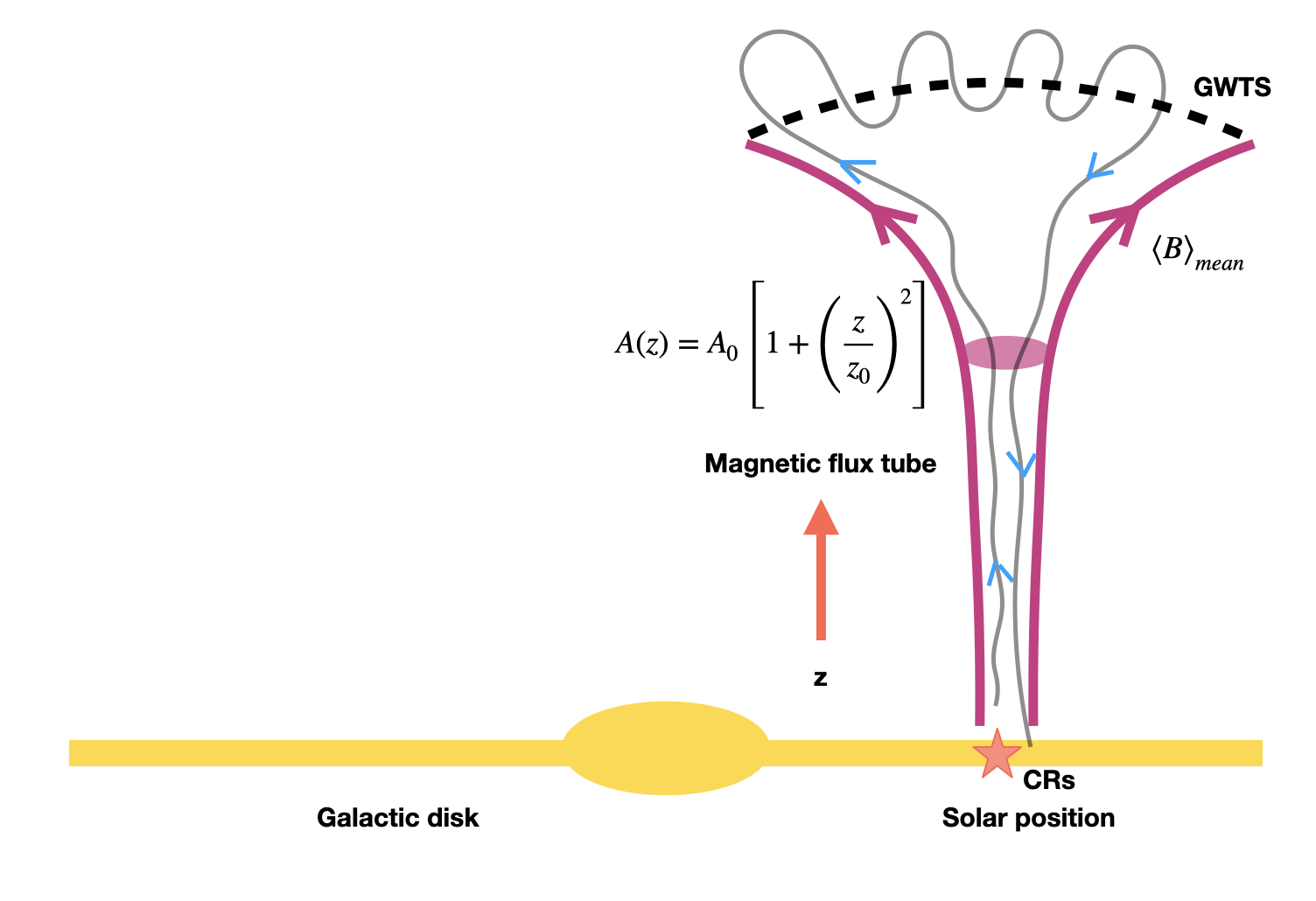}
    \caption{Sketch of the problem under consideration. The wind geometry is shown as a magnetic flux tube. Cosmic rays produced by Galactic sources in the disk travel with the wind up to the Galactic wind termination shock (GWTS), where they get reaccelerated to higher energies.}
    \label{fig:sketch}
\end{figure*}

When the streaming speed of the cosmic rays through the ambient medium exceeds the Alfv\'{e}n speed, 
$v_{\mathrm{A}} = {B}/\sqrt{{4\pi \rho}}$
($B$ is the magnetic field and $\rho$ is the ion density of the medium), there is a quasi-linear instability in which the amplitude of the resonant Alfv\'{e}n waves grows exponentially with time until their amplitude is strong enough to limit the cosmic-ray streaming speed to slightly more than the Alfv\'{e}n speed \citep{1967ApJ...147..689L}. See \citet{1987:Blandford} for a review. The cosmic rays, scattering off these resonantly excited Alfv\'{e}n waves, inevitably transfer momentum to the background medium. This ensures that the pressure gradient of the cosmic rays accelerates the background fluid, which accelerates the wind in our model. This mechanism only works if the growth rate of these Alfv\'{e}n waves, $\Gamma_{\rm res}$, is faster than $\frac{v}{z}$, the inverse of the dynamical time scale of the problem. The growth rate of these waves is

\begin{equation}
    \Gamma_{\rm res} = \Omega \left[ \frac{v_{\rm stream}} {v_{\mathrm{A}}} - 1 \right] \frac{n_{\rm cr} }{ n_{\mathrm{i}}},
\end{equation}

\noindent where $n_{\rm cr} \sim 4\pi f p^3$ is an estimate of the total number of resonant comsic rays, and $n_{\mathrm{i}}$ is the background ion density \citep{1987:Blandford}. $v_{\rm stream}$ is the cosmic-ray streaming speed, $v_{A}$ is the Alfv\'{e}n speed, $\Omega$ is the non-relativistic gyrofrequency of the cosmic rays, and $n_{\rm cr}$ is the cosmic-ray number density, which is $\sim 10^{-12}$~cm$^{-3}$ for GeV cosmic rays, the primary drivers of the wind. Using $v_{\rm stream} = 2 ~v_{\rm A}$, $n_{\rm i} \sim 10^{-4}$~cm$^{-3}$, and $B \sim 2~\mu{\rm G}$, which are order-of-magnitude numbers from our calculation at $z \sim 10$~kpc, we find that the condition $\Gamma_{\rm res} \times z/v \sim 10^6 \gg 1$ is easily satisfied for $O($1~GeV) CRs. Accordingly, the cosmic-ray pressure is sufficient to excite Alfv\'{e}n waves to launch the wind, and this instability can grow faster than the dynamical time ($z/v$) of the problem for cosmic rays of energies up to $O(1~\mathrm{TeV})$. We note these waves could be damped by processes such as ion--neutral or nonlinear Landau damping. We neglect the former because a hot outflow plasma with $T \gtrsim 10^5$~K can maintain a high level of ionization. The latter is expected to become important when $\frac{\delta B}{B} \sim 1$, when quasilinear theory breaks down. In this paper, we neglect these damping processes and acknowledge the caveat that these processes could be important.

We assume the magnetic field, which is also the geometry of the flow, has a ``mushroom-type'' geometry \citep{1991:Breitschwerdt}, which emphasizes the concept of isolated open magnetic flux tubes, and can be conveniently represented by a one-dimensional flux tube model, as shown in Fig.~\ref{fig:sketch}. Edge-on observations of spiral galaxies like NGC 5775 reveal large scale magnetic fields of similar topology in their galactic halo \citep{2011A&A...531A.127S}. 3D numerical simulations of the magnetic field evolution in barred galaxies under the influence of cosmic ray driven dynamo find that the halo magnetic field can have strong vertical components, possibly transported by mass outflow from the galactic disk \citep{2011ApJ...733L..18K}.

This formulation accounts for a vertical flow geometry near the disk, which transitions to a spherical geometry adequate to describe the flow at altitudes greater than the disk radius, $R_{\mathrm{G}} \sim 15$~kpc. Following this discussion, the geometry for the system is physically represented by outwardly directed magnetic lines of force:
\begin{align}
    A(z) = A_0 \left[ 1 + \left(\frac{z}{z_0}\right)^2 \right],
\label{eq:hyperbola}
\end{align}
where $z$ is the altitude from the disk, and $z_0$ is the vertical scale beyond which the geometry becomes radial. The parameter $z_0$ is typically expected to be of the order of the Galactic radius \citep{1991:Breitschwerdt,2012:Dorfi}. $A(z)$ represents the area cross section of the flow as a function of $z$. 

The following dynamical equations describing the overall balance of mass, momentum, and energy are solved \citep{1991:Breitschwerdt}:
\begin{align}
&\nabla \cdot \left( \rho \textbf{v}\right) = 0 \label{eq:Br1}\\
&\nabla \cdot \left(\rho \textbf{v} \otimes \textbf{v}  + \left[P_{\mathrm{g}} + P_{\mathrm{c}} + \frac{\langle \left(\delta \textbf{B}\right)^2 \rangle} {8 \pi} \right] \cdot \textbf{I} \right) = - \rho \nabla \Phi \label{eq:Br2}\\
&\nabla \cdot \left( \rho \textbf{v} \left[ \frac{1}{2} v^2 + \frac{\gamma_{\mathrm{g}}}{\gamma_{\mathrm{g}} - 1} \frac{P_{\mathrm{g}}}{\rho} + \Phi \right] + \frac{\langle \left(\delta \textbf{B}\right)^2 \rangle} {4 \pi} \left[ \frac{3}{2} \textbf{v} + \textbf{v}_{\rm A} \right] \nonumber \right.\\
&~~\left.+ \frac{1}{\gamma_{\mathrm{c}} - 1} \left[ \gamma_{\mathrm{c}} P_{\mathrm{c}} (\textbf{v} + \textbf{v}_{\rm A} ) - \overline{D} \nabla P_{\mathrm{c}}\right] \right) = 0 \label{eq:Br3}
\end{align}
\begin{align}
&\nabla \cdot\left( \frac{\gamma_{\mathrm{c}}}{\gamma_{\mathrm{c}} -1 } (\textbf{v} + \textbf{v}_{\rm A}) P_{\mathrm{c}} - \frac{\overline{D}}{\gamma_{\mathrm{c}} -1} \nabla P_{\mathrm{c}}\right) = (\textbf{v} + \textbf{v}_{\rm A}) \nabla P_{\mathrm{c}}\label{eq:Br4}\\
&\nabla \cdot \left( \frac{\langle \left(\delta \textbf{B}\right)^2 \rangle} {4 \pi} \left[ \frac{3}{2} \textbf{v} + \textbf{v}_{\rm A}\right] \right) = \textbf{v} \nabla \left( \frac{\langle \left(\delta \textbf{B}\right)^2 \rangle} {8 \pi} \right) - \textbf{v}_{\rm A} \nabla P_{\mathrm{c}}\label{eq:Br5}\\
&\nabla \cdot\textbf{B} = 0.\label{eq:Br6}
\end{align}
\noindent Here, $\rho$ denotes the gas density, $\textbf{v}$ denotes the outflow gas velocity vector, $\textbf{v}_{\rm A}$ is the Alfv\'{e}n speed, $\textbf{B}$ is the mean magnetic field, $P_{\mathrm{c}}$ and $P_{\mathrm{g}}$ are the pressures of the cosmic rays and gas, respectively, and $\Phi$ is the gravitational potential. $\langle \left(\delta \textbf{B}\right)^2 \rangle/(8 \pi)$ is the pressure contribution of the magnetic field fluctuations. These fluctuations are generated via the resonant streaming instability of cosmic rays streaming at roughly the Alfv\'{e}n speed with respect to the background medium. In the rest of the text, the pressure contribution due to magnetic field fluctuations will be denoted as $P_{\mathrm{w}} = \langle \left(\delta \textbf{B} \right)^2 \rangle /(8 \pi)$. The adiabatic index of the thermal gas is $\gamma_{\mathrm{g}} = 5 / 3$, and the adiabatic index of the cosmic rays is $\gamma_{\mathrm{c}} = 4 / 3$. The first three equations (\ref{eq:Br1}, \ref{eq:Br2}, and \ref{eq:Br3}) are the overall conservation of mass, energy, and momentum. Eq.~\ref{eq:Br4} is the equation for the energy balance of the cosmic ray component. Eq.~\ref{eq:Br5} is the energy balance of the mean-squared fluctuating waves that are assumed to propagate down the cosmic-ray gradient with the Alfv\'{e}n speed, and finally Eq.~\ref{eq:Br6} states that the divergence of the mean magnetic field is zero. $\overline{D}$ in Eqs.~\ref{eq:Br3} and \ref{eq:Br4} represents the effective diffusion coefficient of the cosmic rays. We neglect the diffusion term for finding the wind solutions and assume that the transport is advection dominated for the GeV cosmic rays, providing most of the pressure support for the wind. Finally, the symbol $\otimes$ is the tensor product, and $\textbf{I}$ is the unit tensor. 

With a flux tube geometry, the divergence operator becomes $\nabla\cdot x = \frac{1}{A} \frac{\dd (A\cdot x)}{\dd z}$, where $A$ is the cross-section area function and $z$ is altitude. We define the Alfv\'{e}n Mach number as $M_{\mathrm{A}} = \frac{v}{v_{\mathrm{A}}}$. We simplify the above set of equations as follows:
\begin{align}
\label{eq:wind_eq1}
&\rho v A = \mathrm{const}\\
&\frac{\dd P_{\mathrm{g}}}{\dd z} = \gamma_{\mathrm{g}} \frac{P_{\mathrm{g}}}{\rho} \frac{\dd \rho}{\dd z}\\
&\frac{\dd P_{\mathrm{c}}}{\dd z} = \gamma_{\mathrm{c}} \frac{P_{\mathrm{c}}}{\rho} \left( \frac{M_{\mathrm{A}} + \frac{1}{2}}{M_{\mathrm{A}} + 1}\right)\frac{\dd \rho}{\dd z}\\
&\frac{\dd P_{\mathrm{w}}}{\dd z} = \frac{1}{2(M_{\mathrm{A}} + 1)} \left[ (3 M_{\mathrm{A}} + 1) \frac{P_{\mathrm{w}}}{\rho} \frac{\dd \rho}{\dd z} - \frac{\dd P_{\mathrm{c}}}{\dd z} \right] \\
&B A = \mathrm{const}\\
&\frac{\dd v}{\dd z} = v\frac{ ~c^{2}_* \frac{1}{A} \frac{\dd A}{\dd z} - \frac{\dd \Phi}{\dd z}}  {v^2 - c^{2}_*},
\label{eq:wind_eq2}
\end{align}
where we have defined a so-called composite sound speed $c_*$ defined to be the following \citep{1991:Breitschwerdt}:

\begin{equation}
    c^{2}_* = \gamma_{\mathrm{g}} \frac{P_{\mathrm{g}}}{\rho} + \gamma_{\mathrm{c}} \frac{P_{\mathrm{c}}}{\rho} \frac{(M_{\mathrm{A}} + \frac{1}{2})^2}{(M_{\mathrm{A}} + 1)^2} + \frac{P_{\mathrm{w}}}{\rho} \frac{3 M_{\mathrm{A}} + 1}{2(M_{\mathrm{A}} + 1)}.
\label{eq:composite_sound}
\end{equation}

\begin{figure}
    \centering
    \includegraphics[width=0.5\textwidth]{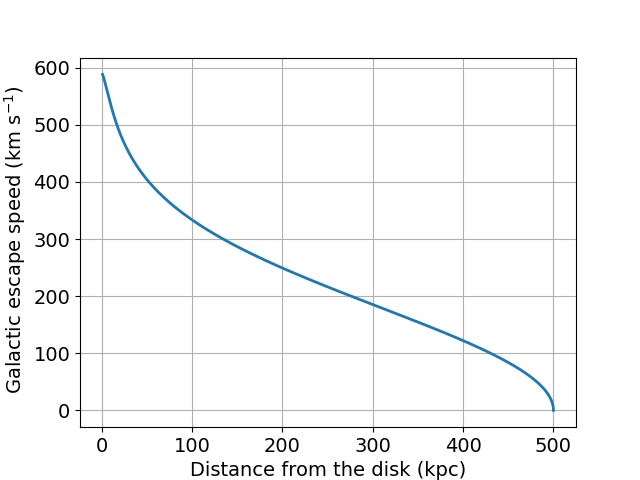}
    \caption{ Milky Way escape-speed profile used in this study. The halo potential assumed is given in Eq.~(\ref{eq:halo_potential}). The bulge and disk potential are modeled according to Eq.~(\ref{eq:disk_bulge}). The escape speed near the solar circle is $\sim$ 580 km/s, consistent with GAIA data.}
    \label{fig:DM_escape_speed}
\end{figure}

The gravitational potential $\Phi$ is modeled by taking inputs from GAIA data \citep{2018:Monari,Necib:2021vxr}. The dark-matter halo potential is taken to be
\begin{equation}
    \Phi_{\mathrm{halo}} = \frac{v_0^2}{2} \ln{\left(1 + \left[\frac{z^2 + R_0^2}{R_{\rm c}^2}\right]\right)},
\label{eq:halo_potential}
\end{equation}
with the characteristic speed $v_0 \sim 180$~km~s$^{-1}$, a radial concentration parameter $R_{\rm c} \sim 5$~kpc \citep{2018:Monari}, and the galactocentric distance $R_0$, which is $\sim$ 8.5~kpc near the solar circle. Since $\Phi_{\mathrm{halo}}$ does not go to 0 for large $z$, the escape speed for this potential is computed by \citet{2018:Monari}:
\begin{equation}
    v_{\mathrm{esc}}(r) = \sqrt{2|\Phi(r) - \Phi(3 r_{340})|},
\label{eq:escape_speed}
\end{equation}
where $r_{340}$ is the spherical radius within which the average density of the whole Galaxy is 340 times the critical density at redshift 0 $\left(\rho_{\rm c} = \frac{3H^2}{8\pi G}\right)$. This definition of the escape speed physically means that when a particle is at large distances $r \gtrsim 3r_{340}$, the particle has essentially `escaped' because the Galaxy is not an isolated system, and at those radii the gravitational potentials of other galaxies, like Andromeda, start to dominate \citep{2014:Piffl,2018:Monari}. The escape speed as a function of disk height $z$ at $R_0 = 8.5$ kpc is shown in Fig.~\ref{fig:DM_escape_speed}. Near the solar circle, the escape speed in our model is $\sim 580$~km~s$^{-1}$, which is consistent with GAIA data \citep{2018:Monari,Necib:2021vxr}. 

The gravitational potential of the bulge--disk component is modelled following Miyamoto and Nagai prescription \citep{1975:Miyamoto}:

\begin{equation}
    \Phi_{\rm B,D}(R_0, z) = - \sum_{i = 1}^2 \frac{G M_{\rm i}}{\sqrt{R_0^2 + \left(a_{\rm i} + \sqrt{z^2 + b_{\rm i}^2} \right)^2}}.
\label{eq:disk_bulge}
\end{equation}
The mass of the disk and bulge are taken to be $M_{\rm D} = 9 \times 10^{10}$~M$_{\odot}$ and $M_{\rm B} = 2 \times 10^{10}$~M$_{\odot}$, respectively. The disk parameters are ($a_1$, $b_1$) = (7.2~kpc, 0.52~kpc), and the bulge parameters ($a_2$, $b_2$) = (0.0~kpc, 0.495~kpc). 

\subsection{Wind solution}
\label{dynamicsCRAW}

\subsubsection{Method of solution --- Boundary conditions and critical points}

As the starting conditions of the problem, we set a cosmic-ray pressure ($P_{\rm c}^0$), a gas pressure ($P_{\rm g}^0$), and a starting value of the wave pressure ($P_{\rm w}^0$) at the base of the wind. We also set an initial value of the gas mass density, $\rho_0$, at the wind base. The wave pressure can be conveniently prescribed by setting an initial value of ${\left< \delta \textbf{B}\right>}/{\textbf{B}}$ at the base of the wind. The magnetic field geometry is also prescribed by setting the geometry parameter $z_0$ defined in Eq.~(\ref{eq:hyperbola}). 

With these conditions, we solve Eqs. (\ref{eq:wind_eq1})--(\ref{eq:wind_eq2}) to obtain a supersonic wind solution. 
The set of equations has a critical point when $v = c_*$ in Eq.~(\ref{eq:wind_eq2}), which leads to both the numerator and the denominator of the momentum conservation equation to go to zero at the sonic point. The continuous solution across the critical point is obtained using L'Hospital's rule by integrating inwards from the critical point to match to the CR and gas pressure at the disk. Integration is also performed from the critical point outwards to get the solution up to $O(100~\mathrm{kpc})$ distances. Details on how to solve for a wind solution can be found in Section~4.1 of the paper by \citet{1991:Breitschwerdt}.

\subsubsection{Galactic wind termination shock}\label{ssec:TS}

The supersonic wind forms a termination shock when the ram pressure of the flow ($\rho v^2$) becomes comparable to the pressure of the ambient medium, which in our case is the pressure of the intergalactic medium ($P_{\mathrm{IGM}}$). This condition holds as long as the other sources of pressure like the thermal gas pressure ($P_{\mathrm{g}}$), magnetic pressure ($\frac{B^2}{8\pi}$), and cosmic-ray pressure ($P_{\rm c}$) are sufficiently smaller than the ram pressure. In our model we make sure that the cosmic-ray pressure (including the pressure of reaccelerated CRs, which is computed separately as a solution to CR transport equation) at the GWTS never exceeds the ram pressure. Our reference model forms a GWTS at an altitude of $z_{\mathrm{s}} \sim 200$~kpc  for $ P_{\mathrm{IGM}} \sim 10^{-15}$ erg~cm$^{-3}$. This value of $P_{\mathrm{IGM}}$ is consistent with observations of the warm--hot IGM \citep{2018:Nicastro} and with $P_{\mathrm{IGM}}$ values assumed in previous works on cosmic-ray acceleration at termination shocks \citep{2006:Zirakashvilli,2017:Bustard,Merten:2018qoa}. The termination shock causes a discontinuity in the speed, temperature, and density of the fluid, and the strength of the shock jump ($\frac{v_2}{v_1}$) is obtained using Rankine--Hugoniot jump conditions \citep{1967:Zeldovich}. The evolution of the flow beyond the termination shock (downstream) is obtained by assuming the downstream flow to be gas dominated with adiabatic index $\gamma_{\mathrm{down}} = \frac{5}{3}$.

\subsubsection{A reference solution} 
\label{subsec:fiducial_wind}

We assume that the wind is launched at a height of $z_{\mathrm{base}} = 1$~kpc from the midplane of the disk. Our results are ultimately not very sensitive to the launching height. The cosmic ray pressure at the base is taken to be $P_{\mathrm{c}}^0 = 2 \times 10^{-13}$ erg~cm$^{-3}$, consistent with the measured values of the CR pressure at Earth. The gas number density is taken to be $n_0 = 2.5 \times 10^{-3}$ \,cm$^{-3}$, consistent with the observations of Milky Way's halo with oxygen lines \citep{2015:Miller}. The gas pressure at the base is set to be $P_{\mathrm{g}}^0 = 3.4 \times 10^{-14}$ erg~cm$^{-3}$, which corresponds to a temperature of $\sim 10^5$~K, characteristic of a warm ionized medium. The geometry parameter $z_0$ in Eq.~(\ref{eq:hyperbola}) is set to be 20~kpc, with the expectation that the flux tube opens up from vertical to spherical geometry at a height of the order of the Galactic radius \citep{1991:Breitschwerdt,2012:Dorfi}. 
 We also set the magnetic field $B_0$ at the base of the wind which is the field along the $z$ direction. The Galactic magnetic field has a rich structure consisting of disk and out-of-plane components \citep{2012ApJ...757...14J}. \citet{2012ApJ...757...14J} estimate an out-of-plane field component of $\sim ~1$ $\mu$G near the Solar circle, which is what we are interested in. Consistent with this result, the magnetic field at the base is set to be $B_0 = 2~\mu$G. Additionally, at the base, we set the wave pressure such that $\frac{\delta B_0}{B_0} = 0.1$ \citep{1991:Breitschwerdt}, corresponding to a small wave pressure of $\sim 2 \times 10^{-15}$ erg cm$^{-3}$. 
With these initial parameters, the wind solution results in a total CR energy flux of $\sim 10^{-5}$ erg cm$^{-2}$ s$^{-1}$, consistent with observations \citep{Murase:2018utn}. 

\begin{figure*}
    \centering
    \includegraphics[width=0.48\textwidth]{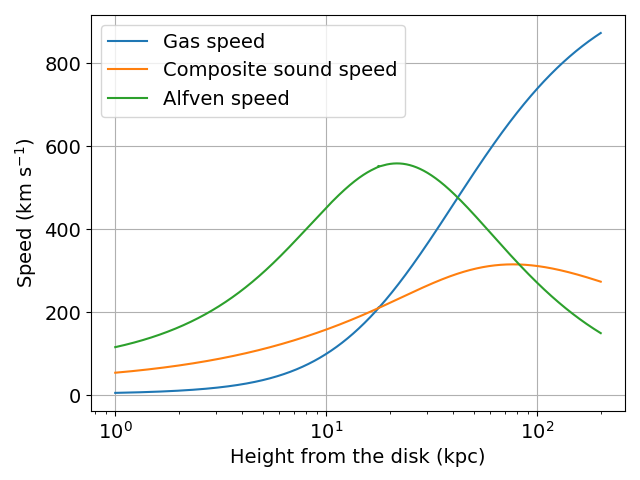}
    \includegraphics[width=0.48\textwidth]{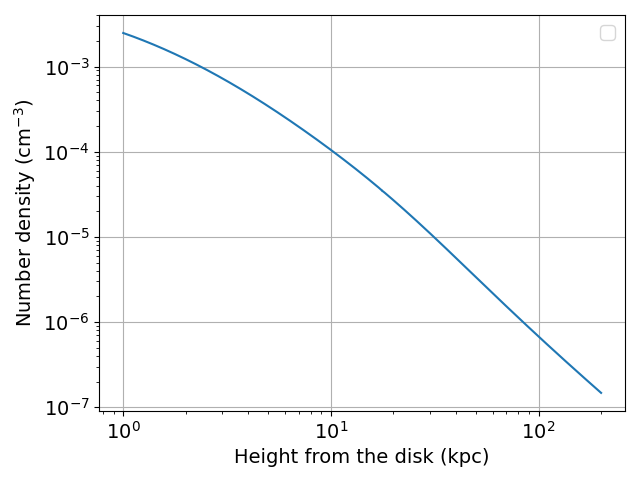} \\
    \includegraphics[width=0.48\textwidth]{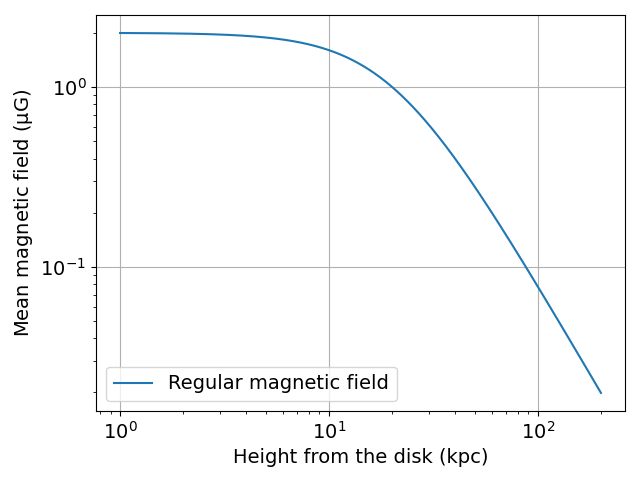}
    \includegraphics[width=0.48\textwidth]{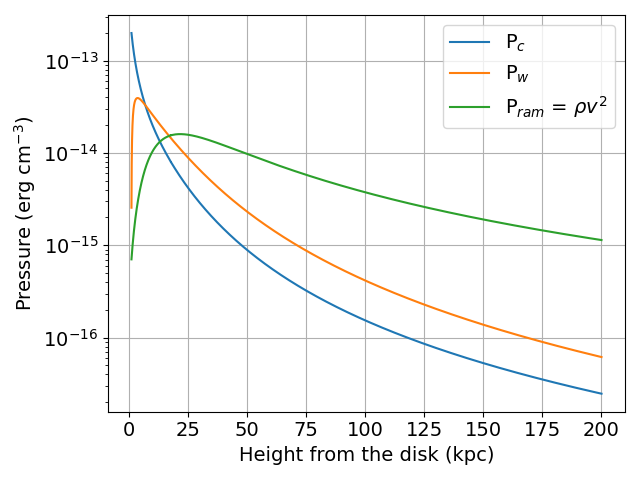}
    
    \caption{Top left: speed profiles of the gas, composite sound speed as defined in Eq.~(\ref{eq:composite_sound}), and Alfv\'{e}n speed. The sonic point occurs at $\sim$ 20~kpc. Top right: baryon number density in the wind profile. Bottom left: mean magnetic field vs. height from the disk. Bottom right: pressure contribution from the cosmic rays, waves, and the ram pressure. These figures are shown up to the location of the GWTS at 200 kpc.}
    \label{fig:speed_profiles}
\end{figure*}

The resulting wind profile as a function of the height from the disk, $z$, is shown in the top-left panel of Fig.~\ref{fig:speed_profiles}. 
In this model, the gas starts with a very small speed of $\sim$ 6~km~s$^{-1}$ and increases to $\sim$ 870~km~s$^{-1}$ at $z \sim 200$~kpc. At the wind base, the magnetic field is, $B_0 \sim 2 ~ \mu{\rm G}$, and the corresponding Alfv\'{e}n speed is \mbox{$v_{\mathrm{A}} = B_0 / \sqrt{4 \pi m_{\rm i} n_{\mathrm{i}}} \sim 100$~km~s$^{-1}$} ($m_{\rm i}$, $n_{\mathrm{i}}$ are the mass and number density of ions in the plasma). This plot shows the velocity profile up to the GWTS. 
Additionally, plots for the number density (top-right panel), mean magnetic field (bottom-left panel) and pressure contributions of the different fluid components (bottom-right panel) are shown. The number density decreases by four orders of magnitude between the wind base up to the GWTS. The mean magnetic field follows the flux tube geometry, decreasing from 2~$\mu{\rm G}$ at the wind base to $\sim 0.02~\mu{\rm G}$ at the GWTS. For the pressure contributions shown in the bottom-right panel of Fig.~\ref{fig:speed_profiles}, we see that the wave pressure starts out much smaller than the cosmic ray pressure at the base of the wind. 
However, it quickly increases and dominates over the CR pressure for most of the wind halo, where accelerates the gas. The mass outflow rate per unit area in this model is $\dot{m} \sim 10^{-21} ~\mathrm{g ~cm^{-2} s^{-1}}$, corresponding to an integrated mass outflow rate from the Galaxy of roughly $\sim$ $0.1$~M$_{\odot}$ ~yr$^{-1}$. This is a non-negligible mass outflow rate, considering that the star formation of the Galaxy is $\sim O(0.1-1~\rm M_{\odot}yr^{-1})$ \citep{2017ApJ...837..150S}, indicating that the cosmic-ray-driven Galactic wind can have a relevant impact on the Galactic ISM and its dynamical evolution.

Due to the pressure of the intergalactic medium ($\sim 10^{-15}~{\rm g~cm}^{-3}$), the GWTS forms at $z_{\mathrm{s}} \sim200$~kpc. At this location, the jump in the speed of the gas is approximately $\frac{u_2}{u_1} = \frac{\gamma_{\mathrm{gas}} - 1}{\gamma_{\mathrm{gas}} + 1} + \frac{2}{\gamma_{\mathrm{gas}} + 1} \frac{1}{M^2}$, where $M = \frac{v_{\rm gas}}{c_*}$.

Beyond the GWTS at 200~kpc, the gas is simply an adiabatically expanding flow approximately leading to a velocity profile that falls like $v \propto z^{-2}$. 

\subsection{Cosmic-ray transport}
\label{sec:transportCR}

The processes involving high-energy particles (most notably the reacceleration) are taking place over typical timescales that are much shorter than the dynamical time of the CR-driven Galactic wind. 
Therefore we assume that the transport of such particles takes place in stationary condition.

We solve the following stationary, linear transport equation: 
\begin{eqnarray}
\begin{aligned}
\label{eq:transport}
\frac{1}{A(z)} \frac{\partial}{\partial z} \left( A(z) D(z,p) \frac{\partial f}{ \partial z}\right) - v(z) \frac{\partial f}{\partial z}  \\
+ \frac{1}{A(z)} \frac{d}{dz}(A(z) v(z)) \frac{1}{3} \frac{\partial f}{\partial \ln{p}} = -Q(z,p),
\end{aligned}
\end{eqnarray}
\noindent where $A(z)$ is the area of the wind defined in Eq.~(\ref{eq:hyperbola}), $v(z)$ is the wind speed, $D(z,p)$ is the diffusion coefficient, $Q(z,p)$ is the source term, and $f = f(z,p)$ is the phase space density of cosmic rays. We note that the diffusion coefficient used in the transport calculation was not incorporated into the wind equations (\ref{eq:wind_eq1})--(\ref{eq:wind_eq2}). Previous calculations have shown that the effect of the diffusion coefficient in the wind solutions starts to become important only for mean diffusion coefficients greater than $\sim 10^{29}$ cm$^2$ s$^{-1}$ \citep{2012:Dorfi}. The mean diffusion coefficients in our model varies between $10^{28}-10^{30}$ cm$^2$s$^{-1}$, depending on the location in the halo. Therefore, the effect of diffusion in the wind profile could in general, not be neglected. However, the primary effect of diffusion is to produce lighter and faster winds with lower mass loss rates \citep{2012:Dorfi}. Having faster winds will only increase the maximum energies achievable in the reacceleration process, as will be described later. Incorporating this effect will therefore only improve the results presented in this paper. The $v(z)$ term that enters the transport equation contains contributions from both the gas velocity and the Alfv\'{e}n speed, $v(z) = v_{\mathrm{gas}}(z) + v_{\mathrm{A}}(z)$ upstream of the GWTS. We note that setting $v(z) = v_{\mathrm{gas}}(z) + v_{\mathrm{A}}(z)$ assumes that the CRs are streaming at the Alfv\'{e}n speed. This assumption, however, only holds for CRs up to $\sim 1$~TeV for which streaming instability grows faster than the dynamical timescale of the problem. For higher energy CRs, the wave growth will not happen, and they will likely not stream at $v_{\mathrm{A}}(z)$. We acknowledge this caveat, but we have checked that our results do not sensitively depend on the value of $v_{\mathrm{A}}(z)$ that enters the transport equation. Therefore, we keep using $v = v_{\mathrm{gas}} + v_{\mathrm{A}}$ for the transport calculations of the entire CR energy range under consideration. For downstream, we assume gas-dominated transport and simply set $v(z) = v_{\mathrm{gas}}(z)$.

We assume a Gaussian source term for the Galactic CRs: 
\begin{equation}
\label{eq:source}
Q(z,p) = Q_0(p) \frac{\exp\left({-\frac{z^2}{2\sigma_z^2}}\right)}{\sqrt{2 \pi \sigma_z^2}}.
\end{equation}
\noindent Here, $\sigma_z \sim 100$~pc is taken as a proxy for the Galactic disk thickness; however, we verify that the results do not strongly depend on the value of $\sigma_z$ because the scale of the problem, given by the size of the GWTS, is ${\cal O}(10^2 ~\mathrm{kpc})$, which is much larger than the size of the the central source. The source normalization $Q_0(p)$ in Eq.~(\ref{eq:source}) is given by:
\begin{equation}
    Q_0(p) = \frac{\mathcal{F}_{\mathrm{SN}} \mathcal{R}_{\mathrm{SN}}}{\pi R_{\rm d}^2},
\label{eq:source_norm}
\end{equation}
\noindent where $\mathcal{R}_{\mathrm{SN}}$ is the supernova rate and $R_{\rm d}$ is the Galactic disk radius. $\mathcal{F}_{\mathrm{SN}}$ is the spectrum of cosmic rays injected by a single supernova. 
We assume that supernova remnants (SNRs) are the sources of Galactic cosmic rays,
although the calculations presented here do not depend crucially on such an assumption. The spectrum of an SNR can be written as:
\begin{equation}
    \mathcal{F}_{\rm SN}(p) = \frac{\xi_{\mathrm{CR}} E_{\rm SN}}{I(\alpha)} \left( \frac{p}{m}\right)^{-\alpha} \exp{\left( - \frac{p}{p_{\rm c,SNR}}\right)}
\label{eq:SNR_spectrum},
\end{equation}
\noindent where $\xi_{\mathrm{CR}}$ is the cosmic-ray injection efficiency by the SNR, $E_{\rm SN}$ is the kinetic energy released in a supernova explosion, and $p_{\rm c,SNR}$ is the momentum cutoff of Galactic accelerators. 

Finally, $I(\alpha)$ is a normalization factor chosen such that:
\begin{equation}
    \int_{p_0}^{\infty} \mathcal{F}_{\rm SN}(p) T(p) d^3p = \xi_{\mathrm{CR}} E_{\rm SN},
\end{equation}
\noindent where $p_0$ is the minimum injection momentum for protons and $T(p)$ is the kinetic energy of a particle of momentum $p$. For our reference model, we choose $E_{\rm SN} = 10^{51}$~erg, $\xi_{\mathrm{CR}} = 0.1$, $p_0$ = 1~GeV, $\mathcal{R_{\mathrm{SN}}} = 2$~SN per century. We also take $\alpha = 4.4$ and $p_{\rm c,SNR} = 3 \times 10^6$~GeV for protons, consistent with standard assumptions in the literature \citep{2016:Thoudam,2016:Recchia,2017:Recchia}. For heavier elements with atomic number $Z$, the maximum momentum injected by Galactic sources is $Z$ times the cutoff momentum for protons.  

As discussed in Sec.~\ref{subsec:fiducial_wind}, the base of the wind is set at $z_{\mathrm{base}} = $ 1~kpc. Below this altitude, the advection term in the transport equation is taken to be zero, and the cosmic ray propagation is assumed to be purely diffusive inside this region. Inside the wind base, in agreement with measurements of the B/C ratio \citep{B-over-C-ratio}, a Kolmogorov-type diffusion coefficient is assumed: 

\begin{equation}
    D(p) = 2 \times 10^{28} ~ \left(\frac{p}{Z}\right)^{1/3}~ \mathrm{cm^2~s^{-1}}, ~~~~~~ z < z_{\mathrm{base}}.
\end{equation}

\noindent For $z_{\mathrm{base}} < z < z_{\mathrm{s}}$, i.e, in the upstream wind region up to the GWTS, we study two different prescriptions for the upstream diffusion coefficient. We briefly outline these prescriptions below.

\subsubsection{Diffusion model 1: Quasi-linear theory}
\label{subsubsec:diff1}

Quasi-linear theory (QLT) is a common approach for modeling diffusion coefficients of magnetised particles in a turbulent medium \citep{1966ApJ...146..480J,2004ApJ...616..617S,2009A&A...507..589S}. Within QLT, the particle motion is assumed to be a superposition of the gyromotion of the particle and stochastic motion along the magnetic field lines. This approximation is valid only for weak turbulence, $\delta B < B$. The scattering rate within the QLT formulation is $\nu = 2 \pi^2 \omega_B \frac{k_{\rm res} \epsilon(k_{\rm res})}{B^2}$ \citep{1969ApJ...156..445K,2013PhPl...20e5501Z}. Here, $\omega_B$ denotes the synchotron frequency proportional to the resonant wavenumber $k_{\rm res}$. At wavenumber $k_{\rm res}$, the wave energy contained is $k_{\rm res} \epsilon_{\rm res}$. We assume isotropic turbulence with a Kolmogorov spectrum and assume that energy is injected at an outer scale $l_{\mathrm{max}}$ and dissipates to smaller scales after a cascade of energy from large to small scales without energy loss. The coherence length of the system, $l_{\mathrm{c}}$, is then given by, $l_{\mathrm{c}} = l_{\mathrm{max}} / 5$ for a Kolmogorov spectrum \citep{2002JHEP...03..045H}. 

As a plausible ansatz, we suppose that an isotropic magnetic turbulence spectrum is maintained by essentially external processes (other than CRs generating their own turbulence) that determine the diffusion coefficient, which in turn controls the particle transport. In particular, we assume that the wind itself can generate a turbulence at the level $\delta B$ on top of the mean magnetic field $B$. We define a parameter, $\eta$, such that 

\begin{equation}
\eta \equiv \frac{\delta B^2}{4 \pi\,\rho v^2}\,,
\label{eq:eta}
\end{equation} 

\noindent implying $\eta$ is the fraction of energy used in generating turbulence out of the ram pressure of the wind. The mean magnetic field is obtained from the wind solution described in Sec. \ref{subsec:fiducial_wind}. This turbulence generates a diffusion coefficient parallel to the mean magnetic field given by \citet{2020:Reichherzer}:

\begin{equation}
    D(p, z) = D_{\mathrm{Bohm}}(p_0) \left(\frac{p}{p_0}\right)^{\frac{1}{3}} \left( \frac{B}{\delta B}\right)^2 + D_{\mathrm{Bohm}}(p_1) \left(\frac{p}{p_1}\right)^2\left( \frac{B}{\delta B}\right)^2,
\label{eq:QLT_diff}
 \end{equation}

 \noindent where $r_L(p)$ is the Larmor radius computed with respect to the mean magnetic field. $p_0$ and $p_1$ are such that $2\pi r_L(p_0) = l_{\mathrm{c}}$ and $2\pi r_L(p_1) = \frac{3}{2}l_{\mathrm{c}}$ \citep{2008A&A...479...97G}. The Bohm diffusion coefficient is $D_{\mathrm{Bohm}} = \frac{1}{3} r_L(p) c$. For momenta, $p > p^*$, where $r_L(p^*) = l_{\mathrm{c}}$, the diffusion coefficient changes its energy dependence to $p^2$ due to lack of resonant perturbations \citep{2008A&A...479...97G,Snodin2016,Subedi:2017,Dundovic:2020}. We prescribe $\frac{\delta B}{B} \sim 0.1$, a condition that can be easily satisfied by converting $\eta <$ 5\% of the wind ram pressure into generating magnetic turbulence. Note that we keep $\frac{\delta B}{B} \sim 0.1$ fixed throughout the halo in this case. We also prescribe a coherence length, $l_{\mathrm{c}} \propto z$, i.e, the coherence length of the turbulent field scales with height, as a reasonable guess. The prescription is such that $l_{\mathrm{c}}$ increases from $\sim$ 5~pc to $\sim$ 1~kpc from the wind base to the GWTS. We note that our results are not very sensitive to the details of the prescription of the coherence length in the halo. With this prescription, $D(p,z)$ continuously increases from the wind base up to the GWTS because the coherence length increasing with $z$ implies an increasing mean free path of the CRs with height, causing an increasing diffusion coefficient. We will explain later that this model fails to generate any efficient reacceleration of the galactic CRs at the GWTS because the diffusion coefficient is too high close to the GWTS, and the maximum energies achieved in the acceleration process goes as inverse power of the diffusion coefficient ahead of the shock. In the following, we will see that in order to achieve more efficient acceleration, we need to assume that a larger fraction of the ram pressure is converted into magnetic turbulence close to the GWTS.
 
\subsubsection{Diffusion model 2: QLT halo + fully turbulent field near GWTS}\label{subsubsec:diff2}

In this model, we prescribe a variant of the QLT model presented above that allows CRs to reach energies up to 10--40~PV rigidities in the reacceleration process. This is a phenomenological prescription where we allow the possibility of an increased particle scattering close to the shock, leading to a lower diffusion coefficient near the GWTS. As we will describe later, this lower value of the diffusion coefficient close to the GWTS is crucial for an efficient reacceleration. In this model, we assume the same QLT prescription as outlined in Sec.~\ref{subsubsec:diff1} given by Eq.~(\ref{eq:QLT_diff}) up to $z = 100$~kpc with $\frac{\delta B}{B} = 0.1$ and $l_{\mathrm{c}} \propto z$. Between 100~kpc and up to the GWTS located at 200~kpc, we assume an enhanced turbulence induced by converting $\eta \sim 20\%$ of the ram pressure of the wind into generating turbulence. This criterion generates strong turbulence close to the GWTS such that $\delta B > B$ is satisfied close to the GWTS between 100--200~kpc. We also set a constant coherence length $l_{\mathrm{c}} = 1$~kpc in this region of lower diffusion coefficient in the halo.

In this strong scattering approximation, the diffusion coefficient near the GWTS is given by \citet{2008A&A...479...97G}:
\begin{equation}
    D(p, z) = D_{\mathrm{Bohm}}(p_0) \left(\frac{p}{p_0}\right)^{1/3} + D_{\mathrm{Bohm}}(p_1) \left(\frac{p}{p_1}\right)^2 .
\label{eq:noemie_diff}
\end{equation}
Note that the only difference between Eqs.~(\ref{eq:QLT_diff}) and (\ref{eq:noemie_diff}) is the absence of the $\left( \frac{B}{\delta B} \right)^2$ in the latter. We stress that the only difference between this model and the previous one is the presence of a strong scattering region near the GWTS, which as we will see later helps in efficient reacceleration of the CRs. This strong scattering region is a phenomenological prescription that we use to demonstrate the conditions under which particles can be accelerated to up to $\sim$ 40 PV, which is about an order of magnitude larger than the maximum achievable energies by Galactic CR sources. In principle, the particle scattering close to the shock can be enhanced by a number of mechanisms, such as the non-resonant streaming instability \citep{2004MNRAS.353..550B} and turbulent amplification of magnetic fields by CR-driven pressure gradients \citep{2009ApJ...707.1541B,2012MNRAS.427.2308D}. These processes might even decrease the diffusion coeffficient to the Bohm limit at the GWTS, and such a strong turbulence limit has been assumed in various earlier studies of CR acceleration at the GWTS \citep{2006:Zirakashvilli,2012:Dorfi,2019:dorfi}. These processes, however, need sophisticated numerical simulations to accurately assess whether the diffusion coefficient can be decreased to the Bohm limit close to the GWTS. Our prescription, instead provides a simple mechanism to generate a strongly turbulent region close to the shock, which, as will find later, is important in getting efficient CR reacceleration.

Finally, for the transport equation in the downstream region, we assume as a benchmark scenario that the downstream diffusion coefficient is an order of magnitude smaller than the diffusion coefficient right upstream of the GWTS.  This assumption is justified because one can expect that the GWTS will create strong magnetohydrodynamic (MHD) turbulence downstream towards intergalactic space \citep{2004:Volk,2018SSRv..214..122D}, and the diffusion coefficient there should be small. The resulting diffusion coefficient is small enough to make the diffusion term negligible compared to the advection term in the downstream region. We note, however, that the maximum energies reached in the DSA mechanism decreases with increasing downstream diffusion coefficient because it increases escape of CRs from the downstream region. For completeness, we will also consider the effect of having higher downstream diffusion in Sec.~\ref{sec:param_space}.

\subsubsection{Maximum energy achieved in reacceleration}
\label{subsec:max_energy}

Beyond a certain maximum momentum, $p_{\mathrm{max}}$, the distribution function at the shock, $f_s(p)$, is exponentially suppressed (see Eq.~(\ref{eq:fs})) because the accelerator has a finite size, so it cannot accelerate particles to infinitely large momenta. The maximum momentum achieved in reacceleration, $p_{\mathrm{max}}$, is discussed below.  

As seen from Eq.~(\ref{eq:fs}), the exponential drop-off in the distribution function at the shock, $f_s(p)$, is controlled by the two functions $\Gamma_1(p)$ and $\Gamma_2(p)$. The function $\Gamma_1$ represents the effects of the geometry of the system and the adiabatic energy losses upstream. The suppression term, e$^{-\Gamma_1}$, is appreciably different from unity for momenta close to $p_{\mathrm{max}}$, where the upstream diffusion length, $D_1/v$, becomes comparable to the size of the system, $z_{\mathrm{s}}$ \citep{1997:Berezhko}. For a spatially constant diffusion coefficient upstream, $D_1(p) = \kappa_1 ~ p^{2 - \delta}$~cm$^2$~s$^{-1}$, and a uniform wind velocity, $v$, the maximum energy achievable in the GWTS is \citep{2021:Morlino,Starburst_wind}:
\begin{equation}
p_{\mathrm{max}} \sim \left( \frac{ v ~ z_{\mathrm{s}}}{\kappa_1} \right)^{1 / (2 - \delta)}.
\label{eq:p_max}
\end{equation}
Note that Eq.~(\ref{eq:p_max}) was obtained with the assumption of a spatially constant upstream diffusion coefficient, but in our reference model, the diffusion coefficient varies with $z$, and, therefore, $p_{\mathrm{max}}$ in our model doesn't necessarily agree with the equation above. However, some basic features of the solution, like the dependencies of $p_{\mathrm{max}}$ on the wind speed, shock radius, and diffusion coefficient, are the same as in Eq.~(\ref{eq:p_max}). Specifically, we find that $p_{\mathrm{max}}$ decreases with lower wind speeds, smaller shock sizes, and higher upstream diffusion coefficients. Going back to our prescription of the two upstream diffusion models in Secs.~\ref{subsubsec:diff1} and \ref{subsubsec:diff2}, we find that model~2 provides efficient CR acceleration because, as we describe here, $p_{\mathrm{max}}$ increases with decreasing $D_1$.

The distribution function also depends on $\Gamma_2$, which is a result of the transport in the downstream region. $\Gamma_2$ becomes important when the diffusion length of the particles in the downstream region, $D_2 / v_2$, becomes comparable to the size of the shocked wind region, $z_{\mathrm{b}} - z_{\mathrm{s}}$, where $z_{\mathrm{b}}$ is the location of the outer boundary of the bubble. Typically, for the model scenarios considered here, $\Gamma_2(p) \ll \Gamma_1(p)$, and $p_{\mathrm{max}}$ is effectively set via $\Gamma_1(p_{\mathrm{max}}) \sim 1$, which is equivalent to the condition shown in Eq.~(\ref{eq:p_max}). Therefore, the maximum energy achievable in the GWTS in our model is set by the upstream plasma velocity, diffusion coefficient, and the shock size. For more details on the interplay of $\Gamma_1$ and $\Gamma_2$ for setting the maximum energy of accelerated cosmic rays in the wind termination shock, see \citet{2021:Morlino}. 

\section{Results}
\label{sec:results_CR}

\subsection{Comparison of different diffusion prescriptions}
\label{subsec:paramaters}

\setlength{\tabcolsep}{5pt}
\begin{table}[h]
\centering
\renewcommand{\arraystretch}{1.2}
\begin{tabular}{| l | r |}
\hline
Parameter  &  Value   \\ 
\hline \hline
$P_{\mathrm{CR}}^0$ & 2 $\times 10^{-13}$ erg~cm$^{-3}$\\
$P_{\mathrm{g}}^0$ & $3.4 \times 10^{-14}$ erg~cm$^{-3}$\\
$n_0$ & 2.5 $\times 10^{-3}$ ~cm$^{-3}$\\
$B_0$ & 2 $\mu$G\\
$z_0$  &   20 kpc \\
$z_{\mathrm{s}}$  &   200 kpc \\
$v_{\mathrm{t}}$ & 870 km~s$^{-1}$ \\
\hline \hline
\end{tabular}
\caption{Parameter values for the reference wind model. Values for the base cosmic-ray pressure ($P_{\mathrm{CR}}^0$), gas pressure ($P_{\mathrm{g}}^0$), number density ($n_0$), magnetic field ($B_0$), geometry parameter ($z_0$), shock location ($z_{\mathrm{s}}$), and terminal gas speed ($v_{\mathrm{t}} = v_{\mathrm{gas}}$) are are all described in Sec.~\ref{subsec:fiducial_wind}.} 
\label{table:fiducial}
\end{table}

First, we summarize the key parameters used for the reference wind model in Table~\ref{table:fiducial}. We also summarize the characteristics of the two upstream diffusion models as follows:

\begin{itemize}
    \item Model 1 (QLT) --- assumes isotropic turbulence at a length scale $l_{\mathrm{c}}$. Prescription of a quasi-linear description with $\frac{\delta B}{B} = 0.1$ and $l_{\mathrm{c}} \propto z$ following Eq.~(\ref{eq:QLT_diff}).  The diffusion coefficient increases starting from the wind base up to the GWTS (Sec.~\ref{subsubsec:diff1}). 

    \item Model 2 (QLT + strongly turbulent GWTS) --- we use QLT prescription following Eq.~(\ref{eq:QLT_diff}) up to 100~kpc. Beyond 100~kpc till the shock location at 200~kpc, we allow for strong scattering and a smaller diffusion coefficient, with $l_{\mathrm{c}} = 1$~kpc and $\delta B > B$ (Sec.~\ref{subsubsec:diff2}).
\end{itemize}

\begin{figure}
    \centering
    \includegraphics[width=.50\textwidth]{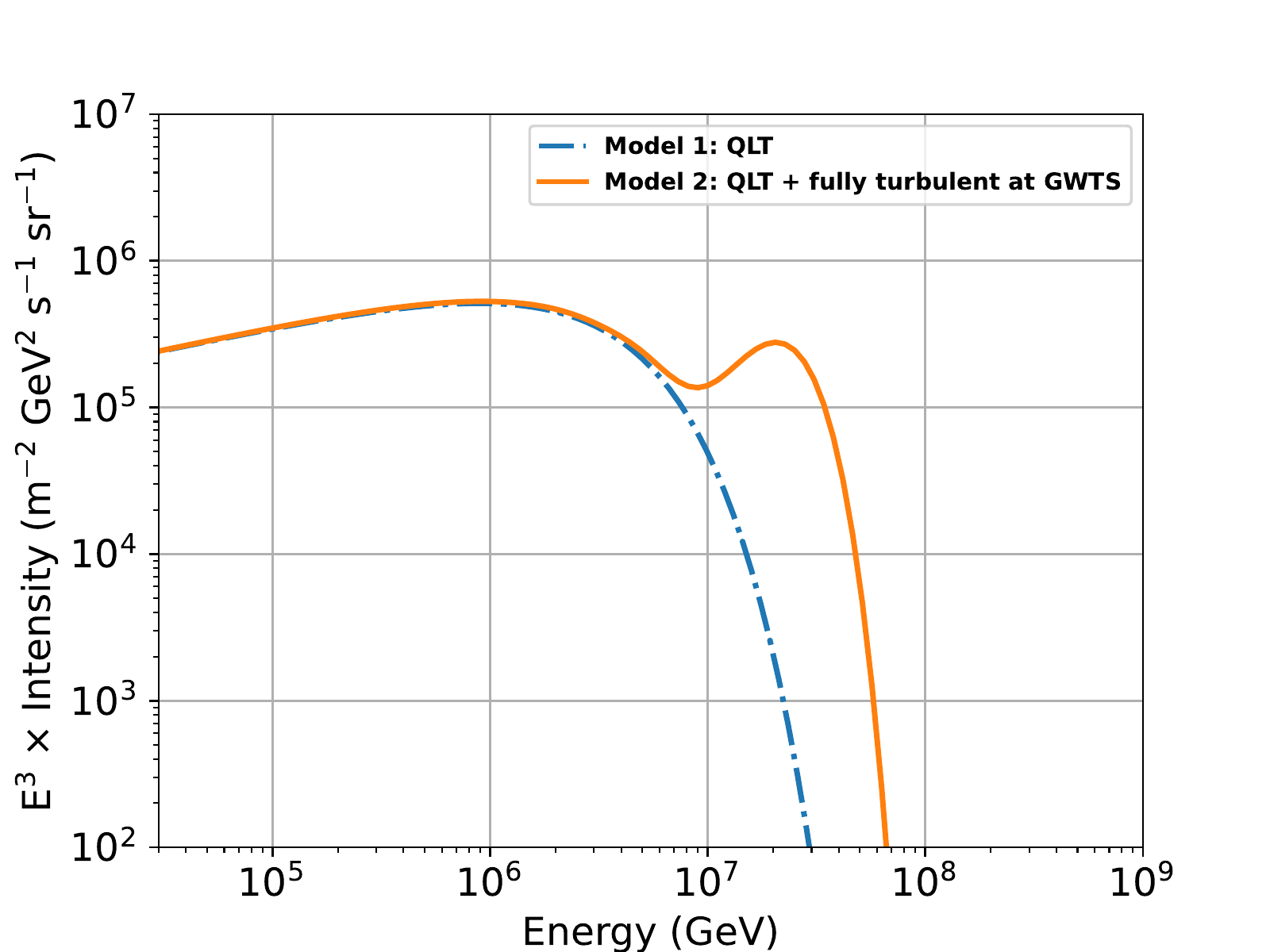}
    \caption{Spectrum of CR protons at the disk for the different diffusion prescriptions. Model~1 gives negligible contribution from the GWTS component, while model~2 is capable of accelerating CRs up to 40~PeV, an order of magnitude energy gain compared to the CRs produced within the disk.}
    \label{fig:diff_compared}
\end{figure}

Fig.~\ref{fig:diff_compared} shows the proton spectrum at the disk for the two upstream diffusion models. The spectrum at the disk is expected to be composed of two different components. One is the Galactic CR contribution with a maximum energy cutoff of 3~PeV, which marks the knee of the CR spectrum at the disk. Another contribution is expected from the backstreaming CRs that have been reaccelerated at the GWTS. The transition between the two components is expected to lead to a "bump" feature when the reacceleration is efficient.

For model~1, the spectrum at the disk consists only of the Galactic component with negligible contribution from the GWTS reacceleration. This is because, for model~1, the diffusion coefficient near the GWTS, which sets the maximum energy (Sec.~\ref{subsec:max_energy}), is $D_{1, \mathrm{GWTS}}^{\mathrm{model~1}}\sim 10^{31} \times p_{\mathrm{GeV}}^{1/3}$~cm$^2$~s$^{-1}$. This diffusion coefficient, combined with a speed of $\sim 870$ km/s and $z_s \sim$ 200 kpc gives a maximum momentum, $\sim 10^2$ GV/c (Eq. \ref{eq:p_max}), which is much smaller than the $\sim$~3~PeV energy reached in the Galactic CR sources of protons. Additionally, these CRs have a harder time propagating back to the disk owing to their lower rigidities. The result is a negligible contribution of the reaccelerated particles to the disk spectrum for model~1.

On the other hand, model~2 can reaccelerate the Galactic protons to up to $\sim$~40~PeV energies, which is an order of magnitude higher than the knee, as is shown in the figure. The lower diffusion coefficient of model~2, $D_{1, \mathrm{GWTS}}^{\mathrm{model~2}}\sim (3 \times 10^{28} \times p_{\mathrm{GeV}}^{1/3} + 10^{17} p_{\rm GeV}^2)$~cm$^2$~s$^{-1}$, allows for a higher maximum energy. In this case, $D_1 / v$ becomes equal to the shock size $z_s$ for an energy $\sim O(10~\rm PeV)$, which sets the maximum energy of the system. In this paragraph and the rest of the paper, ``maximum energy" in reference to a plot (such as Fig.~\ref{fig:diff_compared}) refers to the energy at which the $E^3 \times$ intensity of the reaccelerated component is roughly smaller by a factor of $1/e$ than the peak of the bump feature. The maximum rigidity is that maximum energy divided by charge, $Ze$.

The main difference between model~1 and model~2 is the orders of magnitude lower diffusion coefficient for the latter close to the GWTS. This leads to much greater maximum energies achieved for model~2 in accordance with Eq.~(\ref{eq:p_max}). We note that the values for the diffusion coefficients in the halo at $O(100~\mathrm{kpc})$ distances is completely unconstrained, thereby rendering comparison between these extremely different assumptions reasonable. In the disk, the two models have diffusion coefficient that converge to the value constrained by the B/C ratio.  In the following, we choose diffusion model~2 to demonstrate other properties of our results, such as the elemental composition of the CRs, observational implications, parameter space investigation etc.

\subsection{Elemental spectrum}
\label{subsec:elements}

\begin{figure*}
\centering
\includegraphics[width=.48\textwidth]{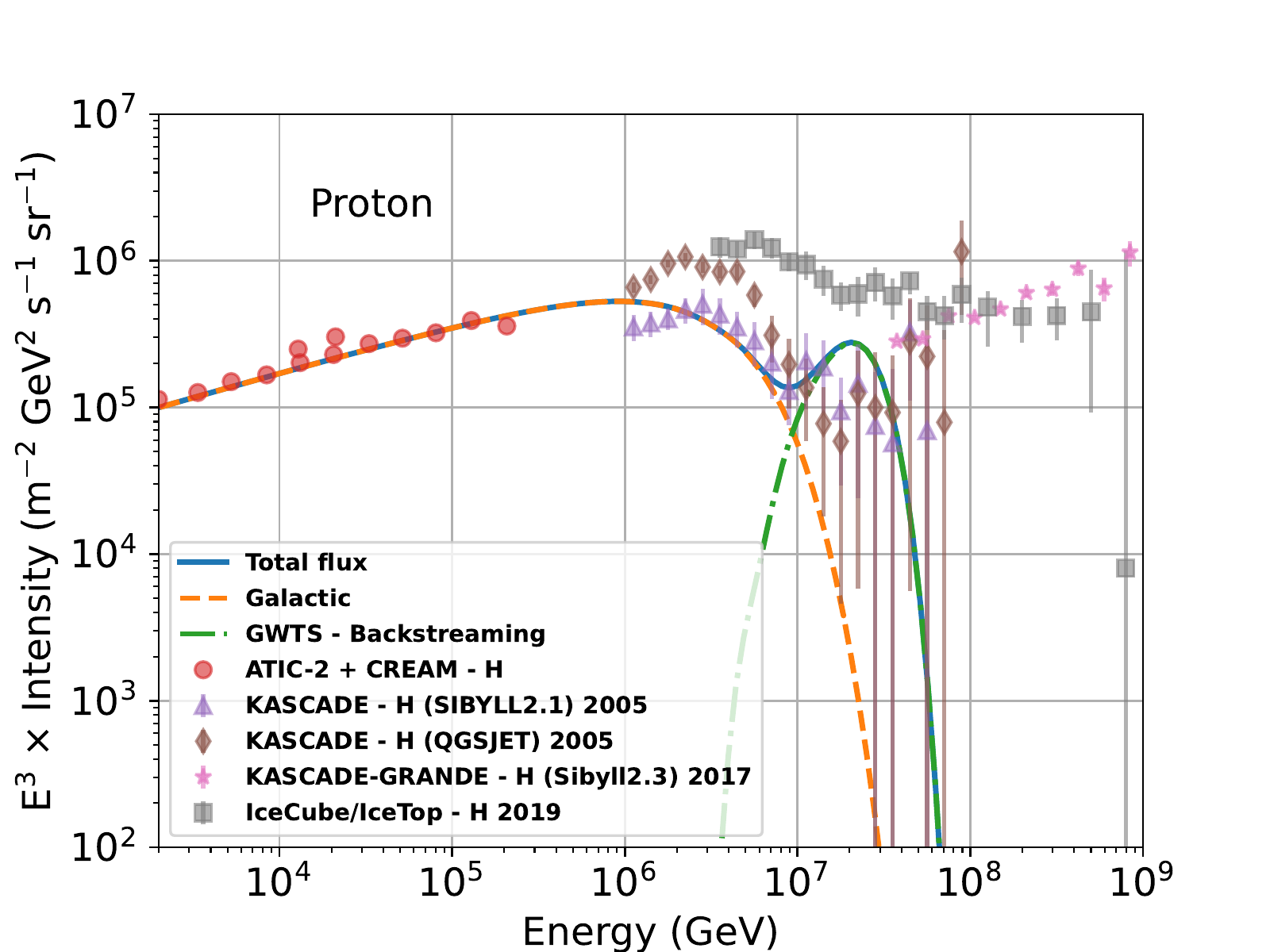}
\includegraphics[width=.48\textwidth]{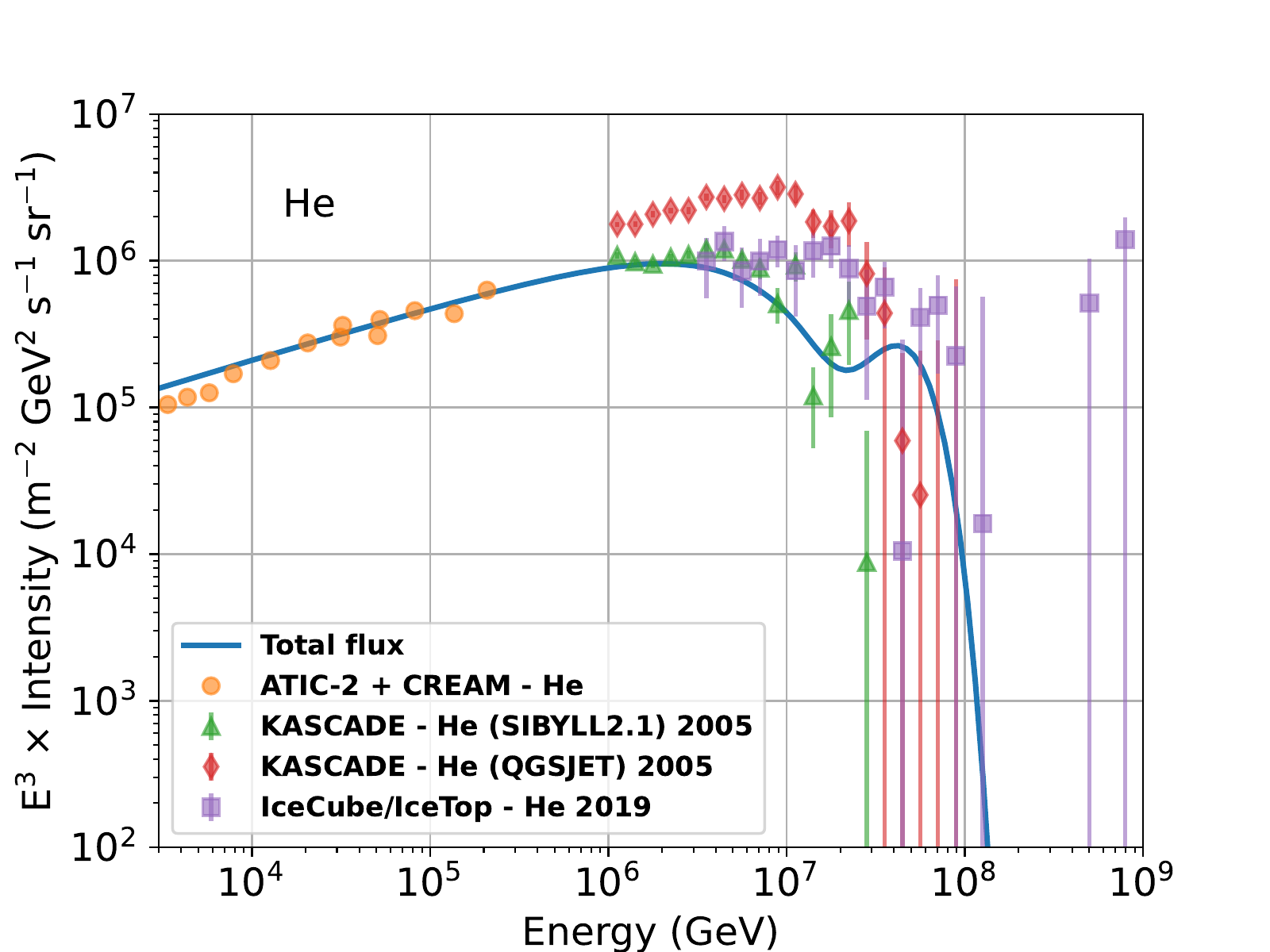} \\
\includegraphics[width=.48\textwidth]{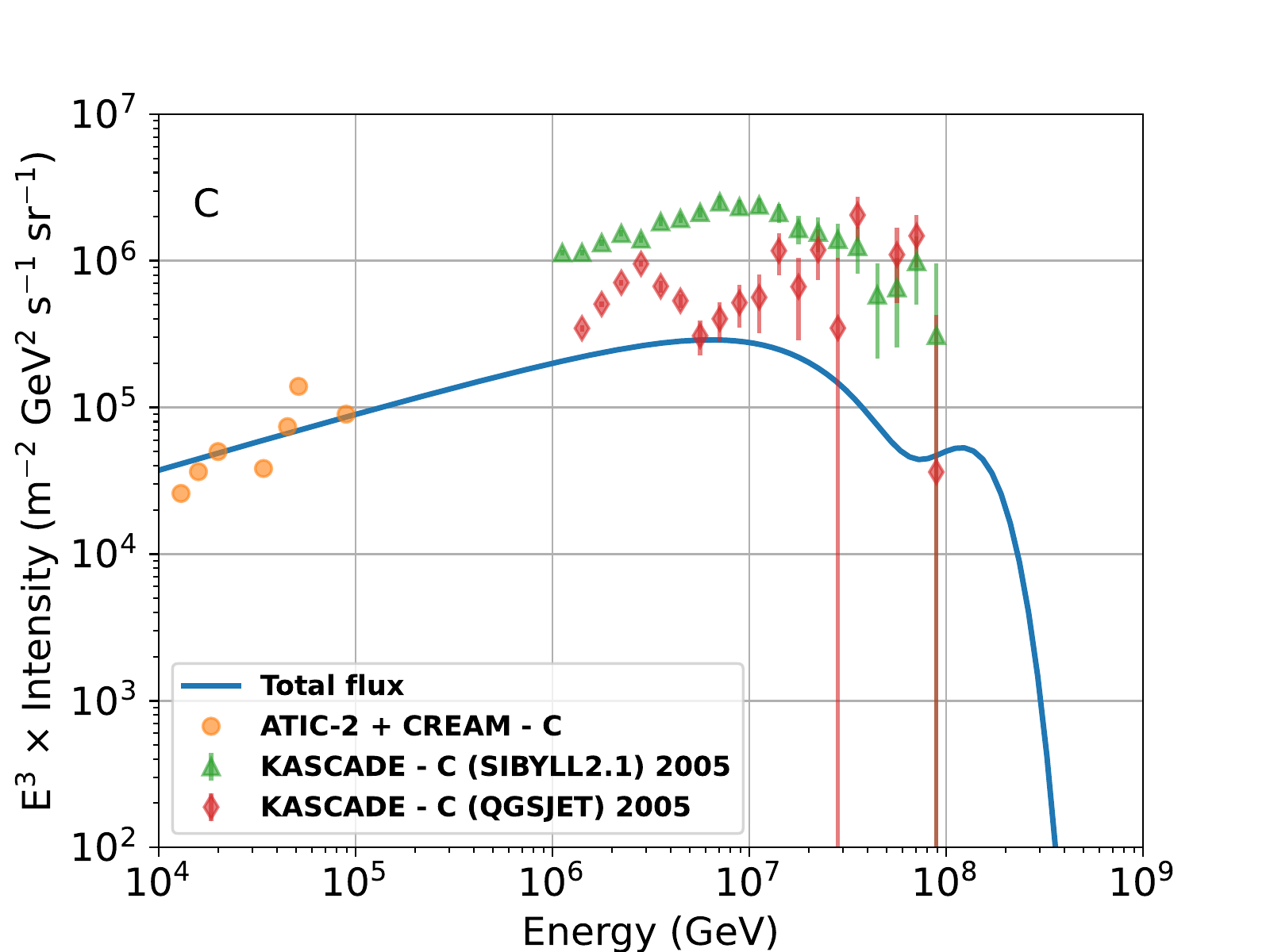}
\includegraphics[width=.48\textwidth]{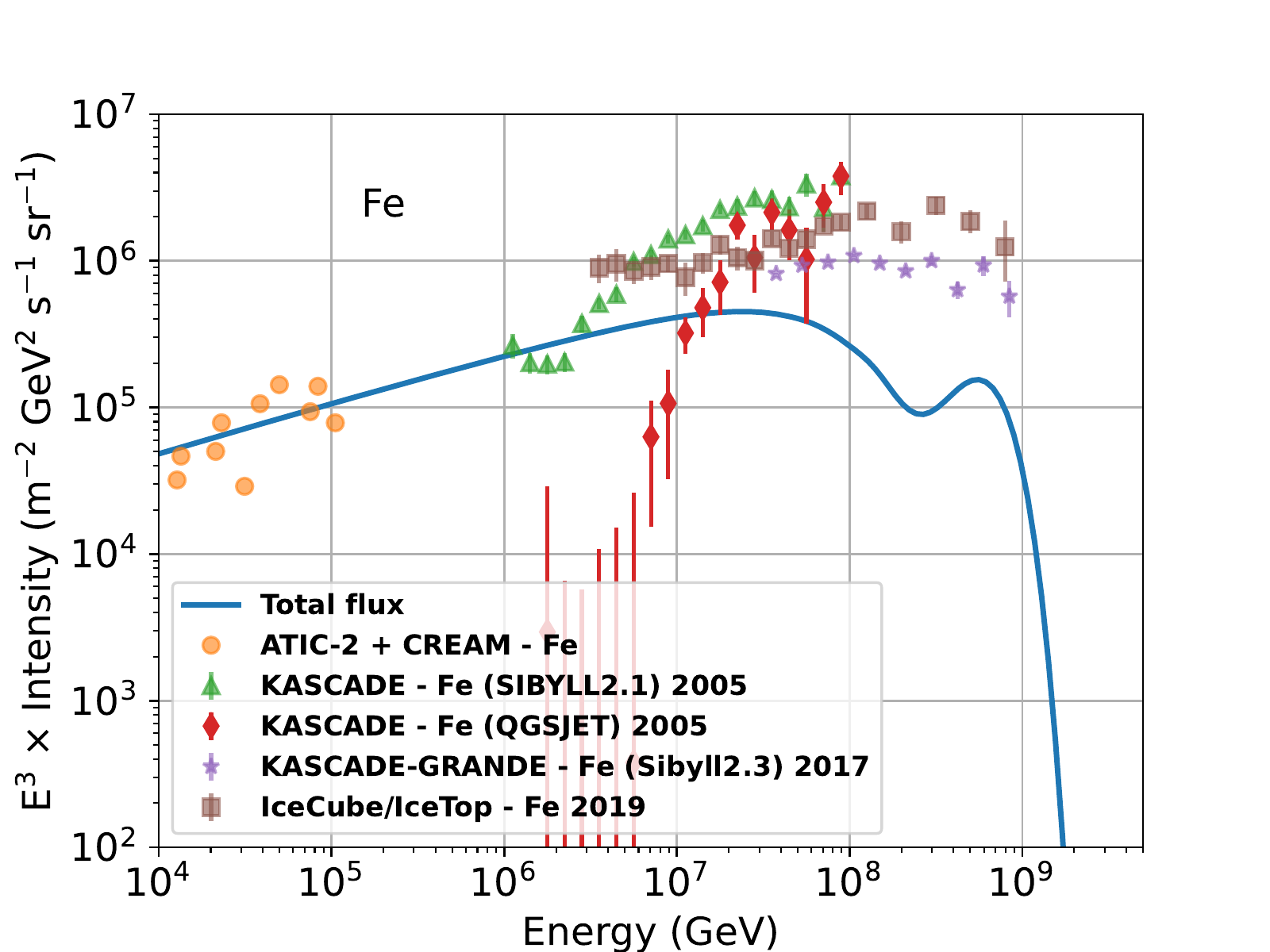} \\
\caption{Calculated cosmic-ray spectra at Earth for different species (proton, helium, carbon, iron) compared with data. The reference model (model~2) results are plotted as a thick blue line. The two different components (Galactic sources and backstreaming) are shown separately only for the proton case because their behavior is qualitatively identical for the other chemical elements. The reaccelerated component forms a bump feature due to the transition from Galactic cosmic rays to the GWTS component. The reaccelerated component of the proton spectrum at Earth is enough to explain the KASCADE (2005) data. Additionally, these protons can contribute to the measured flux at KASCADE-GRANDE and IceCube/IceTop at $\gtrsim$ 10\% between 10-40~PeV, with a maximum contribution of $\sim$ 50\% at the peak of the second bump (marked in green) at $\sim20$~PeV. The behavior of the spectrum at Earth for heavier elements is qualitatively similar to protons except the maximum energies achieved at the Galactic sources and the GWTS are shifted by a factor of $Z$.}
\label{fig:elements}
\end{figure*}

In Fig.~\ref{fig:elements}, we plot the spectra, $E^3 \times \rm intensity$ as a function of energy for some individual elements and compare them to observed elemental spectra at the disk from different experiments. The results for proton, helium, carbon, and iron are shown. The Galactic source injection for the elements is taken to be: $f_Z(p) \propto p^{-\alpha} \exp{\left( - \frac{p}{Z \times p_{\rm c,SNR}} \right)}$, with $\alpha = -4.4$ and $p_{\rm c,SNR} = $ 3~PeV. 
The chemical composition of the elements is chosen in agreement with the composition inferred from detailed studies of Galactic cosmic-ray transport problem \citep[see Table~1 of the paper by][]{2016:Thoudam}. The resulting composition consists of $\sim$ 90\% protons, 9\% helium, and 1\% of remaining heavier nuclei. Note that we do not take into account the additional complications of Galactic transport like transitions in the diffusion coefficient \citep{Genolini:2017dfb}, particle losses due to inelastic interactions with interstellar matter, or stochastic reacceleration of cosmic rays in the interstellar medium which may impact the observed spectrum at Earth \citep{2016:Thoudam}.

The flux produced by the Galactic sources and those contributed by the reaccelerated backstreaming particles are shown separately for protons in Fig.~\ref{fig:elements}. For protons, we find that the reaccelerated component can contribute between $\sim 10-50\%$ of the measured proton spectrum at IceCube \citep{2019:Aartsen} and KASCADE-GRANDE experiments \citep{2017:KG} for $E \sim 10-40$~PeV, while it fully explains the flux measured by KASCADE \citep[][]{2005:KASCADE}. This shows the viability of this mechanism to substantially contribute to the observed spectrum. The reaccelerated component is also in agreement with the helium measurements at IceCube \citep{2019:Aartsen} at $E \sim 30-60$~PeV, although these measurements have large uncertainties as indicated by the large error bars. Carbon nuclei are reaccelerated up to $\sim 100-200$ PeV, but their overall contribution to the carbon spectrum at the disk is largely unconstrained due to the lack of data in this energy range. Iron nuclei are also reaccelerated to up to $\sim 700$ PeV, and contributes at a level of $\sim$ 10\% to the spectra measured by KASCADE-GRANDE at the disk.

\begin{figure} 
    \centering
    \includegraphics[width=.50\textwidth]{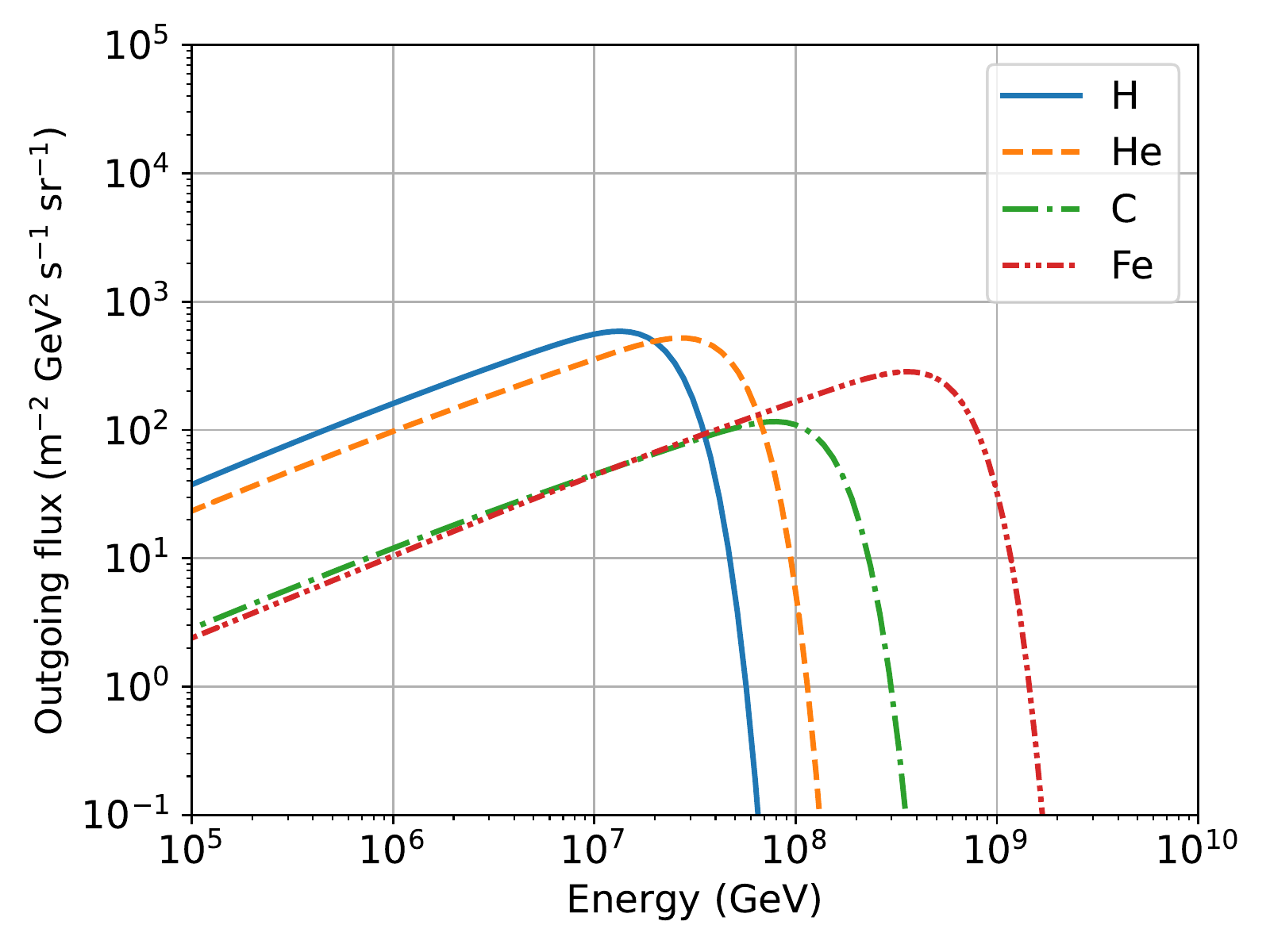}
    \caption{Fluxes of cosmic rays escaping out into the intergalactic medium for different elements. The plot shows that reacceleration at the GWTS can be a source of seeding the intergalactic medium with intermediate-energy cosmic rays, reaching up to $O(1 ~ \mathrm{EeV})$ for iron.}
    \label{fig:outgoing}
\end{figure}

In Fig.~\ref{fig:outgoing}, we plot the escaping flux into the IGM, $E^3 \times D \left.\frac{\partial f}{\partial z}\right|_{z_{\mathrm{b}}} = \frac{E^3 \times \phi_{esc}(E)}{A(z_{\mathrm{b}})}$ (Eq.~(\ref{eq:phi_esc})) for different elements. 
The plot shows that protons and helium can seed the IGM with particles having energies up to $\sim$ $O(100~\mathrm{PeV})$. Heavier nuclei, such as iron, can escape with energies as high as $\sim$ $O(1 ~\mathrm{EeV})$. Therefore, particles reaccelerated at the GWTS can play an important role in releasing intermediate-energy particles into the intergalactic medium. Active star-forming galaxies, with faster winds will reaccelerate CRs up to even higher energies of more than $\sim 100$ PV \citep[see e.g., the paper by][]{Starburst_wind}, and can dominantly contribute to seeding the IGM with intermediate-energy CRs. 

\subsection{Parameter space investigation}
\label{sec:param_space}

In this section we perform a parameter space exploration in order to understand how the contribution of the reaccelerated component at the disk changes with changing the main physical parameters of the problem. We focus in particular on 1) the wind speed and, 2) the diffusion coefficient downstream of the GWTS. These parameters have no observational constraints and, as we will see shortly, can sensitively impact the particle (re)acceleration at the GWTS. In this section, all the parameter space results presented are only for cosmic ray protons. The conclusions are valid for other heavier elements as well, albeit with maximum momentum of $Z \times p_{\rm max}^{\rm proton}$.

\begin{figure}
    \centering
    \includegraphics[width=.5\textwidth]{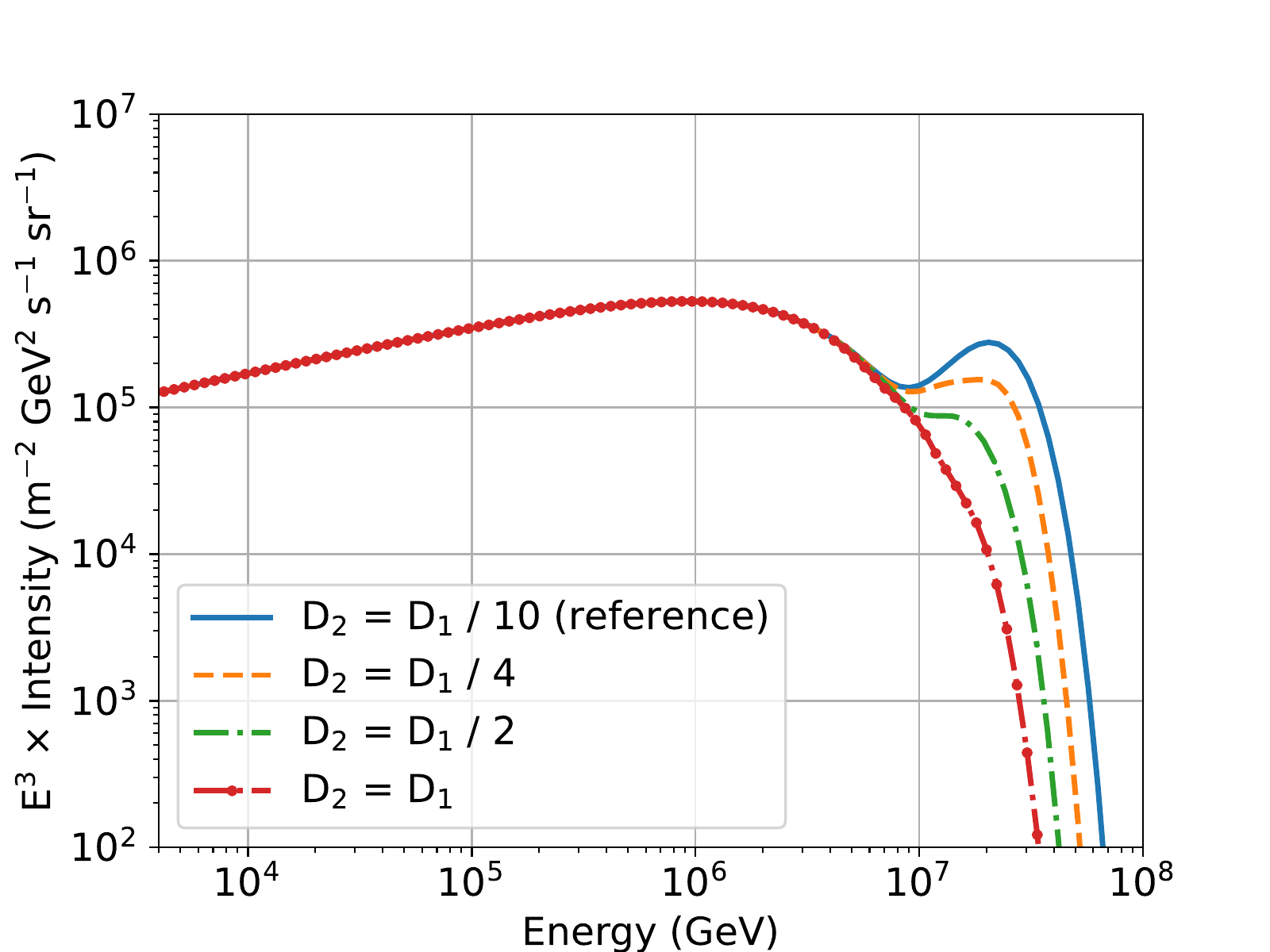}
    \caption{Proton intensity at the disk as a function of energy as the diffusion coefficient downstream of the GWTS is increased. The reference model assumes the downstream diffusion ($D_2$) is an order of magnitude smaller than the diffusion coefficient right upstream ($D_1$) of the shock. As the downstream diffusion coefficient is increased, the maximum energies achieved at the GWTS decreases, and the overall flux of reaccelerated particles back to the disk also decreases. }
    \label{fig:D2_D1}
\end{figure}

In Fig.~\ref{fig:D2_D1}, we show how the proton flux at the disk changes by parametrically changing the downstream diffusion coefficient $D_2$ from the reference assumption. In the reference case (blue curve), the downstream diffusion coefficient is assumed to be an order of magnitude smaller than the upstream diffusion coefficient ($D_2 = D_1 / 10$). We show a few more curves as the downstream diffusion coefficient is increased. The red curve describes a very pessimistic scenario in which the downstream diffusion coefficient is equal to the upstream one. We find that increasing the downstream diffusion coefficient systematically suppresses the overall flux of backstreaming particles. For a given $D_1$, as the downstream diffusion coefficient ($D_2$) increases, the maximum energies achieved at the GWTS decreases (this comes from the exponential suppression factor $\Gamma_2$ in Eq.~\ref{eq:fs}). Another effect of increasing the downstream diffusion is an increase in the escaping flux of particles, as defined in Eq.~(\ref{eq:phi_esc}), thereby suppressing the flux of the particles backstreaming to the disk. If, instead of treating the downstream diffusion coefficient parametrically (as we have done so far), one assumes that the decrease in the downstream diffusion coefficient is due to a simple magnetic field compression at the GWTS, then the compression factor is $\sim 4$, i.e., $B_2 \sim 4 B_1$ (typical for strong shocks), where $B_{1,2}$ are the magnetic fields right ahead and behind the shock, respectively. For this compression factor\footnote{The ratio $D_2 / D_1$ depends on whether the cosmic rays are in the diffusive regime or the weak-scattering regime in Eq~\ref{eq:noemie_diff}. In model 2 the upstream magnetic field at the shock is about 70~nG, and 10~PV cosmic rays are in the middle of the two diffusion regimes, where $D$ is roughly $\propto B^{-1}$.}, $D_2 / D_1 \sim B_1 / B_2$, and the maximum proton energy achieved is similar to our reference model, i.e, $\sim 30$~PV. We additionally explored scenarios of extreme suppression of the downstream diffusion coefficient, namely $D_2 \lesssim D_1/100$. In these configurations we noticed that the spectrum of backstreaming particles does not change substantially from our reference scenario since the transport in the downstream region is practically advection dominated.

\begin{figure} [t]
    \centering
    \includegraphics[width=.5\textwidth]{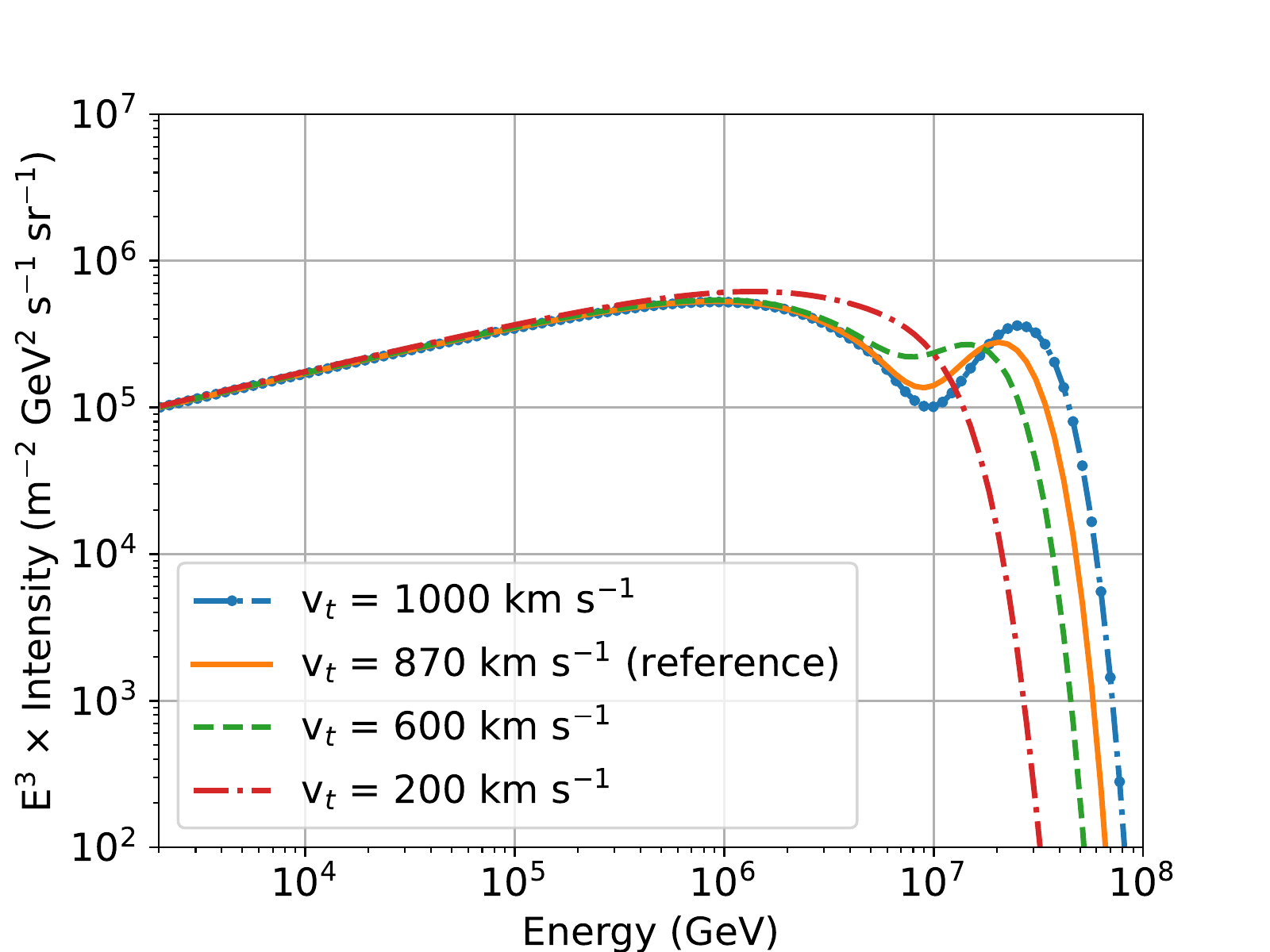}
    \caption{Proton intensity at the disk as a function of energy as the terminal speed ($v_{\mathrm{t}} = v_{\mathrm{gas}}$) is varied. The reference model has $v_{\mathrm{t}} \sim 870$~km~s$^{-1}$. Decreasing this speed to 200~km~s$^{-1}$ makes the contribution of the reaccelerated particles to be negligible at the disk. Increasing the speed to 1000~km~s$^{-1}$ increases the maximum energy achieved in the reacceleration mechanism. }
    \label{fig:speed_var}
\end{figure}

Fig.~\ref{fig:speed_var} illustrates the change in the reaccelerated component at the disk as the terminal speed (defined as the gas velocity at the shock) is varied by changing the geometry parameter $z_0$. The reference model has a terminal wind speed of $\sim$ 870~km~s$^{-1}$, as obtained in Sec. \ref{subsec:fiducial_wind} (also see Table~\ref{table:fiducial}). Decreasing the speed to 200~km~s$^{-1}$ decreases the maximum proton energy achieved at the GWTS to $\sim 5$ PeV, in accordance with Eq.~(\ref{eq:p_max}). By contrast, increasing the wind speed to $1000$~km~s$^{-1}$ increases the maximum energy and shifts the GWTS bump to higher energies. A precise measurement of the spectrum above the knee can therefore constrain the wind speed in our model. We have also checked the dependence of the overall flux at the disk on other parameters, such as the shock radius and slope of the diffusion coefficients. The effect of changing the shock radius can be qualitatively understood as follows. As the radius of the shock decreases, the size of the system decreases, and the maximum energy achieved at the GWTS also decreases in accordance with Eq.~(\ref{eq:p_max}). The results remain similar if, instead of the Kolmogorov term (slope $\frac{1}{3}$) in Eq.~(\ref{eq:QLT_diff}), we assume Kraichnan diffusion with a slope of $\frac{1}{2}$. Kraichnan diffusion also gives acceleration up to $\sim 40$~PeV for protons (equivalently, 40~PV rigidity for all elements) and can similarly contribute to the observed spectrum at the disk.

\section{Summary and Conclusions}
\label{conclusions_CRs}
We propose that $\sim$ GeV Galactic cosmic rays drive a wind out into the halo which then passes through a Galactic wind termination shock (GWTS) at a radius $\sim200\,{\rm kpc}$, where the wind ram pressure matches the pressure of the circumgalactic medium. We have applied the theory of diffusive shock acceleration at this shock and considered the possibility that some of these cosmic rays are reaccelerated to energies up to $\sim 40$~PeV and diffuse upstream back to the disk, thereby contributing to the observed spectrum in the shin region.

In carrying out this investigation, we solved the wind hydrodynamic equations to find a cosmic-ray-driven wind solution and then used a semi-analytic iteration technique developed for transport calculations by \citet{2021:Morlino} and \citet{Starburst_wind} for the case of spherical termination shocks in star clusters and starburst galaxies, respectively, to solve for cosmic ray transport in a Galactic wind. Our procedure takes into account cosmic-ray injection at the disk, diffusion, advection, adiabatic energy losses, and reacceleration at the GWTS. For the reacceleration, we considered two models of  diffusive CR transport upstream of the GWTS:  ``model 1" where $\frac{\delta B}{B} \sim 0.1$  throughout the halo (weak turbulence introduced at an outer scale which increases with $z$, as described in Sec. \ref{subsubsec:diff1}), and  ``model 2" where $\frac{\delta B}{B} \sim 1$ close to the GWTS (fully developed magnetic turbulence at an outer scale $l_{\rm max} \sim 5~\rm kpc$, corresponding to a coherence length, $l_{\rm c} \sim 1~\rm kpc$, as described in Sec. \ref{subsubsec:diff2} ).

Some important points about the model and our main results are:

\begin{itemize}

\item The existence of a Galactic wind is strongly suggested by the requirement that local cosmic rays escape in less than a hundred million years \citep{2014arXiv1407.5223L}. The cosmic-ray energy flux from the local Galactic disk is adequate to drive the wind and most of the cosmic rays out of the Galactic gravitational potential.

\item The presence of a wind carrying off mass, angular momentum, and energy from the Galactic disk should be an important feature of models of the interstellar medium because these winds can have mass loss rates of $O(0.1~\rm M_{\odot} yr^{-1})$, comparable to the star formation rate of the Galaxy \citep{2017ApJ...837..150S}.

\item We tested our results with two different prescriptions for the upstream diffusion coefficient. We found that if the diffusion coefficient ahead of the GWTS is too high, we get negligible reacceleration at the shock. The diffusion coefficient ahead of the GWTS needs to be sufficiently small to give enough reaccelerationup to $O(10~\rm PV)$ range.

\item In the latter case, cosmic rays can be efficiently reaccelerated at the Galactic wind termination shock up to rigidities ranging from 10~PV to 40~PV. This is the first time, to the best of our knowledge, that the maximum energy is derived self-consistently based on a transport model that derived from the physical parameters of the wind in the context of reacceleration at the GWTS.

\item We note that even in the limiting case when the turbulent pressure near the shock is comparable with the ram pressure, we do not reach rigidities higher than $\sim$ 70~PV in our model. In particular, they cannot account for the observed cosmic rays all the way up to the ankle in the spectrum at energy $\sim4\,{\rm EeV}$. If there are additional sources of turbulence ahead of the GWTS, such as turbulence generated by CRs themselves via resonant or non-resonant streaming instabilities, and if these can cause the diffusion coefficient to decrease close to the Bohm limit ahead of the shock, one could possibly get reacceleration up to higher rigidities.

\item Even though the wind strongly suppresses the absolute number of reaccelerated particles that can diffuse back to the Galactic disk, we find that, for our benchmark scenarios, the flux of backstreaming protons can be as high as $\sim 50 \%$ of the observed flux in the 10--40~PeV region measured in IceCube/IceTop and KASCADE-Grande experiments. The backstreaming flux at the disk is marked by a characteristic `bump' feature formed by the reaccelerated proton component at $\sim$ 10~PeV that could be detected in future observations. 

\item The location of the outer boundary of the downstream flow behind the termination shock at radius $\sim$ 500~kpc is chosen so as to roughly mimic the distance corresponding to three times the virial radius of the Galaxy from which cosmic rays can escape freely. We have verified that our results are relatively robust with respect to changes in the location of the outer boundary as long as it is located sufficiently far from the Galactic wind termination shock ($z_{\mathrm{b}} \gtrsim 2 \,z_{\mathrm{s}}$).

\item A fraction of the reaccelerated cosmic rays can seed the intergalactic medium with intermediate energy particles. Protons with energies up to $\sim$ 40~PeV and heavier nuclei like iron with energies $\sim$ $O(1~\mathrm{EeV})$ can be transmitted into the intergalactic medium. These reaccelerated cosmic rays that escape downstream from termination shocks around active star forming galaxies can undergo further reacceleration to rigidities $\sim$~10~EV by even larger intergalactic shock fronts, which may ultimately account for the spectrum of ultra-high-energy cosmic rays seen at Earth.

\end{itemize}

Our investigation opens up several directions for future studies. In order to fully assess the velocity profile of the Galactic wind and the characteristics of the termination shock, one needs to perform full MHD calculations of the wind along with the transport equation taking into account cosmic-ray pressure, thermal pressure, and magnetic stresses. Moreover, the reaccelerated cosmic rays can cause feedback effects on the GWTS, which can only be realistically modeled in the MHD framework. Additionally, the reaccelerated particles can produce multi-messenger signals during propagation, like gamma rays and neutrinos, which will be investigated elsewhere. \\

\section*{Acknowledgements}
We thank Chad Bustard, Lukas Merten, Vladimir Zirakashvilli, and Satyendra Thoudam for answering various questions related to particle acceleration at termination shocks. We thank Patrick Reichherzer for answering questions on values diffusion coefficients. We thank Giovanni Morlino for fruitful discussions. We also thank Sarah Recchia for many insightful comments on the nature of galactic winds and cosmic-ray transport therein. The research activity of E.P. was supported by Villum Fonden (project n. 18994) and by the European Union’s Horizon 2020 research and innovation program under the Marie Sklodowska-Curie grant agreement No. 847523 ‘INTERACTIONS’. N.G.’s research is supported by the Simons Foundation, the Chancellor Fellowship at UCSC and the Vera Rubin Presidential Chair.

\newpage
\bibliography{bibliography_GWTS.bib}

\begin{thebibliography}{}
\expandafter\ifx\csname natexlab\endcsname\relax\def\natexlab#1{#1}\fi
\providecommand{\url}[1]{\href{#1}{#1}}
\providecommand{\dodoi}[1]{doi:~\href{http://doi.org/#1}{\nolinkurl{#1}}}
\providecommand{\doeprint}[1]{\href{http://ascl.net/#1}{\nolinkurl{http://ascl.net/#1}}}
\providecommand{\doarXiv}[1]{\href{https://arxiv.org/abs/#1}{\nolinkurl{https://arxiv.org/abs/#1}}}

\bibitem[{{Aab} {et~al.}(2018){Aab}, {Abreu}, {Aglietta}, {Albuquerque},
  {Albury}, {Allekotte}, {Almela}, {Alvarez Castillo}, {Alvarez-Mu{\~n}iz},
  {Anastasi}, {Anchordoqui}, {Andrada}, {Andringa}, {Aramo}, {Asorey}, {Assis},
  {Avila}, {Badescu}, {Balaceanu}, {Barbato}, {Barreira Luz}, {Baur}, {Becker},
  {Bellido}, {Berat}, {Bertaina}, {Bertou}, {Biermann}, {Biteau}, {Blaess},
  {Blanco}, {Blazek}, {Bleve}, {Boh{\'a}{\v{c}}ov{\'a}}, {Bonifazi}, {Borodai},
  {Botti}, {Brack}, {Bretz}, {Bridgeman}, {Briechle}, {Buchholz}, {Bueno},
  {Buitink}, {Buscemi}, {Caballero-Mora}, {Caccianiga}, {Calcagni}, {Cancio},
  {Canfora}, {Carceller}, {Caruso}, {Castellina}, {Catalani}, {Cataldi},
  {Cazon}, {Chinellato}, {Chudoba}, {Chytka}, {Clay}, {Cobos Cerutti},
  {Colalillo}, {Coleman}, {Coluccia}, {Concei{\c{c}}{\~a}o}, {Consolati},
  {Contreras}, {Cooper}, {Coutu}, {Covault}, {Daniel}, {Dasso}, {Daumiller},
  {Dawson}, {Day}, {de Almeida}, {de Jong}, {De Mauro}, {de Mello Neto}, {De
  Mitri}, {de Oliveira}, {de Souza}, {Debatin}, {Deligny}, {Dhital}, {D{\'\i}az
  Castro}, {Diogo}, {Dobrigkeit}, {D'Olivo}, {Dorosti}, {dos Anjos}, {Dova},
  {Dundovic}, {Ebr}, {Engel}, {Erdmann}, {Escobar}, {Etchegoyen}, {Falcke},
  {Farmer}, {Farrar}, {Fauth}, {Fazzini}, {Feldbusch}, {Fenu}, {Ferreyro},
  {Figueira}, {Filip{\v{c}}i{\v{c}}}, {Freire}, {Fujii}, {Fuster},
  {Garc{\'\i}a}, {Gemmeke}, {Gherghel-Lascu}, {Ghia}, {Giaccari}, {Giammarchi},
  {Giller}, {G{\l}as}, {Glombitza}, {Golup}, {G{\'o}mez Berisso}, {G{\'o}mez
  Vitale}, {Gonz{\'a}lez}, {Goos}, {G{\'o}ra}, {Gorgi}, {Gottowik}, {Grubb},
  {Guarino}, {Guedes}, {Guido}, {Halliday}, {Hampel}, {Hansen}, {Harari},
  {Harrison}, {Harvey}, {Haungs}, {Hebbeker}, {Heck}, {Heimann}, {Hill},
  {Hojvat}, {Holt}, {Homola}, {H{\"o}randel}, {Horvath}, {Hrabovsk{\'y}},
  {Huege}, {Hulsman}, {Insolia}, {Isar}, {Jandt}, {Johnsen}, {Josebachuili},
  {Jurysek}, {K{\"a}{\"a}p{\"a}}, {Kampert}, {Keilhauer}, {Kemmerich}, {Kemp},
  {Klages}, {Kleifges}, {Kleinfeller}, {Krause}, {Kuempel}, {Kukec Mezek},
  {Kuotb Awad}, {Lago}, {LaHurd}, {Lang}, {Legumina}, {Leigui de Oliveira},
  {Lenok}, {Letessier-Selvon}, {Lhenry-Yvon}, {Lo Presti}, {Lopes},
  {L{\'o}pez}, {L{\'o}pez Casado}, {Lorek}, {Luce}, {Lucero}, {Malacari},
  {Mallamaci}, {Mancarella}, {Mandat}, {Mantsch}, {Mariazzi}, {Mari{\c{s}}},
  {Marsella}, {Martello}, {Martinez}, {Mart{\'\i}nez Bravo}, {Mathes},
  {Mathys}, {Matthews}, {Matthiae}, {Mayotte}, {Mazur}, {Medina-Tanco}, {Melo},
  {Menshikov}, {Merenda}, {Michal}, {Micheletti}, {Middendorf}, {Miramonti},
  {Mitrica}, {Mockler}, {Mollerach}, {Montanet}, {Morello}, {Morlino},
  {Mostaf{\'a}}, {M{\"u}ller}, {Muller}, {M{\"u}ller}, {Mussa}, {Nellen},
  {Nguyen}, {Niculescu-Oglinzanu}, {Niechciol}, {Nitz}, {Nosek}, {Novotny},
  {No{\v{z}}ka}, {Nucita}, {N{\'u}{\~n}ez}, {Olinto}, {Palatka}, {Pallotta},
  {Papenbreer}, {Parente}, {Parra}, {Pech}, {Pedreira}, {P{\c{e}}kala},
  {Pelayo}, {Pe{\~n}a-Rodriguez}, {Pereira}, {Perlin}, {Perrone}, {Peters},
  {Petrera}, {Phuntsok}, {Pierog}, {Pimenta}, {Pirronello}, {Platino}, {Poh},
  {Pont}, {Porowski}, {Prado}, {Privitera}, {Prouza}, {Puyleart}, {Querchfeld},
  {Quinn}, {Ramos-Pollan}, {Rautenberg}, {Ravignani}, {Reininghaus}, {Ridky},
  {Riehn}, {Risse}, {Ristori}, {Rizi}, {Rodrigues de Carvalho}, {Rodriguez
  Rojo}, {Roncoroni}, {Roth}, {Roulet}, {Rovero}, {Ruehl}, {Saffi}, {Saftoiu},
  {Salamida}, {Salazar}, {Saleh}, {Salina}, {S{\'a}nchez}, {Santos}, {Santos},
  {Sarazin}, {Sarmento}, {Sarmiento-Cano}, {Sato}, {Savina}, {Schauer},
  {Scherini}, {Schieler}, {Schimassek}, {Schimp}, {Schmidt}, {Scholten},
  {Schov{\'a}nek}, {Schr{\"o}der}, {Schr{\"o}der}, {Schumacher}, {Sciutto},
  {Shellard}, {Sigl}, {Silli}, {Sima}, {{\v{S}}m{\'\i}da}, {Snow}, {Sommers},
  {Soriano}, {Souchard}, {Squartini}, {Stanca}, {Stani{\v{c}}}, {Stasielak},
  {Stassi}, {Stolpovskiy}, {Streich}, {Suarez}, {Su{\'a}rez-Dur{\'a}n},
  {Sudholz}, {Suomij{\"a}rvi}, {Supanitsky}, {{\v{S}}up{\'\i}k}, {Szadkowski},
  {Taboada}, {Taborda}, {Tapia}, {Timmermans}, {Todero Peixoto}, {Tom{\'e}},
  {Torralba Elipe}, {Travnicek}, {Trini}, {Tueros}, {Ulrich}, {Unger}, {Urban},
  {Vald{\'e}s Galicia}, {Vali{\~n}o}, {Valore}, {van Bodegom}, {van den Berg},
  {van Vliet}, {Varela}, {Vargas C{\'a}rdenas}, {V{\'a}zquez}, {Veberi{\v{c}}},
  {Ventura}, {Vergara Quispe}, {Verzi}, {Vicha}, {Villase{\~n}or}, {Vorobiov},
  {Wahlberg}, {Wainberg}, {Watson}, {Weber}, {Weindl}, {Wiede{\'n}ski},
  {Wiencke}, {Wilczy{\'n}ski}, {Wirtz}, {Wittkowski}, {Wundheiler}, {Yang},
  {Yushkov}, {Zas}, {Zavrtanik}, {Zavrtanik}, {Zehrer}, {Zepeda}, {Zimmermann},
  {Ziolkowski}, {Zong}, {Zuccarello}, \& {Pierre Auger
  Collaboration}}]{PAO_2018}
{Aab}, A., {Abreu}, P., {Aglietta}, M., {et~al.} 2018, \apj, 868, 4,
  \dodoi{10.3847/1538-4357/aae689}

\bibitem[{{Aartsen} {et~al.}(2019){Aartsen}, {Ackermann}, {Adams}, {Aguilar},
  {Ahlers}, {Ahrens}, {Alispach}, {Andeen}, {Anderson}, {Ansseau}, {Anton},
  {Arg{\"u}elles}, {Auffenberg}, {Axani}, {Backes}, {Bagherpour}, {Bai},
  {Barbano}, {Barwick}, {Baum}, {Baur}, {Bay}, {Beatty}, {Becker}, {Becker
  Tjus}, {BenZvi}, {Berley}, {Bernardini}, {Besson}, {Binder}, {Bindig},
  {Blaufuss}, {Blot}, {Bohm}, {B{\"o}rner}, {B{\"o}ser}, {Botner},
  {B{\"o}ttcher}, {Bourbeau}, {Bourbeau}, {Bradascio}, {Braun}, {Bretz},
  {Bron}, {Brostean-Kaiser}, {Burgman}, {Buscher}, {Busse}, {Carver}, {Chen},
  {Cheung}, {Chirkin}, {Clark}, {Classen}, {Collin}, {Conrad}, {Coppin},
  {Correa}, {Cowen}, {Cross}, {Dave}, {de Andr{\'e}}, {De Clercq}, {DeLaunay},
  {Dembinski}, {Deoskar}, {De Ridder}, {Desiati}, {de Vries}, {de Wasseige},
  {de With}, {DeYoung}, {Diaz}, {D{\'\i}az-V{\'e}lez}, {Dujmovic}, {Dunkman},
  {Dvorak}, {Eberhardt}, {Ehrhardt}, {Eller}, {Evenson}, {Fahey}, {Fazely},
  {Felde}, {Feusels}, {Filimonov}, {Finley}, {Franckowiak}, {Friedman},
  {Fritz}, {Gaisser}, {Gallagher}, {Ganster}, {Garrappa}, {Gerhardt},
  {Ghorbani}, {Glauch}, {Gl{\"u}senkamp}, {Goldschmidt}, {Gonzalez}, {Grant},
  {Griffith}, {G{\"u}nder}, {G{\"u}nd{\"u}z}, {Haack}, {Hallgren}, {Halve},
  {Halzen}, {Hanson}, {Hebecker}, {Heereman}, {Heix}, {Helbing}, {Hellauer},
  {Henningsen}, {Hickford}, {Hignight}, {Hill}, {Hoffman}, {Hoffmann},
  {Hoinka}, {Hokanson-Fasig}, {Hoshina}, {Huang}, {Huber}, {Hultqvist},
  {H{\"u}nnefeld}, {Hussain}, {In}, {Iovine}, {Ishihara}, {Jacobi},
  {Japaridze}, {Jeong}, {Jero}, {Jones}, {Jonske}, {Joppe}, {Kang}, {Kappes},
  {Kappesser}, {Karg}, {Karl}, {Karle}, {Katz}, {Kauer}, {Kelley},
  {Kheirandish}, {Kim}, {Kintscher}, {Kiryluk}, {Kittler}, {Klein}, {Koirala},
  {Kolanoski}, {K{\"o}pke}, {Kopper}, {Kopper}, {Koskinen}, {Kowalski},
  {Krings}, {Kr{\"u}ckl}, {Kulacz}, {Kunwar}, {Kurahashi}, {Kyriacou},
  {Labare}, {Lanfranchi}, {Larson}, {Lauber}, {Lazar}, {Leonard}, {Leuermann},
  {Liu}, {Lohfink}, {Lozano Mariscal}, {Lu}, {Lucarelli}, {L{\"u}nemann},
  {Luszczak}, {Madsen}, {Maggi}, {Mahn}, {Makino}, {Mallik}, {Mallot},
  {Mancina}, {Mari{\c{s}}}, {Maruyama}, {Mase}, {Maunu}, {Meagher}, {Medici},
  {Medina}, {Meier}, {Meighen-Berger}, {Menne}, {Merino}, {Meures}, {Miarecki},
  {Micallef}, {Moment{\'e}}, {Montaruli}, {Moore}, {Morse}, {Moulai}, {Muth},
  {Nagai}, {Nahnhauer}, {Nakarmi}, {Naumann}, {Neer}, {Niederhausen},
  {Nowicki}, {Nygren}, {Obertacke Pollmann}, {Olivas}, {O'Murchadha},
  {O'Sullivan}, {Palczewski}, {Pandya}, {Pankova}, {Park}, {Peiffer},
  {P{\'e}rez de los Heros}, {Philippen}, {Pieloth}, {Pinat}, {Pizzuto}, {Plum},
  {Porcelli}, {Price}, {Przybylski}, {Raab}, {Raissi}, {Rameez}, {Rauch},
  {Rawlins}, {Rea}, {Reimann}, {Relethford}, {Renzi}, {Resconi}, {Rhode},
  {Richman}, {Robertson}, {Rongen}, {Rott}, {Ruhe}, {Ryckbosch}, {Rysewyk},
  {Safa}, {Sanchez Herrera}, {Sandrock}, {Sandroos}, {Santander}, {Sarkar},
  {Sarkar}, {Satalecka}, {Schaufel}, {Schlunder}, {Schmidt}, {Schneider},
  {Schneider}, {Schumacher}, {Sclafani}, {Seckel}, {Seunarine}, {Shefali},
  {Silva}, {Snihur}, {Soedingrekso}, {Soldin}, {Song}, {Spiczak}, {Spiering},
  {Stachurska}, {Stamatikos}, {Stanev}, {Stasik}, {Stein}, {Stettner},
  {Steuer}, {Stezelberger}, {Stokstad}, {St{\"o}{\ss}l}, {Strotjohann},
  {St{\"u}rwald}, {Stuttard}, {Sullivan}, {Sutherland}, {Taboada}, {Tenholt},
  {Ter-Antonyan}, {Terliuk}, {Tilav}, {Tomankova}, {T{\"o}nnis}, {Toscano},
  {Tosi}, {Tselengidou}, {Tung}, {Turcati}, {Turcotte}, {Turley}, {Ty},
  {Unger}, {Unland Elorrieta}, {Usner}, {Vandenbroucke}, {Van Driessche}, {van
  Eijk}, {van Eijndhoven}, {Vanheule}, {van Santen}, {Vraeghe}, {Walck},
  {Wallace}, {Wallraff}, {Wandkowsky}, {Watson}, {Weaver}, {Weiss}, {Weldert},
  {Wendt}, {Werthebach}, {Westerhoff}, {Whelan}, {Whitehorn}, {Wiebe},
  {Wiebusch}, {Wille}, {Williams}, {Wills}, {Wolf}, {Wood}, {Wood},
  {Woschnagg}, {Wrede}, {Xu}, {Xu}, {Xu}, {Yanez}, {Yodh}, {Yoshida}, {Yuan},
  {Z{\"o}cklein}, \& {IceCube Collaboration}}]{2019:Aartsen}
{Aartsen}, M.~G., {Ackermann}, M., {Adams}, J., {et~al.} 2019, \prd, 100,
  082002, \dodoi{10.1103/PhysRevD.100.082002}

\bibitem[{Abeysekara {et~al.}(2021{\natexlab{a}})}]{Abeysekara:2021yum}
Abeysekara, A.~U., {et~al.} 2021{\natexlab{a}}, Nature Astron., 5, 465,
  \dodoi{10.1038/s41550-021-01318-y}

\bibitem[{Abeysekara {et~al.}(2021{\natexlab{b}})}]{HAWC:2021zmq}
---. 2021{\natexlab{b}}, PoS, ICRC2021, 811, \dodoi{10.22323/1.395.0811}

\bibitem[{Adriani {et~al.}(2019)Adriani, Akaike, Asano, Asaoka, Bagliesi,
  Berti, Bigongiari, Binns, Bonechi, Bongi, Brogi, Bruno, Buckley, Cannady,
  Castellini, Checchia, Cherry, Collazuol, Di~Felice, Ebisawa, Fuke, Guzik,
  Hams, Hasebe, Hibino, Ichimura, Ioka, Ishizaki, Israel, Kasahara, Kataoka,
  Kataoka, Katayose, Kato, Kawanaka, Kawakubo, Kohri, Krawczynski, Krizmanic,
  Lomtadze, Maestro, Marrocchesi, Messineo, Mitchell, Miyake, Moiseev, Mori,
  Mori, Mori, Motz, Munakata, Murakami, Nakahira, Nishimura, de~Nolfo, Okuno,
  Ormes, Ozawa, Pacini, Palma, Papini, Penacchioni, Rauch, Ricciarini, Sakai,
  Sakamoto, Sasaki, Shimizu, Shiomi, Sparvoli, Spillantini, Stolzi, Suh, Sulaj,
  Takahashi, Takayanagi, Takita, Tamura, Terasawa, Tomida, Torii, Tsunesada,
  Uchihori, Ueno, Vannuccini, Wefel, Yamaoka, Yanagita, Yoshida, \&
  Yoshida}]{2019:CALET}
Adriani, O., Akaike, Y., Asano, K., {et~al.} 2019, Phys. Rev. Lett., 122,
  181102, \dodoi{10.1103/PhysRevLett.122.181102}

\bibitem[{Adriani {et~al.}(2021)Adriani, Akaike, Asano, Asaoka, Berti,
  Bigongiari, Binns, Bongi, Brogi, Bruno, Buckley, Cannady, Castellini,
  Checchia, Cherry, Collazuol, Ebisawa, Fuke, Gonzi, Guzik, Hams, Hibino,
  Ichimura, Ioka, Ishizaki, Israel, Kasahara, Kataoka, Kataoka, Katayose, Kato,
  Kawanaka, Kawakubo, Kobayashi, Kohri, Krawczynski, Krizmanic, Link, Maestro,
  Marrocchesi, Messineo, Mitchell, Miyake, Moiseev, Mori, Mori, Motz, Munakata,
  Nakahira, Nishimura, de~Nolfo, Okuno, Ormes, Ospina, Ozawa, Pacini, Papini,
  Rauch, Ricciarini, Sakai, Sakamoto, Sasaki, Shimizu, Shiomi, Spillantini,
  Stolzi, Sugita, Sulaj, Takita, Tamura, Terasawa, Torii, Tsunesada, Uchihori,
  Vannuccini, Wefel, Yamaoka, Yanagita, Yoshida, \& Yoshida}]{2021:CALET}
---. 2021, Phys. Rev. Lett., 126, 241101,
  \dodoi{10.1103/PhysRevLett.126.241101}

\bibitem[{{Aguilar} {et~al.}(2015){Aguilar}, {Aisa}, {Alpat}, {Alvino},
  {Ambrosi}, {Andeen}, {Arruda}, {Attig}, {Azzarello}, {Bachlechner}, {Barao},
  {Barrau}, {Barrin}, {Bartoloni}, {Basara}, {Battarbee}, {Battiston}, {Bazo},
  {Becker}, {Behlmann}, {Beischer}, {Berdugo}, {Bertucci}, {Bigongiari},
  {Bindi}, {Bizzaglia}, {Bizzarri}, {Boella}, {de Boer}, {Bollweg},
  {Bonnivard}, {Borgia}, {Borsini}, {Boschini}, {Bourquin}, {Burger}, {Cadoux},
  {Cai}, {Capell}, {Caroff}, {Casaus}, {Cascioli}, {Castellini}, {Cernuda},
  {Cerreta}, {Cervelli}, {Chae}, {Chang}, {Chen}, {Chen}, {Cheng}, {Chen},
  {Cheng}, {Chou}, {Choumilov}, {Choutko}, {Chung}, {Clark}, {Clavero},
  {Coignet}, {Consolandi}, {Contin}, {Corti}, {Gil}, {Coste}, {Creus},
  {Crispoltoni}, {Cui}, {Dai}, {Delgado}, {Della Torre}, {Demirk{\"o}z},
  {Derome}, {Di Falco}, {Di Masso}, {Dimiccoli}, {D{\'\i}az}, {von Doetinchem},
  {Donnini}, {Du}, {Duranti}, {D'Urso}, {Eline}, {Eppling}, {Eronen}, {Fan},
  {Farnesini}, {Feng}, {Fiandrini}, {Fiasson}, {Finch}, {Fisher},
  {Galaktionov}, {Gallucci}, {Garc{\'\i}a}, {Garc{\'\i}a-L{\'o}pez},
  {Gargiulo}, {Gast}, {Gebauer}, {Gervasi}, {Ghelfi}, {Gillard}, {Giovacchini},
  {Goglov}, {Gong}, {Goy}, {Grabski}, {Grandi}, {Graziani}, {Guandalini},
  {Guerri}, {Guo}, {Haas}, {Habiby}, {Haino}, {Han}, {He}, {Heil}, {Hoffman},
  {Hsieh}, {Huang}, {Huh}, {Incagli}, {Ionica}, {Jang}, {Jinchi}, {Kanishev},
  {Kim}, {Kim}, {Kirn}, {Kossakowski}, {Kounina}, {Kounine}, {Koutsenko},
  {Krafczyk}, {La Vacca}, {Laudi}, {Laurenti}, {Lazzizzera}, {Lebedev}, {Lee},
  {Lee}, {Leluc}, {Levi}, {Li}, {Li}, {Li}, {Li}, {Li}, {Li}, {Li}, {Li}, {Li},
  {Lim}, {Lin}, {Lipari}, {Lippert}, {Liu}, {Liu}, {Lolli}, {Lomtadze}, {Lu},
  {Lu}, {Lu}, {Luebelsmeyer}, {Luo}, {Lv}, {Majka}, {Ma{\~n}{\'a}},
  {Mar{\'\i}n}, {Martin}, {Mart{\'\i}nez}, {Masi}, {Maurin}, {Menchaca-Rocha},
  {Meng}, {Mo}, {Morescalchi}, {Mott}, {M{\"u}ller}, {Ni}, {Nikonov},
  {Nozzoli}, {Nunes}, {Obermeier}, {Oliva}, {Orcinha}, {Palmonari},
  {Palomares}, {Paniccia}, {Papi}, {Pauluzzi}, {Pedreschi}, {Pensotti},
  {Pereira}, {Picot-Clemente}, {Pilo}, {Piluso}, {Pizzolotto}, {Plyaskin},
  {Pohl}, {Poireau}, {Postaci}, {Putze}, {Quadrani}, {Qi}, {Qin}, {Qu},
  {R{\"a}ih{\"a}}, {Rancoita}, {Rapin}, {Ricol}, {Rodr{\'\i}guez},
  {Rosier-Lees}, {Rozhkov}, {Rozza}, {Sagdeev}, {Sandweiss}, {Saouter},
  {Sbarra}, {Schael}, {Schmidt}, {von Dratzig}, {Schwering}, {Scolieri}, {Seo},
  {Shan}, {Shan}, {Shi}, {Shi}, {Shi}, {Siedenburg}, {Son}, {Spada},
  {Spinella}, {Sun}, {Sun}, {Tacconi}, {Tang}, {Tang}, {Tang}, {Tao},
  {Tescaro}, {Ting}, {Ting}, {Tomassetti}, {Torsti}, {T{\"u}rko{\v{g}}lu},
  {Urban}, {Vagelli}, {Valente}, {Vannini}, {Valtonen}, {Vaurynovich},
  {Vecchi}, {Velasco}, {Vialle}, {Vitale}, {Vitillo}, {Wang}, {Wang}, {Wang},
  {Wang}, {Wang}, {Wang}, {Weng}, {Whitman}, {Wienkenh{\"o}ver}, {Wu}, {Wu},
  {Xia}, {Xie}, {Xie}, {Xiong}, {Xin}, {Xu}, {Xu}, {Yan}, {Yang}, {Yang}, {Ye},
  {Yi}, {Yu}, {Yu}, {Zeissler}, {Zhang}, {Zhang}, {Zhang}, {Zhang}, {Zheng},
  {Zhuang}, {Zhukov}, {Zichichi}, {Zimmermann}, {Zuccon}, {Zurbach}, \& {AMS
  Collaboration}}]{2015:AMS}
{Aguilar}, M., {Aisa}, D., {Alpat}, B., {et~al.} 2015, \prl, 114, 171103,
  \dodoi{10.1103/PhysRevLett.114.171103}

\bibitem[{Aguilar {et~al.}(2016)Aguilar, Ali~Cavasonza, Ambrosi, Arruda, Attig,
  Aupetit, Azzarello, Bachlechner, Barao, Barrau, Barrin, Bartoloni, Basara,
  Ba\ifmmode \mbox{\c{s}}\else \c{s}\fi{}e\ifmmode \breve{g}\else
  \u{g}\fi{}mez-du Pree, Battarbee, Battiston, Becker, Behlmann, Beischer,
  Berdugo, Bertucci, Bindel, Bindi, Boella, de~Boer, Bollweg, Bonnivard,
  Borgia, Boschini, Bourquin, Bueno, Burger, Cadoux, Cai, Capell, Caroff,
  Casaus, Castellini, Cervelli, Chae, Chang, Chen, Chen, Chen, Cheng, Chou,
  Choumilov, Choutko, Chung, Clark, Clavero, Coignet, Consolandi, Contin,
  Corti, Creus, Crispoltoni, Cui, Dai, Delgado, Della~Torre, Demakov,
  Demirk\"oz, Derome, Di~Falco, Dimiccoli, D\'{\i}az, von Doetinchem, Dong,
  Donnini, Duranti, D'Urso, Egorov, Eline, Eronen, Feng, Fiandrini, Finch,
  Fisher, Formato, Galaktionov, Gallucci, Garc\'{\i}a, Garc\'{\i}a-L\'opez,
  Gargiulo, Gast, Gebauer, Gervasi, Ghelfi, Giovacchini, Goglov, G\'omez-Coral,
  Gong, Goy, Grabski, Grandi, Graziani, Guo, Haino, Han, He, Heil, Hoffman,
  Hsieh, Huang, Huang, Huh, Incagli, Ionica, Jang, Jinchi, Kang, Kanishev, Kim,
  Kim, Kirn, Konak, Kounina, Kounine, Koutsenko, Krafczyk, La~Vacca, Laudi,
  Laurenti, Lazzizzera, Lebedev, Lee, Lee, Leluc, Li, Li, Li, Li, Li, Li, Li,
  Li, Li, Lim, Lin, Lipari, Lippert, Liu, Liu, Lordello, Lu, Lu, Luebelsmeyer,
  Luo, Luo, Lv, Machate, Majka, Ma\~n\'a, Mar\'{\i}n, Martin, Mart\'{\i}nez,
  Masi, Maurin, Menchaca-Rocha, Meng, Mikuni, Mo, Morescalchi, Mott, Nelson,
  Ni, Nikonov, Nozzoli, Oliva, Orcinha, Palmonari, Palomares, Paniccia,
  Pauluzzi, Pensotti, Pereira, Picot-Clemente, Pilo, Pizzolotto, Plyaskin,
  Pohl, Poireau, Putze, Quadrani, Qi, Qin, Qu, R\"aih\"a, Rancoita, Rapin,
  Ricol, Rosier-Lees, Rozhkov, Rozza, Sagdeev, Sandweiss, Saouter, Schael,
  Schmidt, Schulz~von Dratzig, Schwering, Seo, Shan, Shi, Siedenburg, Son,
  Song, Sun, Tacconi, Tang, Tang, Tao, Tescaro, Ting, Ting, Tomassetti, Torsti,
  T\"urko\ifmmode~\breve{g}\else \u{g}\fi{}lu, Urban, Vagelli, Valente,
  Vannini, Valtonen, V\'azquez~Acosta, Vecchi, Velasco, Vialle, Vitale,
  Vitillo, Wang, Wang, Wang, Wang, Wang, Wang, Wei, Weng, Whitman,
  Wienkenh\"over, Wu, Wu, Xia, Xiong, Xu, Yan, Yang, Yang, Yang, Yi, Yu, Yu,
  Zeissler, Zhang, Zhang, Zhang, Zhang, Zhang, Zhang, Zheng, Zhu, Zhuang,
  Zhukov, Zichichi, Zimmermann, \& Zuccon}]{B-over-C-ratio}
Aguilar, M., Ali~Cavasonza, L., Ambrosi, G., {et~al.} 2016, Phys. Rev. Lett.,
  117, 231102, \dodoi{10.1103/PhysRevLett.117.231102}

\bibitem[{{Aguilar} {et~al.}(2017){Aguilar}, {Ali Cavasonza}, {Alpat},
  {Ambrosi}, {Arruda}, {Attig}, {Aupetit}, {Azzarello}, {Bachlechner}, {Barao},
  {Barrau}, {Barrin}, {Bartoloni}, {Basara}, {Ba{\c{s}}e{\v{g}}mez-du Pree},
  {Battarbee}, {Battiston}, {Becker}, {Behlmann}, {Beischer}, {Berdugo},
  {Bertucci}, {Bindel}, {Bindi}, {de Boer}, {Bollweg}, {Bonnivard}, {Borgia},
  {Boschini}, {Bourquin}, {Bueno}, {Burger}, {Burger}, {Cadoux}, {Cai},
  {Capell}, {Caroff}, {Casaus}, {Castellini}, {Cervelli}, {Chae}, {Chang},
  {Chen}, {Chen}, {Chen}, {Cheng}, {Chou}, {Choumilov}, {Choutko}, {Chung},
  {Clark}, {Clavero}, {Coignet}, {Consolandi}, {Contin}, {Corti}, {Creus},
  {Crispoltoni}, {Cui}, {Dadzie}, {Dai}, {Datta}, {Delgado}, {Della Torre},
  {Demakov}, {Demirk{\"o}z}, {Derome}, {Di Falco}, {Dimiccoli}, {D{\'\i}az},
  {von Doetinchem}, {Dong}, {Donnini}, {Duranti}, {D'Urso}, {Egorov}, {Eline},
  {Eronen}, {Feng}, {Fiandrini}, {Fisher}, {Formato}, {Galaktionov},
  {Gallucci}, {Garc{\'\i}a-L{\'o}pez}, {Gargiulo}, {Gast}, {Gebauer},
  {Gervasi}, {Ghelfi}, {Giovacchini}, {G{\'o}mez-Coral}, {Gong}, {Goy},
  {Grabski}, {Grandi}, {Graziani}, {Guo}, {Haino}, {Han}, {He}, {Heil},
  {Hoffman}, {Hsieh}, {Huang}, {Huang}, {Huh}, {Incagli}, {Ionica}, {Jang},
  {Jia}, {Jinchi}, {Kang}, {Kanishev}, {Khiali}, {Kim}, {Kim}, {Kirn}, {Konak},
  {Kounina}, {Kounine}, {Koutsenko}, {Kulemzin}, {La Vacca}, {Laudi},
  {Laurenti}, {Lazzizzera}, {Lebedev}, {Lee}, {Lee}, {Leluc}, {Li}, {Li}, {Li},
  {Li}, {Li}, {Li}, {Li}, {Lim}, {Lin}, {Lipari}, {Lippert}, {Liu}, {Liu},
  {Lordello}, {Lu}, {Lu}, {Luebelsmeyer}, {Luo}, {Luo}, {Lyu}, {Machate},
  {Ma{\~n}{\'a}}, {Mar{\'\i}n}, {Martin}, {Mart{\'\i}nez}, {Masi}, {Maurin},
  {Menchaca-Rocha}, {Meng}, {Mikuni}, {Mo}, {Mott}, {Nelson}, {Ni}, {Nikonov},
  {Nozzoli}, {Oliva}, {Orcinha}, {Palmonari}, {Palomares}, {Paniccia},
  {Pauluzzi}, {Pensotti}, {Perrina}, {Phan}, {Picot-Clemente}, {Pilo},
  {Pizzolotto}, {Plyaskin}, {Pohl}, {Poireau}, {Quadrani}, {Qi}, {Qin}, {Qu},
  {R{\"a}ih{\"a}}, {Rancoita}, {Rapin}, {Ricol}, {Rosier-Lees}, {Rozhkov},
  {Rozza}, {Sagdeev}, {Schael}, {Schmidt}, {Schulz von Dratzig}, {Schwering},
  {Seo}, {Shan}, {Shi}, {Siedenburg}, {Son}, {Song}, {Tacconi}, {Tang}, {Tang},
  {Tescaro}, {Ting}, {Ting}, {Tomassetti}, {Torsti}, {T{\"u}rko{\v{g}}lu},
  {Urban}, {Vagelli}, {Valente}, {Valtonen}, {V{\'a}zquez Acosta}, {Vecchi},
  {Velasco}, {Vialle}, {Vitale}, {Vitillo}, {Wang}, {Wang}, {Wang}, {Wang},
  {Wang}, {Wang}, {Wei}, {Weng}, {Whitman}, {Wu}, {Wu}, {Xiong}, {Xu}, {Yan},
  {Yang}, {Yang}, {Yang}, {Yi}, {Yu}, {Yu}, {Zannoni}, {Zeissler}, {Zhang},
  {Zhang}, {Zhang}, {Zhang}, {Zhang}, {Zhang}, {Zheng}, {Zhuang}, {Zhukov},
  {Zichichi}, {Zimmermann}, {Zuccon}, \& {AMS Collaboration}}]{2017:AMS}
{Aguilar}, M., {Ali Cavasonza}, L., {Alpat}, B., {et~al.} 2017, \prl, 119,
  251101, \dodoi{10.1103/PhysRevLett.119.251101}

\bibitem[{Aguilar {et~al.}(2020)Aguilar, Ali~Cavasonza, Ambrosi, Arruda, Attig,
  Barao, Barrin, Bartoloni, Ba\ifmmode \mbox{\c{s}}\else \c{s}\fi{}e\ifmmode
  \breve{g}\else \u{g}\fi{}mez-du Pree, Battiston, Becker, Behlmann, Beischer,
  Berdugo, Bertucci, Bindi, de~Boer, Bollweg, Borgia, Boschini, Bourquin,
  Bueno, Burger, Burger, Burmeister, Cai, Capell, Casaus, Castellini, Cervelli,
  Chang, Chen, Chen, Chen, Cheng, Chou, Chouridou, Choutko, Chung, Clark,
  Coignet, Consolandi, Contin, Corti, Cui, Dadzie, Dai, Delgado, Della~Torre,
  Demirk\"oz, Derome, Di~Falco, Di~Felice, D\'{\i}az, Dimiccoli, von
  Doetinchem, Dong, Donnini, Duranti, Egorov, Eline, Feng, Fiandrini, Fisher,
  Formato, Freeman, Galaktionov, G\'amez, Garc\'{\i}a-L\'opez, Gargiulo, Gast,
  Gebauer, Gervasi, Giovacchini, G\'omez-Coral, Gong, Goy, Grabski, Grandi,
  Graziani, Guo, Haino, Han, Hashmani, He, Heber, Hsieh, Hu, Huang, Incagli,
  Jang, Jia, Jinchi, Kanishev, Khiali, Kim, Kirn, Konyushikhin, Kounina,
  Kounine, Koutsenko, Kuhlman, Kulemzin, La~Vacca, Laudi, Laurenti, Lazzizzera,
  Lebedev, Lee, Lee, Li, Li, Li, Li, Li, Li, Light, Lin, Lippert, Liu, Lu, Lu,
  Luebelsmeyer, Luo, Lyu, Machate, Ma\~n\'a, Mar\'{\i}n, Marquardt, Martin,
  Mart\'{\i}nez, Masi, Maurin, Menchaca-Rocha, Meng, Mo, Molero, Mott,
  Mussolin, Ni, Nikonov, Nozzoli, Oliva, Orcinha, Palermo, Palmonari, Paniccia,
  Pashnin, Pauluzzi, Pensotti, Phan, Piandani, Plyaskin, Poluianov, Qi, Qin,
  Qu, Quadrani, Rancoita, Rapin, Reina~Conde, Rosier-Lees, Rozhkov, Rozza,
  Sagdeev, Schael, Schmidt, Schulz~von Dratzig, Schwering, Seo, Shan, Shi,
  Siedenburg, Solano, Sonnabend, Song, Sun, Sun, Tacconi, Tang, Tang, Tian,
  Ting, Ting, Tomassetti, Torsti, T\"uys\"uz, Urban, Usoskin, Vagelli, Vainio,
  Valente, Valtonen, V\'azquez~Acosta, Vecchi, Velasco, Vialle, Wallmann, Wang,
  Wang, Wang, Wang, Wang, Wang, Wei, Weng, Wu, Xiong, Xu, Yan, Yang, Yi, Yu,
  Yu, Zannoni, Zhang, Zhang, Zhang, Zhang, Zhang, Zhao, Zheng, Zhuang, Zhukov,
  Zichichi, Zimmermann, \& Zuccon}]{2020:AMS}
Aguilar, M., Ali~Cavasonza, L., Ambrosi, G., {et~al.} 2020, Phys. Rev. Lett.,
  124, 211102, \dodoi{10.1103/PhysRevLett.124.211102}

\bibitem[{{Aharonian} {et~al.}(2007){Aharonian}, {Akhperjanian}, {Bazer-Bachi},
  {Beilicke}, {Benbow}, {Berge}, {Bernl{\"o}hr}, {Boisson}, {Bolz}, {Borrel},
  {Braun}, {Brion}, {Brown}, {B{\"u}hler}, {B{\"u}sching}, {Carrigan},
  {Chadwick}, {Chounet}, {Coignet}, {Cornils}, {Costamante}, {Degrange},
  {Dickinson}, {Djannati-Ata{\"\i}}, {O'C. Drury}, {Dubus}, {Egberts},
  {Emmanoulopoulos}, {Espigat}, {Feinstein}, {Ferrero}, {Fiasson}, {Fontaine},
  {Funk}, {Funk}, {F{\"u}{\ss}ling}, {Gallant}, {Giebels}, {Glicenstein},
  {Gl{\"u}ck}, {Goret}, {Hadjichristidis}, {Hauser}, {Hauser}, {Heinzelmann},
  {Henri}, {Hermann}, {Hinton}, {Hoffmann}, {Hofmann}, {Holleran}, {Hoppe},
  {Horns}, {Jacholkowska}, {de Jager}, {Kendziorra}, {Kerschhaggl},
  {Kh{\'e}lifi}, {Komin}, {Konopelko}, {Kosack}, {Lamanna}, {Latham}, {Le
  Gallou}, {Lemi{\`e}re}, {Lemoine-Goumard}, {Lohse}, {Martin},
  {Martineau-Huynh}, {Marcowith}, {Masterson}, {Maurin}, {McComb}, {Moulin},
  {de Naurois}, {Nedbal}, {Nolan}, {Noutsos}, {Olive}, {Orford}, {Osborne},
  {Panter}, {Pelletier}, {Pita}, {P{\"u}hlhofer}, {Punch}, {Ranchon},
  {Raubenheimer}, {Raue}, {Rayner}, {Reimer}, {Reimer}, {Ripken}, {Rob},
  {Rolland}, {Rosier-Lees}, {Rowell}, {Sahakian}, {Santangelo}, {Saug{\'e}},
  {Schlenker}, {Schlickeiser}, {Schr{\"o}der}, {Schwanke}, {Schwarzburg},
  {Schwemmer}, {Shalchi}, {Sol}, {Spangler}, {Spanier}, {Steenkamp},
  {Stegmann}, {Superina}, {Tam}, {Tavernet}, {Terrier}, {Tluczykont}, {van
  Eldik}, {Vasileiadis}, {Venter}, {Vialle}, {Vincent}, {V{\"o}lk}, {Wagner},
  \& {Ward}}]{2007:HESS_SNR}
{Aharonian}, F., {Akhperjanian}, A.~G., {Bazer-Bachi}, A.~R., {et~al.} 2007,
  \aap, 464, 235, \dodoi{10.1051/0004-6361:20066381}

\bibitem[{Ahlers(2019)}]{Ahlers:2019}
Ahlers, M. 2019, Astrophys. J. Lett., 886, L18,
  \dodoi{10.3847/2041-8213/ab552f}

\bibitem[{{Albert} {et~al.}(2020){Albert}, {Alfaro}, {Alvarez}, {Camacho},
  {Arteaga-Vel{\'a}zquez}, {Arunbabu}, {Avila Rojas}, {Ayala Solares},
  {Baghmanyan}, {Belmont-Moreno}, {BenZvi}, {Brisbois}, {Caballero-Mora},
  {Capistr{\'a}n}, {Carrami{\~n}ana}, {Casanova}, {Cotti}, {Cotzomi},
  {Couti{\~n}o de Le{\'o}n}, {De la Fuente}, {Diaz-Cruz}, {Dingus},
  {DuVernois}, {D{\'\i}az-V{\'e}lez}, {Ellsworth}, {Engel}, {Espinoza}, {Fan},
  {Fang}, {Fern{\'a}ndez Alonso}, {Fleischhack}, {Fraija},
  {Galv{\'a}n-G{\'a}mez}, {Garcia}, {Garc{\'\i}a-Gonz{\'a}lez}, {Garfias},
  {Giacinti}, {Gonz{\'a}lez}, {Goodman}, {Harding}, {Hernandez}, {Hinton},
  {Hona}, {Huang}, {Hueyotl-Zahuantitla}, {H{\"u}ntemeyer}, {Iriarte},
  {Jardin-Blicq}, {Joshi}, {Lee}, {Le{\'o}n Vargas}, {Linnemann}, {Longinotti},
  {Luis-Raya}, {Lundeen}, {Malone}, {Marinelli}, {Martinez},
  {Martinez-Castellanos}, {Mart{\'\i}nez-Castro}, {Matthews},
  {Miranda-Romagnoli}, {Morales-Soto}, {Moreno}, {Mostaf{\'a}}, {Nayerhoda},
  {Nellen}, {Newbold}, {Nisa}, {Noriega-Papaqui}, {Omodei}, {Peisker},
  {P{\'e}rez Araujo}, {P{\'e}rez-P{\'e}rez}, {Rho}, {Rosa-Gonz{\'a}lez},
  {Ruiz-Velasco}, {Salazar}, {Salesa Greus}, {Sandoval}, {Schneider},
  {Schoorlemmer}, {Serna Franco}, {Sinnis}, {Smith}, {Springer}, {Surajbali},
  {Tabachnick}, {Tanner}, {Tibolla}, {Tollefson}, {Torres}, {Torres-Escobedo},
  {Ure{\~n}a-Mena}, {Villase{\~n}or}, {Weisgarber}, {Zepeda}, {Zhou}, {de
  Le{\'o}n}, {{\'A}lvarez}, \& {HAWC Collaboration}}]{2020:hawcsnr}
{Albert}, A., {Alfaro}, R., {Alvarez}, C., {et~al.} 2020, \apjl, 896, L29,
  \dodoi{10.3847/2041-8213/ab96cc}

\bibitem[{{Alemanno} {et~al.}(2021){Alemanno}, {An}, {Azzarello}, {Barbato},
  {Bernardini}, {Bi}, {Cai}, {Catanzani}, {Chang}, {Chen}, {Chen}, {Chen},
  {Cui}, {Cui}, {Cui}, {Dai}, {D'Amone}, {de Benedittis}, {de Mitri}, {de
  Palma}, {Deliyergiyev}, {di Santo}, {Dong}, {Dong}, {Donvito}, {Droz},
  {Duan}, {Duan}, {D'Urso}, {Fan}, {Fan}, {Fang}, {Fang}, {Feng}, {Feng},
  {Fusco}, {Gao}, {Gargano}, {Gong}, {Gong}, {Guo}, {Guo}, {Guo}, {Han}, {Hu},
  {Huang}, {Huang}, {Huang}, {Ionica}, {Jiang}, {Kong}, {Kotenko}, {Kyratzis},
  {Lei}, {Li}, {Li}, {Li}, {Li}, {Liang}, {Liu}, {Liu}, {Liu}, {Liu}, {Liu},
  {Liu}, {Loparco}, {Luo}, {Ma}, {Ma}, {Ma}, {Ma}, {Marsella}, {Mazziotta},
  {Mo}, {Niu}, {Pan}, {Parenti}, {Peng}, {Peng}, {Perrina}, {Qiao}, {Rao},
  {Ruina}, {Salinas}, {Shang}, {Shen}, {Shen}, {Shen}, {Silveri}, {Song},
  {Stolpovskiy}, {Su}, {Su}, {Sun}, {Surdo}, {Teng}, {Tykhonov}, {Wang},
  {Wang}, {Wang}, {Wang}, {Wang}, {Wang}, {Wang}, {Wang}, {Wang}, {Wei}, {Wei},
  {Wei}, {Wen}, {Wu}, {Wu}, {Wu}, {Wu}, {Wu}, {Xia}, {Xu}, {Xu}, {Xu}, {Xu},
  {Xue}, {Yang}, {Yang}, {Yang}, {Yao}, {Yu}, {Yuan}, {Yuan}, {Yue}, {Zang},
  {Zhang}, {Zhang}, {Zhang}, {Zhang}, {Zhang}, {Zhang}, {Zhang}, {Zhang},
  {Zhang}, {Zhang}, {Zhao}, {Zhao}, {Zhao}, {Zhou}, {Zhu}, \& {Dampe
  Collaboration}}]{Alemanno:2021gpb}
{Alemanno}, F., {An}, Q., {Azzarello}, P., {et~al.} 2021, \prl, 126, 201102,
  \dodoi{10.1103/PhysRevLett.126.201102}

\bibitem[{Amenomori {et~al.}(2019)Amenomori, Bao, Bi, Chen, Chen, Chen, Chen,
  Chen, Cirennima, Cui, Danzengluobu, Ding, Fang, Fang, Feng, Feng, Feng, Gao,
  Gou, Guo, He, He, Hibino, Hotta, Hu, Hu, Huang, Jia, Jiang, Jin, Kajino,
  Kasahara, Katayose, Kato, Kato, Kawata, Kozai, Labaciren, Le, Li, Li, Li,
  Lin, Liu, Liu, Liu, Liu, Lou, Lu, Meng, Mitsui, Munakata, Nakamura, Nanjo,
  Nishizawa, Ohnishi, Ohta, Ozawa, Qian, Qu, Saito, Sakata, Sako, Sengoku,
  Shao, Shibata, Shiomi, Sugimoto, Takita, Tan, Tateyama, Torii, Tsuchiya, Udo,
  Wang, Wu, Xue, Yagisawa, Yamamoto, Yang, Yuan, Zhai, Zhang, Zhang, Zhang,
  Zhang, Zhang, Zhang, Zhang, Zhaxisangzhu, \& Zhou}]{2021:Amenomori}
Amenomori, M., Bao, Y.~W., Bi, X.~J., {et~al.} 2019, Phys. Rev. Lett., 123,
  051101, \dodoi{10.1103/PhysRevLett.123.051101}

\bibitem[{{Andeen} \& {Plum}(2019)}]{andeen2019}
{Andeen}, K., \& {Plum}, M. 2019, in International Cosmic Ray Conference,
  Vol.~36, 36th International Cosmic Ray Conference (ICRC2019), 172,
  \dodoi{10.22323/1.358.0172}

\bibitem[{{Antoni} {et~al.}(2005){Antoni}, {Apel}, {Badea}, {Bekk}, {Bercuci},
  {Bl{\"u}mer}, {Bozdog}, {Brancus}, {Chilingarian}, {Daumiller}, {Doll},
  {Engel}, {Engler}, {Fe{\ss}ler}, {Gils}, {Glasstetter}, {Haungs}, {Heck},
  {H{\"o}randel}, {Kampert}, {Klages}, {Maier}, {Mathes}, {Mayer}, {Milke},
  {M{\"u}ller}, {Obenland}, {Oehlschl{\"a}ger}, {Ostapchenko}, {Petcu},
  {Rebel}, {Risse}, {Risse}, {Roth}, {Schatz}, {Schieler}, {Scholz}, {Thouw},
  {Ulrich}, {van Buren}, {Vardanyan}, {Weindl}, {Wochele}, \&
  {Zabierowski}}]{2005:KASCADE}
{Antoni}, T., {Apel}, W.~D., {Badea}, A.~F., {et~al.} 2005, Astroparticle
  Physics, 24, 1, \dodoi{10.1016/j.astropartphys.2005.04.001}

\bibitem[{{Apel} {et~al.}(2011){Apel}, {Arteaga-Vel{\'a}zquez}, {Bekk},
  {Bertaina}, {Bl{\"u}mer}, {Bozdog}, {Brancus}, {Buchholz}, {Cantoni},
  {Chiavassa}, {Cossavella}, {Daumiller}, {de Souza}, {di Pierro}, {Doll},
  {Engel}, {Engler}, {Finger}, {Fuhrmann}, {Ghia}, {Gils}, {Glasstetter},
  {Grupen}, {Haungs}, {Heck}, {H{\"o}randel}, {Huber}, {Huege}, {Isar},
  {Kampert}, {Kang}, {Klages}, {Link}, {{\L}uczak}, {Ludwig}, {Mathes},
  {Mayer}, {Melissas}, {Milke}, {Mitrica}, {Morello}, {Navarra},
  {Oehlschl{\"a}ger}, {Ostapchenko}, {Over}, {Palmieri}, {Petcu}, {Pierog},
  {Rebel}, {Roth}, {Schieler}, {Schr{\"o}der}, {Sima}, {Toma}, {Trinchero},
  {Ulrich}, {Weindl}, {Wochele}, {Wommer}, \& {Zabierowski}}]{apel2011}
{Apel}, W.~D., {Arteaga-Vel{\'a}zquez}, J.~C., {Bekk}, K., {et~al.} 2011, \prl,
  107, 171104, \dodoi{10.1103/PhysRevLett.107.171104}

\bibitem[{{Apel} {et~al.}(2013){Apel}, {Arteaga-Vel{\`a}zquez}, {Bekk},
  {Bertaina}, {Bl{\"u}mer}, {Bozdog}, {Brancus}, {Cantoni}, {Chiavassa},
  {Cossavella}, {Daumiller}, {de Souza}, {Di Pierro}, {Doll}, {Engel},
  {Engler}, {Finger}, {Fuchs}, {Fuhrmann}, {Gils}, {Glasstetter}, {Grupen},
  {Haungs}, {Heck}, {H{\"o}randel}, {Huber}, {Huege}, {Kampert}, {Kang},
  {Klages}, {Link}, {{\L}uczak}, {Ludwig}, {Mathes}, {Mayer}, {Melissas},
  {Milke}, {Mitrica}, {Morello}, {Oehlschl{\"a}ger}, {Ostapchenko}, {Palmieri},
  {Petcu}, {Pierog}, {Rebel}, {Roth}, {Schieler}, {Schoo}, {Schr{\"o}der},
  {Sima}, {Toma}, {Trinchero}, {Ulrich}, {Weindl}, {Wochele}, {Wommer}, \&
  {Zabierowski}}]{apel2013}
{Apel}, W.~D., {Arteaga-Vel{\`a}zquez}, J.~C., {Bekk}, K., {et~al.} 2013, \prd,
  87, 081101, \dodoi{10.1103/PhysRevD.87.081101}

\bibitem[{Arteaga-Vel\'azquez {et~al.}(2018)}]{2017:KG}
Arteaga-Vel\'azquez, C.~J., {et~al.} 2018, PoS, ICRC2017, 316,
  \dodoi{10.22323/1.301.0316}

\bibitem[{{Bell}(2004)}]{2004MNRAS.353..550B}
{Bell}, A.~R. 2004, \mnras, 353, 550, \dodoi{10.1111/j.1365-2966.2004.08097.x}

\bibitem[{{Beresnyak} {et~al.}(2009){Beresnyak}, {Jones}, \&
  {Lazarian}}]{2009ApJ...707.1541B}
{Beresnyak}, A., {Jones}, T.~W., \& {Lazarian}, A. 2009, \apj, 707, 1541,
  \dodoi{10.1088/0004-637X/707/2/1541}

\bibitem[{{Berezhko} \& {V{\"o}lk}(1997)}]{1997:Berezhko}
{Berezhko}, E.~G., \& {V{\"o}lk}, H.~J. 1997, Astroparticle Physics, 7, 183,
  \dodoi{10.1016/S0927-6505(97)00016-9}

\bibitem[{{Blandford} \& {Eichler}(1987)}]{1987:Blandford}
{Blandford}, R., \& {Eichler}, D. 1987, \physrep, 154, 1,
  \dodoi{10.1016/0370-1573(87)90134-7}

\bibitem[{{Blasi}(2002)}]{2002:Blasi}
{Blasi}, P. 2002, Astroparticle Physics, 16, 429,
  \dodoi{10.1016/S0927-6505(01)00127-X}

\bibitem[{{Blasi}(2004)}]{2004:Blasi}
---. 2004, Astroparticle Physics, 21, 45,
  \dodoi{10.1016/j.astropartphys.2003.10.008}

\bibitem[{{Blasi}(2013)}]{2013:Blasi}
---. 2013, \aapr, 21, 70, \dodoi{10.1007/s00159-013-0070-7}

\bibitem[{{Breitschwerdt} {et~al.}(1991){Breitschwerdt}, {McKenzie}, \&
  {Voelk}}]{1991:Breitschwerdt}
{Breitschwerdt}, D., {McKenzie}, J.~F., \& {Voelk}, H.~J. 1991, \aap, 245, 79

\bibitem[{{Burke}(1968)}]{1968:Burke}
{Burke}, J.~A. 1968, \mnras, 140, 241, \dodoi{10.1093/mnras/140.2.241}

\bibitem[{{Bustard} {et~al.}(2017){Bustard}, {Zweibel}, \&
  {Cotter}}]{2017:Bustard}
{Bustard}, C., {Zweibel}, E.~G., \& {Cotter}, C. 2017, \apj, 835, 72,
  \dodoi{10.3847/1538-4357/835/1/72}

\bibitem[{{Cao} {et~al.}(2021){Cao}, {Aharonian}, {An}, {Axikegu}, {Bai},
  {Bao}, {Bastieri}, {Bi}, {Bi}, {Cai}, {Cai}, {Cao}, {Chang}, {Chang},
  {Chang}, {Chen}, {Chen}, {Chen}, {Chen}, {Chen}, {Chen}, {Chen}, {Chen},
  {Chen}, {Chen}, {Chen}, {Chen}, {Chen}, {Cheng}, {Cheng}, {Cui}, {Cui},
  {Cui}, {Dai}, {Dai}, {Dai}, {Danzengluobu}, {della Volpe}, {D'Ettorre
  Piazzoli}, {Dong}, {Fan}, {Fan}, {Fan}, {Fang}, {Fang}, {Feng}, {Feng},
  {Feng}, {Feng}, {Gao}, {Gao}, {Gao}, {Gao}, {Ge}, {Geng}, {Gong}, {Gou},
  {Gu}, {Guo}, {Guo}, {Guo}, {Guo}, {Han}, {He}, {He}, {He}, {He}, {He}, {He},
  {Heller}, {Hor}, {Hou}, {Hou}, {Hu}, {Hu}, {Hu}, {Hu}, {Huang}, {Huang},
  {Huang}, {Huang}, {Huang}, {Ji}, {Ji}, {Jia}, {Jiang}, {Jiang}, {Jin},
  {Kuleshov}, {Levochkin}, {Li}, {Li}, {Li}, {Li}, {Li}, {Li}, {Li}, {Li},
  {Li}, {Li}, {Li}, {Li}, {Li}, {Li}, {Li}, {Li}, {Li}, {Liang}, {Liang},
  {Lin}, {Liu}, {Liu}, {Liu}, {Liu}, {Liu}, {Liu}, {Liu}, {Liu}, {Liu}, {Liu},
  {Liu}, {Liu}, {Liu}, {Liu}, {Liu}, {Long}, {Lu}, {Lv}, {Ma}, {Ma}, {Ma},
  {Mao}, {Masood}, {Mitthumsiri}, {Montaruli}, {Nan}, {Pang},
  {Pattarakijwanich}, {Pei}, {Qi}, {Ruffolo}, {Rulev}, {S{\'a}iz}, {Shao},
  {Shchegolev}, {Sheng}, {Shi}, {Song}, {Stenkin}, {Stepanov}, {Sun}, {Sun},
  {Sun}, {Tam}, {Tang}, {Tian}, {Wang}, {Wang}, {Wang}, {Wang}, {Wang}, {Wang},
  {Wang}, {Wang}, {Wang}, {Wang}, {Wang}, {Wang}, {Wang}, {Wang}, {Wang},
  {Wang}, {Wang}, {Wang}, {Wang}, {Wang}, {Wang}, {Wei}, {Wei}, {Wei}, {Wen},
  {Wu}, {Wu}, {Wu}, {Wu}, {Wu}, {Xi}, {Xia}, {Xia}, {Xiang}, {Xiao}, {Xiao},
  {Xin}, {Xin}, {Xing}, {Xu}, {Xu}, {Xue}, {Yan}, {Yang}, {Yang}, {Yang},
  {Yang}, {Yang}, {Yang}, {Yang}, {Yao}, {Yao}, {Ye}, {Yin}, {Yin}, {You},
  {You}, {Yu}, {Yuan}, {Zeng}, {Zeng}, {Zeng}, {Zeng}, {Zha}, {Zhai}, {Zhang},
  {Zhang}, {Zhang}, {Zhang}, {Zhang}, {Zhang}, {Zhang}, {Zhang}, {Zhang},
  {Zhang}, {Zhang}, {Zhang}, {Zhang}, {Zhang}, {Zhang}, {Zhang}, {Zhang},
  {Zhang}, {Zhang}, {Zhao}, {Zhao}, {Zhao}, {Zhao}, {Zhao}, {Zheng}, {Zheng},
  {Zhou}, {Zhou}, {Zhou}, {Zhou}, {Zhou}, {Zhou}, {Zhu}, {Zhu}, {Zhu}, {Zhu},
  \& {Zuo}}]{2021:lhasso}
{Cao}, Z., {Aharonian}, F.~A., {An}, Q., {et~al.} 2021, \nat, 594, 33,
  \dodoi{10.1038/s41586-021-03498-z}

\bibitem[{{Chevalier} \& {Clegg}(1985)}]{Chevalier:1985}
{Chevalier}, R.~A., \& {Clegg}, A.~W. 1985, \nat, 317, 44,
  \dodoi{10.1038/317044a0}

\bibitem[{{Donnert} {et~al.}(2018){Donnert}, {Vazza}, {Br{\"u}ggen}, \&
  {ZuHone}}]{2018SSRv..214..122D}
{Donnert}, J., {Vazza}, F., {Br{\"u}ggen}, M., \& {ZuHone}, J. 2018, \ssr, 214,
  122, \dodoi{10.1007/s11214-018-0556-8}

\bibitem[{{Dorfi} \& {Breitschwerdt}(2012)}]{2012:Dorfi}
{Dorfi}, E.~A., \& {Breitschwerdt}, D. 2012, \aap, 540, A77,
  \dodoi{10.1051/0004-6361/201118082}

\bibitem[{{Dorfi} {et~al.}(2019){Dorfi}, {Steiner}, {Ragossnig}, \&
  {Breitschwerdt}}]{2019:dorfi}
{Dorfi}, E.~A., {Steiner}, D., {Ragossnig}, F., \& {Breitschwerdt}, D. 2019,
  \aap, 630, A107, \dodoi{10.1051/0004-6361/201935940}

\bibitem[{{Drury} \& {Downes}(2012)}]{2012MNRAS.427.2308D}
{Drury}, L.~O., \& {Downes}, T.~P. 2012, \mnras, 427, 2308,
  \dodoi{10.1111/j.1365-2966.2012.22106.x}

\bibitem[{{Dundovic} {et~al.}(2020){Dundovic}, {Pezzi}, {Blasi}, {Evoli}, \&
  {Matthaeus}}]{Dundovic:2020}
{Dundovic}, A., {Pezzi}, O., {Blasi}, P., {Evoli}, C., \& {Matthaeus}, W.~H.
  2020, \prd, 102, 103016, \dodoi{10.1103/PhysRevD.102.103016}

\bibitem[{{Everett} {et~al.}(2008){Everett}, {Zweibel}, {Benjamin}, {McCammon},
  {Rocks}, \& {Gallagher}}]{2008:Everett}
{Everett}, J.~E., {Zweibel}, E.~G., {Benjamin}, R.~A., {et~al.} 2008, \apj,
  674, 258, \dodoi{10.1086/524766}

\bibitem[{G\'enolini {et~al.}(2017)}]{Genolini:2017dfb}
G\'enolini, Y., {et~al.} 2017, Phys. Rev. Lett., 119, 241101,
  \dodoi{10.1103/PhysRevLett.119.241101}

\bibitem[{{Globus} {et~al.}(2008){Globus}, {Allard}, \&
  {Parizot}}]{2008A&A...479...97G}
{Globus}, N., {Allard}, D., \& {Parizot}, E. 2008, \aap, 479, 97,
  \dodoi{10.1051/0004-6361:20078653}

\bibitem[{{Globus} {et~al.}(2015){Globus}, {Allard}, \& {Parizot}}]{globus2015}
---. 2015, \prd, 92, 021302, \dodoi{10.1103/PhysRevD.92.021302}

\bibitem[{{Gottlieb} \& {Globus}(2021)}]{2021:Ore}
{Gottlieb}, O., \& {Globus}, N. 2021, \apjl, 915, L4,
  \dodoi{10.3847/2041-8213/ac05c5}

\bibitem[{{Harari} {et~al.}(2002){Harari}, {Mollerach}, {Roulet}, \&
  {S{\'a}nchez}}]{2002JHEP...03..045H}
{Harari}, D., {Mollerach}, S., {Roulet}, E., \& {S{\'a}nchez}, F. 2002, Journal
  of High Energy Physics, 2002, 045, \dodoi{10.1088/1126-6708/2002/03/045}

\bibitem[{{HESS Collaboration} {et~al.}(2016){HESS Collaboration},
  {Abramowski}, {Aharonian}, {Benkhali}, {Akhperjanian}, {Ang{\"u}ner},
  {Backes}, {Balzer}, {Becherini}, {Tjus}, {Berge}, {Bernhard}, {Bernl{\"o}hr},
  {Birsin}, {Blackwell}, {B{\"o}ttcher}, {Boisson}, {Bolmont}, {Bordas},
  {Bregeon}, {Brun}, {Brun}, {Bryan}, {Bulik}, {Carr}, {Casanova},
  {Chakraborty}, {Chalme-Calvet}, {Chaves}, {Chen}, {Chr{\'e}tien},
  {Colafrancesco}, {Cologna}, {Conrad}, {Couturier}, {Cui}, {Davids},
  {Degrange}, {Deil}, {Dewilt}, {Djannati-Ata{\"\i}}, {Domainko}, {Donath},
  {Drury}, {Dubus}, {Dutson}, {Dyks}, {Dyrda}, {Edwards}, {Egberts}, {Eger},
  {Ernenwein}, {Espigat}, {Farnier}, {Fegan}, {Feinstein}, {Fernandes},
  {Fernandez}, {Fiasson}, {Fontaine}, {F{\"o}rster}, {F{\"u}{\ss}ling},
  {Gabici}, {Gajdus}, {Gallant}, {Garrigoux}, {Giavitto}, {Giebels},
  {Glicenstein}, {Gottschall}, {Goyal}, {Grondin}, {Grudzi{\'n}ska}, {Hadasch},
  {H{\"a}ffner}, {Hahn}, {Hawkes}, {Heinzelmann}, {Henri}, {Hermann}, {Hervet},
  {Hillert}, {Hinton}, {Hofmann}, {Hofverberg}, {Hoischen}, {Holler}, {Horns},
  {Ivascenko}, {Jacholkowska}, {Jamrozy}, {Janiak}, {Jankowsky},
  {Jung-Richardt}, {Kastendieck}, {Katarzy{\'n}ski}, {Katz}, {Kerszberg},
  {Kh{\'e}lifi}, {Kieffer}, {Klepser}, {Klochkov}, {Klu{\'z}niak}, {Kolitzus},
  {Komin}, {Kosack}, {Krakau}, {Krayzel}, {Kr{\"u}ger}, {Laffon}, {Lamanna},
  {Lau}, {Lefaucheur}, {Lefranc}, {Lemi{\'e}re}, {Lemoine-Goumard}, {Lenain},
  {Lohse}, {Lopatin}, {Lu}, {Lui}, {Marandon}, {Marcowith}, {Mariaud}, {Marx},
  {Maurin}, {Maxted}, {Mayer}, {Meintjes}, {Menzler}, {Meyer}, {Mitchell},
  {Moderski}, {Mohamed}, {Mor{\r{a}}}, {Moulin}, {Murach}, {de Naurois},
  {Niemiec}, {Oakes}, {Odaka}, {{\"O}ttl}, {Ohm}, {Opitz}, {Ostrowski}, {Oya},
  {Panter}, {Parsons}, {Arribas}, {Pekeur}, {Pelletier}, {Petrucci}, {Peyaud},
  {Pita}, {Poon}, {Prokoph}, {P{\"u}hlhofer}, {Punch}, {Quirrenbach}, {Raab},
  {Reichardt}, {Reimer}, {Reimer}, {Renaud}, {de Los Reyes}, {Rieger},
  {Romoli}, {Rosier-Lees}, {Rowell}, {Rudak}, {Rulten}, {Sahakian}, {Salek},
  {Sanchez}, {Santangelo}, {Sasaki}, {Schlickeiser}, {Sch{\"u}ssler}, {Schulz},
  {Schwanke}, {Schwemmer}, {Seyffert}, {Simoni}, {Sol}, {Spanier}, {Spengler},
  {Spies}, {Stawarz}, {Steenkamp}, {Stegmann}, {Stinzing}, {Stycz}, {Sushch},
  {Tavernet}, {Tavernier}, {Taylor}, {Terrier}, {Tluczykont}, {Trichard},
  {Tuffs}, {Valerius}, {van der Walt}, {van Eldik}, {van Soelen},
  {Vasileiadis}, {Veh}, {Venter}, {Viana}, {Vincent}, {Vink}, {Voisin},
  {V{\"o}lk}, {Vuillaume}, {Wagner}, {Wagner}, {Wagner}, {Weidinger},
  {Weitzel}, {White}, {Wierzcholska}, {Willmann}, {W{\"o}rnlein}, {Wouters},
  {Yang}, {Zabalza}, {Zaborov}, {Zacharias}, {Zdziarski}, {Zech}, {Zefi}, \&
  {{\.Z}ywucka}}]{2016:HESS}
{HESS Collaboration}, {Abramowski}, A., {Aharonian}, F., {et~al.} 2016, \nat,
  531, 476, \dodoi{10.1038/nature17147}

\bibitem[{{Ipavich}(1975)}]{1975:Ipavich}
{Ipavich}, F.~M. 1975, \apj, 196, 107, \dodoi{10.1086/153397}

\bibitem[{{Jansson} \& {Farrar}(2012)}]{2012ApJ...757...14J}
{Jansson}, R., \& {Farrar}, G.~R. 2012, \apj, 757, 14,
  \dodoi{10.1088/0004-637X/757/1/14}

\bibitem[{{Johnson} \& {Axford}(1971)}]{1971:Johnson}
{Johnson}, H.~E., \& {Axford}, W.~I. 1971, \apj, 165, 381,
  \dodoi{10.1086/150903}

\bibitem[{{Jokipii}(1966)}]{1966ApJ...146..480J}
{Jokipii}, J.~R. 1966, \apj, 146, 480, \dodoi{10.1086/148912}

\bibitem[{{Jokipii} \& {Morfill}(1987)}]{198:Jokipii}
{Jokipii}, J.~R., \& {Morfill}, G. 1987, \apj, 312, 170, \dodoi{10.1086/164857}

\bibitem[{{Jokipii} \& {Morfill}(1985)}]{1985:Jokipii}
{Jokipii}, J.~R., \& {Morfill}, G.~E. 1985, \apjl, 290, L1,
  \dodoi{10.1086/184430}

\bibitem[{Kang {et~al.}(2021)}]{kang2019}
Kang, D., {et~al.} 2021, PoS, ICRC2019, 306, \dodoi{10.22323/1.358.0306}

\bibitem[{{Kulpa-Dybe{\l}} {et~al.}(2011){Kulpa-Dybe{\l}}, {Otmianowska-Mazur},
  {Kulesza-{\.Z}ydzik}, {Hanasz}, {Kowal}, {W{\'o}lta{\'n}ski}, \&
  {Kowalik}}]{2011ApJ...733L..18K}
{Kulpa-Dybe{\l}}, K., {Otmianowska-Mazur}, K., {Kulesza-{\.Z}ydzik}, B.,
  {et~al.} 2011, \apjl, 733, L18, \dodoi{10.1088/2041-8205/733/2/L18}

\bibitem[{{Kulsrud} \& {Pearce}(1969)}]{1969ApJ...156..445K}
{Kulsrud}, R., \& {Pearce}, W.~P. 1969, \apj, 156, 445, \dodoi{10.1086/149981}

\bibitem[{{Lerche}(1967)}]{1967ApJ...147..689L}
{Lerche}, I. 1967, Apj, 147, 689, \dodoi{10.1086/149045}

\bibitem[{{Lipari}(2014)}]{2014arXiv1407.5223L}
{Lipari}, P. 2014, arXiv e-prints, arXiv:1407.5223.
\newblock \doarXiv{1407.5223}

\bibitem[{{Mao} \& {Ostriker}(2018)}]{2018:Mao}
{Mao}, S.~A., \& {Ostriker}, E.~C. 2018, \apj, 854, 89,
  \dodoi{10.3847/1538-4357/aaa88e}

\bibitem[{Merten {et~al.}(2018)Merten, Bustard, Zweibel, \&
  Becker~Tjus}]{Merten:2018qoa}
Merten, L., Bustard, C., Zweibel, E.~G., \& Becker~Tjus, J. 2018, Astrophys.
  J., 859, 63, \dodoi{10.3847/1538-4357/aabfdd}

\bibitem[{{Miller} \& {Bregman}(2015)}]{2015:Miller}
{Miller}, M.~J., \& {Bregman}, J.~N. 2015, \apj, 800, 14,
  \dodoi{10.1088/0004-637X/800/1/14}

\bibitem[{{Miyamoto} \& {Nagai}(1975)}]{1975:Miyamoto}
{Miyamoto}, M., \& {Nagai}, R. 1975, \pasj, 27, 533

\bibitem[{{Monari} {et~al.}(2018){Monari}, {Famaey}, {Carrillo}, {Piffl},
  {Steinmetz}, {Wyse}, {Anders}, {Chiappini}, \& {Jan{\ss}en}}]{2018:Monari}
{Monari}, G., {Famaey}, B., {Carrillo}, I., {et~al.} 2018, \aap, 616, L9,
  \dodoi{10.1051/0004-6361/201833748}

\bibitem[{{Morlino} {et~al.}(2021){Morlino}, {Blasi}, {Peretti}, \&
  {Cristofari}}]{2021:Morlino}
{Morlino}, G., {Blasi}, P., {Peretti}, E., \& {Cristofari}, P. 2021, \mnras,
  504, 6096, \dodoi{10.1093/mnras/stab690}

\bibitem[{Murase \& Fukugita(2019)}]{Murase:2018utn}
Murase, K., \& Fukugita, M. 2019, Phys. Rev. D, 99, 063012,
  \dodoi{10.1103/PhysRevD.99.063012}

\bibitem[{Necib \& Lin(2022)}]{Necib:2021vxr}
Necib, L., \& Lin, T. 2022, Astrophys. J., 926, 189,
  \dodoi{10.3847/1538-4357/ac4244}

\bibitem[{{Nicastro} {et~al.}(2018){Nicastro}, {Kaastra}, {Krongold},
  {Borgani}, {Branchini}, {Cen}, {Dadina}, {Danforth}, {Elvis}, {Fiore},
  {Gupta}, {Mathur}, {Mayya}, {Paerels}, {Piro}, {Rosa-Gonzalez}, {Schaye},
  {Shull}, {Torres-Zafra}, {Wijers}, \& {Zappacosta}}]{2018:Nicastro}
{Nicastro}, F., {Kaastra}, J., {Krongold}, Y., {et~al.} 2018, \nat, 558, 406,
  \dodoi{10.1038/s41586-018-0204-1}

\bibitem[{{Peretti} {et~al.}(2022){Peretti}, {Morlino}, {Blasi}, \&
  {Cristofari}}]{Starburst_wind}
{Peretti}, E., {Morlino}, G., {Blasi}, P., \& {Cristofari}, P. 2022, \mnras,
  511, 1336, \dodoi{10.1093/mnras/stac084}

\bibitem[{{Piffl} {et~al.}(2014){Piffl}, {Scannapieco}, {Binney}, {Steinmetz},
  {Scholz}, {Williams}, {de Jong}, {Kordopatis}, {Matijevi{\v{c}}},
  {Bienaym{\'e}}, {Bland-Hawthorn}, {Boeche}, {Freeman}, {Gibson}, {Gilmore},
  {Grebel}, {Helmi}, {Munari}, {Navarro}, {Parker}, {Reid}, {Seabroke},
  {Watson}, {Wyse}, \& {Zwitter}}]{2014:Piffl}
{Piffl}, T., {Scannapieco}, C., {Binney}, J., {et~al.} 2014, \aap, 562, A91,
  \dodoi{10.1051/0004-6361/201322531}

\bibitem[{{Recchia} {et~al.}(2016){Recchia}, {Blasi}, \&
  {Morlino}}]{2016:Recchia}
{Recchia}, S., {Blasi}, P., \& {Morlino}, G. 2016, \mnras, 462, 4227,
  \dodoi{10.1093/mnras/stw1966}

\bibitem[{{Recchia} {et~al.}(2017){Recchia}, {Blasi}, \&
  {Morlino}}]{2017:Recchia}
---. 2017, \mnras, 470, 865, \dodoi{10.1093/mnras/stx1214}

\bibitem[{{Reichherzer} {et~al.}(2020){Reichherzer}, {Becker Tjus}, {Zweibel},
  {Merten}, \& {Pueschel}}]{2020:Reichherzer}
{Reichherzer}, P., {Becker Tjus}, J., {Zweibel}, E.~G., {Merten}, L., \&
  {Pueschel}, M.~J. 2020, \mnras, 498, 5051, \dodoi{10.1093/mnras/staa2533}

\bibitem[{{Scoville} {et~al.}(2017){Scoville}, {Lee}, {Vanden Bout},
  {Diaz-Santos}, {Sanders}, {Darvish}, {Bongiorno}, {Casey}, {Murchikova},
  {Koda}, {Capak}, {Vlahakis}, {Ilbert}, {Sheth}, {Morokuma-Matsui}, {Ivison},
  {Aussel}, {Laigle}, {McCracken}, {Armus}, {Pope}, {Toft}, \&
  {Masters}}]{2017ApJ...837..150S}
{Scoville}, N., {Lee}, N., {Vanden Bout}, P., {et~al.} 2017, \apj, 837, 150,
  \dodoi{10.3847/1538-4357/aa61a0}

\bibitem[{{Shalchi} {et~al.}(2004){Shalchi}, {Bieber}, {Matthaeus}, \&
  {Qin}}]{2004ApJ...616..617S}
{Shalchi}, A., {Bieber}, J.~W., {Matthaeus}, W.~H., \& {Qin}, G. 2004, \apj,
  616, 617, \dodoi{10.1086/424839}

\bibitem[{{Shalchi} {et~al.}(2009){Shalchi}, {Skoda}, {Tautz}, \&
  {Schlickeiser}}]{2009A&A...507..589S}
{Shalchi}, A., {Skoda}, T., {Tautz}, R.~C., \& {Schlickeiser}, R. 2009, \aap,
  507, 589, \dodoi{10.1051/0004-6361/200912755}

\bibitem[{Snodin {et~al.}(2016)Snodin, Shukurov, Sarson, Bushby, \&
  Rodrigues}]{Snodin2016}
Snodin, A.~P., Shukurov, A., Sarson, G.~R., Bushby, P.~J., \& Rodrigues, L.
  F.~S. 2016, Monthly Notices of the Royal Astronomical Society, 457, 3975,
  \dodoi{10.1093/mnras/stw217}

\bibitem[{{Soida} {et~al.}(2011){Soida}, {Krause}, {Dettmar}, \&
  {Urbanik}}]{2011A&A...531A.127S}
{Soida}, M., {Krause}, M., {Dettmar}, R.~J., \& {Urbanik}, M. 2011, \aap, 531,
  A127, \dodoi{10.1051/0004-6361/200810763}

\bibitem[{{Subedi} {et~al.}(2017){Subedi}, {Sonsrettee}, {Blasi}, {Ruffolo},
  {Matthaeus}, {Montgomery}, {Chuychai}, {Dmitruk}, {Wan}, {Parashar}, \&
  {Chhiber}}]{Subedi:2017}
{Subedi}, P., {Sonsrettee}, W., {Blasi}, P., {et~al.} 2017, \apj, 837, 140,
  \dodoi{10.3847/1538-4357/aa603a}

\bibitem[{{Thoudam} {et~al.}(2016){Thoudam}, {Rachen}, {van Vliet},
  {Achterberg}, {Buitink}, {Falcke}, \& {H{\"o}randel}}]{2016:Thoudam}
{Thoudam}, S., {Rachen}, J.~P., {van Vliet}, A., {et~al.} 2016, \aap, 595, A33,
  \dodoi{10.1051/0004-6361/201628894}

\bibitem[{{Tibet AS{\ensuremath{\gamma}} Collaboration} {et~al.}(2021){Tibet
  AS{\ensuremath{\gamma}} Collaboration}, {Amenomori}, {Bao}, {Bi}, {Chen},
  {Chen}, {Chen}, {Chen}, {Chen}, {Cirennima}, {Danzengluobu}, {Fang}, {Fang},
  {Feng}, {Feng}, {Feng}, {Gao}, {Gou}, {Guo}, {Guo}, {He}, {He}, {Hibino},
  {Hotta}, {Hu}, {Hu}, {Huang}, {Jia}, {Jiang}, {Jin}, {Kasahara}, {Katayose},
  {Kato}, {Kato}, {Kawata}, {Kihara}, {Ko}, {Kozai}, {Labaciren}, {Li}, {Li},
  {Li}, {Lin}, {Liu}, {Liu}, {Liu}, {Liu}, {Liu}, {Lou}, {Lu}, {Meng},
  {Munakata}, {Nakada}, {Nakamura}, {Nanjo}, {Nishizawa}, {Ohnishi}, {Ohura},
  {Ozawa}, {Qian}, {Qu}, {Saito}, {Sakata}, {Sako}, {Shao}, {Shibata},
  {Shiomi}, {Sugimoto}, {Takano}, {Takita}, {Tan}, {Tateyama}, {Torii},
  {Tsuchiya}, {Udo}, {Wang}, {Wu}, {Xue}, {Yamamoto}, {Yang}, {Yokoe}, {Yuan},
  {Zhai}, {Zhang}, {Zhang}, {Zhang}, {Zhang}, {Zhang}, {Zhang}, {Zhang},
  {Zhao}, \& {Zhaxisangzhu}}]{2021:TIBET}
{Tibet AS{\ensuremath{\gamma}} Collaboration}, {Amenomori}, M., {Bao}, Y.~W.,
  {et~al.} 2021, Nature Astronomy, 5, 460, \dodoi{10.1038/s41550-020-01294-9}

\bibitem[{{V{\"o}lk} \& {Zirakashvili}(2004)}]{2004:Volk}
{V{\"o}lk}, H.~J., \& {Zirakashvili}, V.~N. 2004, \aap, 417, 807,
  \dodoi{10.1051/0004-6361:20040018}

\bibitem[{{Wentzel}(1974)}]{1974:Wentzel}
{Wentzel}, D.~G. 1974, \araa, 12, 71,
  \dodoi{10.1146/annurev.aa.12.090174.000443}

\bibitem[{{Zel'dovich} \& {Raizer}(1967)}]{1967:Zeldovich}
{Zel'dovich}, Y.~B., \& {Raizer}, Y.~P. 1967, {Physics of Shock Waves and
  High-Temperature Hydrodynamic Phenomena}

\bibitem[{{Zhang} {et~al.}(2020){Zhang}, {Murase}, \&
  {M{\'e}sz{\'a}ros}}]{Starburst_Zhang}
{Zhang}, Z., {Murase}, K., \& {M{\'e}sz{\'a}ros}, P. 2020, \mnras, 492, 2250,
  \dodoi{10.1093/mnras/staa022}

\bibitem[{{Zirakashvili} \& {V{\"o}lk}(2006)}]{2006:Zirakashvilli}
{Zirakashvili}, V.~N., \& {V{\"o}lk}, H.~J. 2006, Advances in Space Research,
  37, 1923, \dodoi{10.1016/j.asr.2005.06.013}

\bibitem[{{Zweibel}(2013)}]{2013PhPl...20e5501Z}
{Zweibel}, E.~G. 2013, Physics of Plasmas, 20, 055501,
  \dodoi{10.1063/1.4807033}

\end{thebibliography}

\newpage
\onecolumngrid
\appendix

\section{Iteration Technique}
\label{subsec:technique}

To solve Eq.~(\ref{eq:transport}), we use an iteration technique first introduced in \citet{2002:Blasi,2004:Blasi}, which was subsequently applied for spherical termination shocks in \citet{2021:Morlino}. 
We recast Eq.~(\ref{eq:transport}) into the following form:

\begin{equation}
\begin{aligned}
\label{eq:transport_recast}
\frac{\partial}{\partial z} \left( A(z) \,D(z,p) \frac{\partial f}{ \partial z} - v(z) \,A(z) f\right) + \frac{\dd}{\dd z}(A(z) \,v(z)) \left(f + \frac{p}{3} \frac{\partial f}{\partial p}\right) + A(z) \,Q(z,p) = 0.
\end{aligned}
\end{equation}

We define a variable $q$ such that

\begin{equation}
\label{eq:q_def}
    q = - \left( 3 + \frac{\partial \ln{f}}{\partial \ln{p}} \right).
\end{equation}

 The wind velocity profile $v(z)$ is obtained from the solution to the wind equations described in Section~\ref{subsec:fiducial_wind}. Note, crucially, that we do not solve the wind equations and the transport equation self-consistently. The wind calculation is done separately, and from that the velocity profile is plugged into the transport equation. It is assumed that the wind $v(z)$ is launched from a height $z_0 = 1$~kpc. Within this boundary ($z < $ 1~kpc), the wind is assumed to quickly drop to zero velocity so that the propagation within the Galactic disk and the inner magnetized halo is assumed to be diffusion dominated. 
This assumption is in agreement with standard CR transport modeling in a thin disk of 100-parsec height surrounded by a kiloparsec-sized magnetized halo where CRs are diffusing \citep{2013:Blasi}. 

The solution of the transport equation is obtained by first solving the equation in the unshocked region (upstream) and the shocked region (downstream) and then by joining the two solutions across the shock surface where the solution is $f_s(z,p)$. The wind velocity and the diffusion coefficient $D(z,p)$ are spatially varying quantities. We now describe the iterative procedure in detail. 

In the upstream region ($z < z_{\mathrm{s}}$), the symmetry of the problem imposes the inner boundary condition at $z = 0$, resulting in a simple zero-flux requirement:

\begin{equation}
    \left(\frac{\partial f}{\partial z}\right)_{z = 0} = 0 ~~~~~ (\mathrm{inner ~B.C}).
\label{eq:innerBC}
\end{equation}

\noindent Integrating from $z = 0$ to some upstream $z$, we obtain, 

\begin{equation}
\left( A(z)\,D(z,p) \frac{\partial f}{\partial z} - v(z) \,A(z) f \right) - \int_0^z \frac{\dd}{\dd z}(A(z) \,v(z)) \frac{f}{3}\, q \, \dd z + \int_{0}^{z} A(z')\, Q(z',p) \,\dd z' = 0.
\label{eq:simplify1}
\end{equation}

\noindent We define a new quantity $G(z,p)$ such that,

\begin{equation}
    G(z,p) = \int_0^z \frac{\dd}{\dd z} (A(z)\, v(z)) \frac{f}{3} q \,\dd z.
\label{eq:G_define}
\end{equation}

\noindent Simplifying Equation~(\ref{eq:simplify1}), we obtain,

\begin{equation}
    \frac{\partial f}{\partial z} = \frac{v(z) f(z,p)}{D(z,p)} + \frac{G(z,p)}{A(z) \,D(z,p)} - \frac{\int_{0}^{z} A(z')\, Q(z',p) \dd z'}{A(z) \,D(z,p)}.
\label{eq:simplify2}
\end{equation}

\noindent We define two new quantities:

\begin{equation}
    V_{\rm eff}(z,p) = v(z) + \frac{G(z,p)}{A(z) f}, ~~~ I(z,p) = \int_{0}^{z} A(z') \,Q(z',p) \,\dd z'.
\label{eq:V_eff}
\end{equation}

\noindent Using Eqs.~(\ref{eq:simplify2}) and (\ref{eq:V_eff}), we get

\begin{equation}
    \frac{\partial f}{\partial z} = \frac{V_{\rm eff}(z,p)}{D(z,p)}f - \frac{I}{D(z,p) \,A(z,p)}.
\label{eq:upstream_simplify}
\end{equation}

\noindent Eq.~(\ref{eq:upstream_simplify}) has the following implicit solution:

\begin{equation}
    f_1(z,p) \exp{\left(- \int_0^z \frac{V_{\rm eff}}{D(z',p)} \dd z'\right)} = C_1 - \int_0^{z} \frac{I}{D(z',p)\,A(z')} \exp{\left( - \int_0^{z'} \frac{V_{\rm eff}}{D(z'',p)} \dd z''\right)}\, \dd z',
\label{eq:iterative_solution_upstream}
\end{equation}

\noindent where $C_1$ is a constant to be determined. Equation~(\ref{eq:iterative_solution_upstream}) gives the implicit solution for the upstream distribution $f(z,p)$. The condition that at the shock ($z = z_{\mathrm{s}}$) that $f = f_s(p)$ fixes $C_1$. The final solution for the upstream distribution after imposing this condition gives the upstream solution in an implicit form:

\begin{equation}
    f_1(z,p) = f_s \exp{\left( -\int_z^{z_{\mathrm{s}}} \frac{V_{\rm eff}(z,p)}{D(z,p)}\dd z' \right)} + \int_z^{z_{\mathrm{s}}} \frac{I}{D(z',p)\,A(z')} \exp{\left( - \int_z^{z'} \frac{V_{\rm eff}}{D(z'',p)} \dd z''\right)}\, \dd z'.
\label{eq:upstream_final}
\end{equation}

In the downstream region, $z_{\mathrm{s}}<z<z_{\mathrm{b}}$, where $z_{\mathrm{b}}$ is the escape height, the solution is found by integrating the transport equation from $z_{\mathrm{s}}$ to $z < z_{\mathrm{b}}$:

\begin{equation}
 \left[A(z)\,D(z,p)\frac{\partial f}{\partial z}\right]_{z_{s^+}}^z - v(z)\,A(z)f + v_2 A_s f_s + G_2(z,p) - G_2(z_{\mathrm{s}},p) = 0, 
\label{eq:downstream_simplify1}
\end{equation}

\noindent where we defined

\begin{equation}
    v_2 = v(z_{\mathrm{s}}^+), ~~~~ G_2(z,p) = \int_z^{z_{\mathrm{b}}} \frac{\dd}{\dd z'}(A(z')\,v(z')) \frac{f}{3} q \,\dd z'. 
\end{equation}

\noindent The downstream solution immediately after the shock becomes:

\begin{equation}
     \left(D \frac{\partial f}{\partial z}\right)_{z_{\mathrm{s}}^+} = v_2 f_s - \frac{\phi_{\rm esc}}{A(z)} - \frac{G_2(z_{\mathrm{s}}^+,p)}{A(z)},
\label{eq:immediate_down}
\end{equation}

\noindent where we have defined the quantity $\phi_{\rm esc}$ at the boundary as

\begin{equation}
    \phi_{\rm esc} = -A(z_{\mathrm{b}})\left[D(z,p)\frac{\partial f}{\partial z} - v(z) \,f\right]_{z = z_{\mathrm{b}}}.
\end{equation}

\noindent Using Equations~(\ref{eq:downstream_simplify1}) and (\ref{eq:immediate_down}), we can write the general downstream solution in an implicit form as

\begin{equation}
    f_2(z,p) = \int_{z}^{z_{\mathrm{b}}} \left( \frac{G_2(z',p) + \phi_{\rm esc}}{A(z') \,D(z',p)}\right) \times \exp{\left(-\int_z^{z'} \frac{v(z'')}{D(z'',p)} \dd z'' \right)}\dd z'.
\label{eq:f2_simplify1}
\end{equation}

\noindent In terms of $f_{\rm s}$, the above equation becomes

\begin{equation}
\begin{split}
f_2(z,p) = \left[ f_{\rm s}(p) - \int_{z_{\mathrm{s}}}^z \left( \frac{G_2(z',p) + \phi_{\rm esc}}{A(z')\,D(z',p)}\right) \times \exp{\left( - \int_{z_{\mathrm{s}}}^{z'} \frac{v(z'')}{D(z'',p)} \dd z'' \right)} \dd z'\right] \exp{\left( \int_{z_{\mathrm{s}}}^z \frac{v(z')}{D(z',p)} \dd z'\right)}.
\label{eq:f2_final}
\end{split}
\end{equation}

\noindent Equations~(\ref{eq:upstream_final}) and (\ref{eq:f2_final}) give the general solutions for the upstream ($f_1$) and downstream ($f_2$) distribution function in an implicit form. The quantity $\phi_{\rm esc}$ has to be obtained from the outer boundary condition to be described later. The two solution for $f_1$ and $f_2$ needs to be joined at the shock front.   

At the shock front, the plasma speed discontinuously changes $v(z) \simeq v_2 + (v_1 - v_2) \Theta(z_{\mathrm{s}} - z)$, where $\Theta$ is the Heaviside theta function and $u_1, u_2$ are the speeds just before and after the shock front. The derivative of a Heaviside function is the Dirac delta function, leading to $\left(\frac{\dd}{\dd z}(A(z)\,v(z))\right)_{z_{\mathrm{s}}} = - A_{\rm s} (v_1 - v_2)\, \delta(z - z_{\mathrm{s}}) $, where $A_{\rm s}$ is the area at the shock front. Integrating the transport equation (Eq.~\ref{eq:transport}) between $z_{\mathrm{s}}^-$ to $z_{\mathrm{s}}^+$, we get

\begin{equation}
(v_1 - v_2)\frac{p}{3} \frac{\partial f_{\rm s}}{\partial p} = \left( D \frac{\partial f}{\partial z}\right)_{z_{\mathrm{s}}^+} -  \left( D \frac{\partial f}{\partial z}\right)_{z_{\mathrm{s}}^-}.
\label{eq:shock_solution}
\end{equation}

\noindent The terms on the RHS of the above equation can be obtained from Equations~(\ref{eq:upstream_simplify}) and (\ref{eq:immediate_down}). Equation~(\ref{eq:shock_solution}) can then be simplified to

\begin{equation}
(v_1 - v_2)\frac{p}{3} \frac{\partial f_{\rm s}}{\partial p}  = - (v_1 - v_2) f_{\rm s} - \left(\frac{\phi_{\rm esc}}{A(z)} + \frac{G_2(z_{\mathrm{s}}^+,p)}{A(z)}\right) - \frac{G_1(z,p)}{A(z)} + \frac{I(z_{\mathrm{s}},p)}{A(z)}.
\label{eq:join_fs}
\end{equation}

\noindent The idea of the iteration technique is to implicitly  solve for Equations~(\ref{eq:upstream_final}), (\ref{eq:f2_final}), and (\ref{eq:join_fs}) in an iterative procedure. 

In our case, for the downstream solution, the speed goes like $v\propto z^{-2}$ in the downstream region as obtained from solving the wind equations. For such large radii $z > z_{\mathrm{s}}$, where $z_{\mathrm{s}} \sim 200$~kpc, the area also approximately becomes quasi-spherical $A(z) \propto z^{2}$. This implies that $\frac{\dd}{\dd z} (A(z) \,v(z)) \sim 0$ for the large downstream radii. Therefore, $G_2(z,p) = 0$, which implies zero adiabatic losses in the downstream region. In this limit, the downstream solution becomes analytical. Since the plasma is heated up to high temperatures by the shock, it is reasonable to assume that the downstream magnetic field as well as the diffusion coefficient $D_2(p)$ are spatially constant. The latter one has therefore only a $p$ dependence. We define

\begin{equation}
    \chi_2(z,p) \equiv \int_{z_{\mathrm{s}}}^z \frac{v(z')}{D_2(p)} dz' = \frac{v_2 \,z_{\mathrm{s}}}{D_2(p)} \left( 1 - \frac{z_{\mathrm{s}}}{z}\right).
\end{equation}

\noindent For the far boundary condition, we assume

\begin{equation}
    f_b << f_{\rm s} ~~~~~(\mathrm{outer~B.C}),
\label{eq:far_boundary}
\end{equation}

\noindent where $f_{\rm b}$ is the distribution function at the outer boundary of the bubble which we take to be $z_{\mathrm{b}} = 500$~kpc, which is approximately the order of three times the virial radius of Milky Way according to Global Astrometric Interferometer for Astrophysics (GAIA) and RAdial Velocity Experiment (RAVE) survey estimates \citep{2014:Piffl,2018:Monari}. This free-escape type outer boundary condition is justified because at the outer boundary of the bubble, the diffusion coefficient in the IGM is expected to increase by many orders of magnitude than the diffusion coefficient in the downstream region. We check that the exact location of the outer boundary does not impact our results as long as the outer boundary is sufficiently far away from the termination shock. Using the far boundary condition and Eq.~(\ref{eq:f2_final}), setting $f_2 = f_{\rm b}$ and using $f_{\rm b} << f_{\rm s}$ we get

\begin{equation}
    \phi_{\rm esc} = \frac{v_2 \,f_{\rm s}(p)\, A(z_{\mathrm{s}})}{1 - e^{-\chi_2(p, z_{\mathrm{b}})}}.
\label{eq:phi_esc}
\end{equation}

\noindent The distribution function at the shock, $f_{\rm s}$ is obtained by inserting $\phi_{\rm esc}$ from Eq.~(\ref{eq:phi_esc}) to Eq.~(\ref{eq:join_fs}) to obtain

\begin{equation}
\begin{split}
p \frac{\partial f_{\rm s}}{\partial p}  = - \frac{3 v_1}{v_1 - v_2} \left( \left(1 - \frac{u_2/u_1}{1 - e^{\chi_2(p', z_{\mathrm{b}})}} + \frac{G(p)}{v_1 A(z_{\mathrm{s}}) f_{\rm s}}\right)f_{\rm s} - \frac{I(z_{\mathrm{s}},p)}{v_1 A(z_{\mathrm{s}})}\right),
\label{eq:shock_finaleq} 
\end{split}
\end{equation}

\noindent where $G(p) = G_1(z_{\mathrm{s}},p)$. The solution of Eq.~(\ref{eq:shock_finaleq}) can be expressed in an implicit form as

\begin{equation}
\begin{split}
    f_{\rm s}(p) = \left[ \frac{s p^{-s}}{v_1 \,A(z_{\mathrm{s}})} \int_{p_0}^{p}  \frac{I(z_{\mathrm{s}},p')}{p'} p^{'s} e^{\Gamma_1(p')} e^{\Gamma_2(p')} dp' \right] \times  ~ e^{-\Gamma_1(p)} e^{-\Gamma_2(p)},
\label{eq:fs}
\end{split}
\end{equation}

\noindent where $s = \frac{3 v_1}{v_1 - v_2}$ and $p_0$ is the injection momentum of the central source. We also define the velocity jump factor across the shock to be $\sigma = \frac{v_1}{v_2}$. The exponential terms $\Gamma_1(p)$ and $\Gamma_2(p)$ are

\begin{equation}
\begin{split}
    \Gamma_1(p) = s \int_{p_0}^p \frac{G(p')}{v_1 A(z_{\mathrm{s}}) f_s(p')} \frac{dp'}{p'},  \\ \Gamma_2(p) = \frac{s}{\sigma} \int_{p_0}^p \frac{1}{e^{\chi_2(p',z_{\mathrm{b}})} - 1} \frac{dp'}{p'}.
\label{eq:Gam1_gam2}
\end{split}
\end{equation}

\noindent Eqs.~(\ref{eq:upstream_final}), (\ref{eq:f2_final}), (\ref{eq:fs}), and (\ref{eq:Gam1_gam2}) form a closed set of equations that can be solved using the iteration technique. The iteration algorithm is described as follows. One starts with a guess function for $f_1$ and $f_s$. A convenient guess function is the solution for a planar shock. Then in each subsequent $k^{th}$ step, $q^k(z,p)$, $G_1^k(z,p)$, and $\Gamma_1^k(p)$ are computed using Eqs.~(\ref{eq:q_def}), (\ref{eq:G_define}), and (\ref{eq:Gam1_gam2}) in sequence. $f^k_1(z,p)$ and $f^k_s(p)$ are then updated at each step using the values of $\Gamma_1^k(p)$, $q^k(z,p)$, and $G_1^k(z,p)$ using Eqs.~(\ref{eq:upstream_final}) and (\ref{eq:fs}). The iteration sequence is then repeated until the solution converges. The downstream escaping flux is then obtained from Eq.~(\ref{eq:phi_esc}) using $D\frac{\partial f}{\partial z} = \frac{\phi_{\rm esc}}{A(z_{\mathrm{b}})}$. 

\end{document}